%% file: main.tex
\documentclass[article]{elsarticle}

\usepackage{lineno,hyperref}
\usepackage{amsmath}
\usepackage{cool}
\usepackage{subcaption}
\usepackage{graphicx}
\usepackage{url}
\usepackage{float}
\usepackage{todonotes}
\usepackage{umoline}
\usepackage{epstopdf}
\usepackage{natbib}
\usepackage{nomencl}
\usepackage{notoccite}
\usepackage[flushleft]{threeparttable}

\journal{Journal of Fluids \& Structures}

\begin{document}

\begin{frontmatter}

\title{A hybrid Lagrangian-Eulerian flow solver applied to elastically mounted cylinders in tandem arrangement}

\author{George Papadakis\corref{mycorrespondingauthor}}
\ead{papis@fluid.mech.ntua.gr}

\author{Vasilis A. Riziotis \corref{}}
\ead{vasilis@fluid.mech.ntua.gr}
\author{Spyros G. Voutsinas\corref{}}
\ead{spyros@fluid.mech.ntua.gr}

\address{National Technical University of Athens (NTUA), \\
	9, Heroon Polytechniou Str, 15780, Athens, Greece}
\cortext[mycorrespondingauthor]{Corresponding author}

\begin{abstract}
The fluid structure interaction of cylinders in tandem arrangement is used as validation basis of a multi-domain Lagrangian-Eulerian hybrid flow solver. blackThe solver is built on a Lagrangian approximation of the entire flow-field using particles, in which grid based Eulerian flow solutions are overlaid, one for every solid body. The Eulerian grids are body fitted but of limited width and may overlap in cases of close proximity of the bodies. The Eulerian and Lagrangian solutions are strongly (implicitly) interconnected in two ways: The Lagrangian solution provides the conditions over the outer boundaries of the Eulerian grids, while the Eulerian solutions update the flow properties of the particles that are within their own domains. Also, implicit is the coupling of the solver with the structural dynamics in case the cylinders are elastically supported. The Lagrangian solver is based on the density-dilatation-vorticity-pressure formulation and makes use of the Particle Mesh method to obtain the flow velocity field while the Eulerian one on the density-velocity-pressure formulation. The hybrid solver is first validated in the case of an isolated rigid cylinder at $Re=100$. Then the case of a single elastically mounted cylinder at $Re=200$ is considered, followed by the case of two cylinders in tandem arrangement that are either rigid or elastically mounted. Good agreement with results produced with spectral element  and immersed boundary methods is found indicating the capabilities of the hybrid predictions. Also the flexibility of the method in handling complex multi-body fluid structure interaction problems is demonstrated by allowing grid-overlapping.  
\end{abstract}


\end{frontmatter}
\newcommand{\Epsilon}{\Pi} 
\newcommand{\normfig}{0.48\linewidth}
\newcommand{\smallfig}{0.32\linewidth}
\newcommand{\bigfig}{0.9\linewidth}

\section{Introduction}

Computational Fluid Dynamics (CFD) is a well established tool in modern engineering. In most cases, Eulerian CFD methods are used that have body-fitted grids covering the entire flow domain. This approach  leads to  complications when several bodies are involved that may also move independently in space, as in  Fluid-Structure-Interaction (FSI) problems. In order to handle such situations, the immersed boundary condition methodology has been formulated that disregards the boundaries and covers the entire flow field with a grid \cite{Borazjani2008}. Another option is to introduce deforming and overset grids that are body fitted and in which the same solution methodology is used throughout the domain. A third option which is here followed, is to make use of Lagrangian CFD, based on particle methods. 

In particular, the hybrid Eulerian-Lagrangian methodology presented in \cite{Papadakis2019} is extended to account for viscosity and almost incompressible low Mach flow conditions through preconditioning. {\color{black} Purely incompressible flows,could be considered by employing the artificial compressibility formulation for the Eulerian part \cite{Chorin1967}, which retains the hyperbolic character of the solver and consequently requires few modifications. However, the same does not hold for the Lagrangian part, in which the changes would have been substantial (the dilatation equation is redundant while the energy equation is substituted by a Poisson equation for the pressure).}
The Eulerian part is restricted within grids, separate, narrow and possibly overlapping  around the solid bodies while the Lagrangian part interlinks these zones and completes the formulation. In this way the necessary conditions on the solid boundaries and on the outer boundary of the Eulerian grids can be accurately satisfied. The same holds for the correct implementation of the far field flow behavior. Also,  in case the bodies are in motion, the coupling with an overall Lagrangian solution allows having an overlap of the Eulerian grids that changes in time, which is a very useful feature. {\color{black} Essentially, the hybrid method proposed resembles to the overset grids method (employed in a strictly Eulerian framework) except that the Lagrangian solver provides the background solution.} In order to {\color{black} demonstrate} the capabilities of the above hybrid method, cases with FSI have been chosen, in which a strong (kinematic and dynamic) coupling between the flow and the dynamic (structural) equations is employed . 

Among the flow problems of this kind, the most challenging examples concern Vortex Induced Vibration (VIV) problems. While many fluid-structure interaction problems can be addressed with lower fidelity models (inviscid assumption and potential flow solvers), VIVs usually involve highly separated flows and thus it becomes necessary to apply high fidelity viscous flow solvers. A heavily scrutinized case in the literature is that of an elastically mounted rigid cylinder in uniform inflow. The cylinder vibrates due to the periodic loading induced by the vortices shed in the wake, while the resulting elastic motion  in turn affects the formation of the wake. As suggested in \cite{govardhan2000modes} the mass and damping parameters of the vibrating cylinder affect its response and thereby dictate the formation of the wake. Different wake patterns (modes) have been identified, depending on the values of the defining parameters. In particular there is the '2S' mode where 2 single vortices are shed per period and the '2P' mode where two pairs of vortices are shed per period. 
 
 As shown in the experimental study of  \cite{khalak1999motions} performed at Reynolds Number  $Re=3500-10000$, there are three different response branches for low mass damping values: the initial branch, the upper branch and the lower branch. In \cite{Leontini2006},a 2D spectral element method  was used to identify the response  of an isolated cylinder at a lower Re number and was shown that two different regimes of synchronization exist. 
 
In\cite{Borazjani2008},\cite{Borazjani2009}, the authors used an immersed boundary method to address the problem of two cylinders in tandem arrangement. They considered one and two degrees of motion (transverse and longitudinal) in laminar flow conditions ($Re=200$). The cylinders were in close proximity and the "gap flow" was identified. As the two cylinders vibrate with respect to each other the transverse offset between them allows the separated flow from the upstream cylinder to pass through the gap. In \cite{Griffith2017b} the authors employed a sharp interface immersed method to study the VIV of a single isolated cylinder while in \cite{Griffith2017a}  a thorough investigation was carried out using the same methodology to study the VIV of two cylinders in tandem and staggered arrangements. In both of the above works the flow past the cylinders was considered laminar at $Re=200$. So the evaluation of the hybrid method is mainly done in comparison to these results. 
 
Flow around a circular cylinder has been also investigated by vortex particles methods. The first work which introduced the vortex method as a concept, was that of Chorin  \cite{chorin1973discretization}. A thorough and complete consideration of the transient flow development of an impulsively starting cylinder was published in \cite{koumoutsakos1995high} in which the no slip condition was satisfied by generating vortices close to the solid boundaries, while diffusion was taken care by the Particle Exchange Method introduced in \cite{degond1989weighted}.In \cite{slaouti1992flow} two cylinders in tandem and staggered arrangement where studied using the particle-in cell method. Recently Gillis et. al. \cite{Gillis2019} developed an immersed-interface vortex particle method (IIVPM) and investigated the impulsively starting  flow past a cylinder at three different Re namely, $Re=550$, $Re=3000$ and  $Re=40000$. The authors managed to accurately capture the wall boundary conditions by using an underlying uniform grid and  introducing discontinuities on the velocity field at the intersection of the grid with the cylinder boundary.
 

 
The  purpose of this work is twofold. On one hand, to verify that the hybrid methodology developed in \cite{Papadakis2019} can accurately simulate low Mach, laminar flows  and on the other hand to demonstrate the handling capabilities of the method in complex multi-body flows. Regarding the verification part, laminar flow around a circular cylinder is very attractive not only because it has been widely studied in the literature but also because of the complex vortex dynamics involved. Finally, regarding multi-body application, vibrating cylinders in close proximity reveal the appealing features of the methodology. More specifically, the boundary layer is resolved using a body fitted grid while the relative motion of each component can be treated easily without relying on techniques such as overset or deforming grids. Additionally, compared to traditional immersed boundary methodologies, simulations  at high Reynolds numbers can be easily treated since the near wall region is handled by a body-fitted grid CFD methodology.

Summarizing, in this work the method presented in \cite{Papadakis2019} is now employed for flows in the incompressible regime. Also, it is extended to account for  viscosity with focus on laminar separated flows around circular cylinders. The hybrid methodology is enhanced to account for rigid body dynamics and the method is applied in the case of vibrating cylinders. Finally, the capability of the method to handle multi-body configurations which can move independently  is exposed by considering the flow around the elastically-mounted circular cylinders.

The  paper is structured as follows: In Section \ref{sec:numframe} the proposed numerical methodology is described with focus on the hybrid algorithm. In Section \ref{sec:numres} the hybrid methodology is employed to obtain numerical results in laminar flow conditions. Initially, the flow around a stationary cylinder at Reynolds (Re) 100 is considered for validation purposes. Afterwards the VIV problem of an isolate cylinder is considered and compared with results available in the literature. {\color{black} Moreover, the case of two stationary/vibrating cylinders is investigated and results are compared with those published in \cite{Borazjani2009} and \cite{Griffith2017a} while in Section \ref{sec:conc} the basic conclusions are summarized. Finally, an analysis of the error of the solution near $S_E$ can be found in the \ref{sec:appendix} where comparison is made for both the L and E solvers at $S_E$ against an analytical solution.}

\section{Description of the hybrid method}\label{sec:numframe}

The present hybrid method couples an Eulerian finite volume solver ("E") with a Lagrangian one ("L"). The description starts with the Lagrangian part to which the Eulerian one is overlaid and correctly provides the effect of the solid boundaries. 

\subsection{The Lagrangian solver} \label{sec:Lagrangian}
The formulation assumes that the flow is approximated by a set of (material) particles that cover the entire flow field. The particles are associated to a volume $V_p$ and carry volume integrals of  $\rho$ (density), $\theta$ (dilatation), $\vec{\omega}=\omega \vec k$ (vorticity), $p$ (pressure) that are denoted as $M_p$ (mass), $\Theta_p, \vec{\Omega}_p, \Pi_p$. In Lagrangian (material) coordinates, the flow equations for laminar conditions are defined with respect to the particle positions $\vec{Z}_p$  and  take the form (see for example \cite{Eldredge2002b}):
\begin{align}
	\label{eq:Lagsyspos}
	&\D{\vec{Z}_p}{t}=\vec{U}_p \\
	&\D{V_p}{t}=V_p \, \theta_p \\
	&\D{M_p}{t}=0 \\
	&\D{\vec{\Omega}}{t} =V_p\left((\vec{\omega} \cdot \nabla) \vec{U} -\frac{1}{\rho^2} \nabla  \rho \times \nabla(-p) - \nu \nabla^2 \vec{\omega}\right)_p\\
	&\D{\Theta}{t}=V_p\left(2 \Vert \nabla\vec{U}\Vert  -\nabla \cdot \frac{\nabla p }{\rho} + \nu \frac{4}{3} \nabla ^2 \theta\right)_p\\
	\label{eq:Lagsyspre}
	&\D{\Pi_p}{t} = V_p\left( (1-\gamma) p \theta + (\gamma-1) \left( \nabla\cdot\left[(\nabla \cdot \overleftrightarrow{\sigma}) \cdot\vec{U}\right)-\vec{U}\cdot \nabla \cdot (\nabla \cdot \overleftrightarrow{\sigma}) \right) \right)_p
\end{align}
 
\noindent In the above, $d/dt$ denotes the material time derivative; $ (\cdot)_p $ indicates evaluation at the position of particle p; $\nabla \cdot \overleftrightarrow{\sigma} =\mu\left(\frac{4}{3}\nabla\theta - \nabla \times \vec{\omega} \right)$ denotes the divergence  of the viscous stress tensor; $\nu=\mu\/\rho$ is the kinematic viscosity which here is assumed constant. These equations are integrated in time using a 4th order Runge-Kutta scheme. In doing so, the derivatives that appear in the Right Hand Side (RHS)  as well as the velocity must be evaluated at the particle positions \cite{Eldredge2002}.

In the above formulation, the flow velocity $\vec{U}$ is a derived quantity, obtained via Helmholtz's decomposition,

\begin{equation}
	\vec{U}=\vec{U}_\infty + \nabla \phi + \nabla \times \vec\psi, \quad  \nabla^2\phi=\theta,\quad \nabla^2 \vec{\psi}=-\vec{{\omega}} 
	\label{eq:Hdec}
\end{equation}
where $\phi$ and $\vec{\psi}=\psi \vec{k}$ are the scalar and vector potentials that correspond to the rot and div-free parts of $\vec{U}$ \cite{Batchelor}, while $\vec{U}_\infty$ is the constant velocity at infinity. For a flow defined in $D$ which is exterior to its boundary $S$ and $\phi, \psi$ admit the following integral representations:
\begin{eqnarray}
	\label{eq:frep}
	\phi(\vec{x}) = \int_{{D}} \theta(\vec{y}) \,G(\vec{r}) \,dD(\vec{y}) 
	+  \int_{S} u_n(\vec y) \, G(\vec{r})dS(\vec{y}) \nonumber \\
	\psi(\vec{x}) = \int_{{D}} -\omega(\vec{y}) \,G(\vec{r}) \,dD(\vec{y}) 
	+  \int_{S} u_{\tau} (\vec{y}) \, G(\vec{r}) dS(\vec{y})
\end{eqnarray}
where $\vec{r}=\vec{x}-\vec{{y}}$ and $G$ is the Green's function for the Laplace operator. In the above expression, $u_n, \, u_\tau$  
denote the normal and tangential disturbance velocity components on $S$. 


The convolutions involved in (\ref{eq:frep}) are expensive operations, if directly processed. An efficient way of reducing this cost is to use the Particle Mesh Method (PMM) {\color{black}\cite{Chatelain2016a},\cite{Parmentier2018},\cite{Gillis2019}\cite{ploumhans2002vortex},\cite{caprace2021flups}} which also facilitates the evaluation of the needed velocity derivatives . In this connection, the flow properties $ \rho, \theta, \vec \omega, p$ (collectively denoted $q_p$) that particles carry, are projected onto a uniform Cartesian grid: 
\begin{equation}
	q(\vec{x}_{I})=q_{I} = \operatorname{Proj}_\text{I}(q_p ; V_p) \equiv
	\frac{\sum_{p}^{} q_p {V}_p W(\vec{x}_{I}-\vec{Z}_p)}{h^2} 
	\label{eq:projop}
\end{equation}
 where $\vec{x}_{I}$ denotes the position of the $I=\{i,j\}$ grid node which also appears as a subscript to "Proj" in order to indicate the point at which the operation refers to. Furthermore, $h$ denotes the PM grid spacing which is the same in both directions and $W$ is the projection function, defined with respect to $\vec{r}=(r_x,r_y)=\vec{x}_{I}-\vec{Z}_p$ as follows:
 \begin{equation*}
 	W(\vec{r})=W_1 (r_x/h) W_1(r_y/h)
 \end{equation*}
 In the above, $W_1$ is the 1D interpolation function that is utilized. In the present work, the ${M_4}'$ function is used (for other options see {\color{black}\cite{monaghan2005smoothed})}.
\begin{equation}
    {M_4}'=\begin{cases}
            0 &\text{ if } |x|>0\\
            \frac{1}{2} (2-|x|)^2 (1-|x|) & \text{ if } 1 \le|x|\le 2\\
             1- \frac{5|x|^2}{2} + 3 \frac {3|x|^3}{2} & \text{ if } |x| \le 1
           \end{cases}
\end{equation}

Having the RHS of the Poisson eqs in (\ref{eq:Hdec}), they are solved using Fast Poisson solvers {\color{black} \cite{caprace2021flups}}. Next, any needed derivative as well as any compound term appearing in the RHS of (\ref{eq:Lagsyspos}-\ref{eq:Lagsyspre} ) are calculated on the grid using finite differences. The final step consists of back interpolating all these quantities from the PM grid to the particle positions. The same function $W$ is also used in this operation.  

\begin{figure}[H]
	\centering
	\begin{minipage}[t]{\linewidth}
		\includegraphics[width=\textwidth]{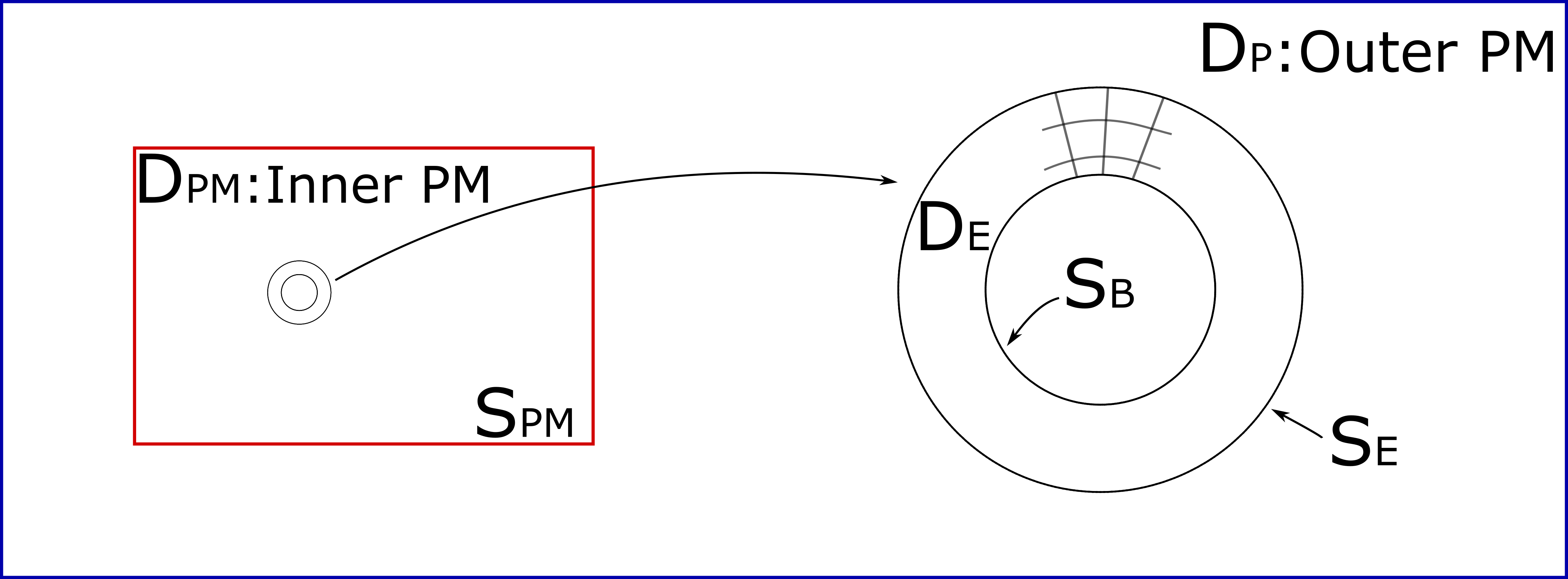}
	\end{minipage}
	\caption{\label{fig:hyb-setup} The E-domain $D_E$ contains the solid boundary $S_B$. The L-domain $D_\infty$ includes $D_{PM}$ and $D_P$, of which $D_{PM}$ denotes the fixed part of $D_\infty$ with fine mesh spacing while $D_P$ denotes the far-field which grows as the flow evolves and has a coarser resolution.}
\end{figure}

 The aim is to define two solutions \cite{Papadakis2019} the E-solution that would be valid close to solid boundaries and the L-solution that would be valid everywhere except within a narrow region around the solid boundaries. The case of one body is shown in Fig. \ref{fig:hyb-setup}, where $D_E$ denotes the domain covered by the Eulerian grid and limited by the solid boundary $S_B$ and the outer boundary $S_E$. $D_E$ is overlaid on $D_{PM}$ which is covered by the L-solver. 
 {\color{black} Resolving the wall boundary  from a Lagrangian numerical framework is not trivial, since, either a very large number of particles are needed (\cite{CottetKoumou}) or an immersed boundary approach must be employed\cite{Gillis2019}. In the presented work this is taken care by coupling the L-solver with  the E-solver}.    

The specific approach is here applied to more than one distinct bodies that may move independently the one from the other. In this case, every body has its own E-solution while all share the same L-solution. So the communication between the E-domains is done indirectly through the common L-solution and therefore the E-domains may overlap and their overlapping may change in time.  In such a case, projection and interpolation operations assure that the update corresponds to a conservative averaging. These features are exploited in simulating the case of two independently moving cylinders that have separate flexible supports. The only limitation is that the Eulerian outflow boundary ($S_{E}$) must keep some safety margin from any solid boundary ($S_{B}$) - this margin depends on the stencil of the interpolation scheme that is used. 

\subsection{The Eulerian solver}
The in-house developed Eulerian solver MaPFlow \cite{Papadakis2014e} is used. MaPFlow solves the  compressible equations in $\rho , \rho\vec{u}, \rho e$ formulation using the finite volume method on unstructured grids.{\color{black} The solver is $2^{nd}$ order accurate in space and time; uses the Roe approximate Riemann solver for the convective fluxes and is equipped with Low Mach preconditioning.}

On every $S_B$ the no-slip condition is applied along with zero Neumann conditions for the pressure and density. On every $S_E$ the complete flow state is specified as provided by the Lagrangian solver. This is the one part of the two-way coupling between the E-parts and the L-solution. The other part corrects the L-solution. Both are detailed next.

\subsection{The two-way coupling procedure}\label{sec:coupling}

\subsubsection{From L-to-E: Provide the Boundary conditions on $S_E$}
\noindent The L-solution, as defined at the PM grid nodes, is interpolated at the ghost nodes of the E-grid situated outside $S_E$. This allows determining the fluxes through $S_E$ from the Riemann invariants that are associated to the flow states on the two sides of $S_E$. 

\subsubsection{From E-to-L: Update the particle flow information in $D_E$}
\begin{figure}[H]
	\begin{minipage}[t]{.3\linewidth}
		\includegraphics[width=0.9\textwidth]{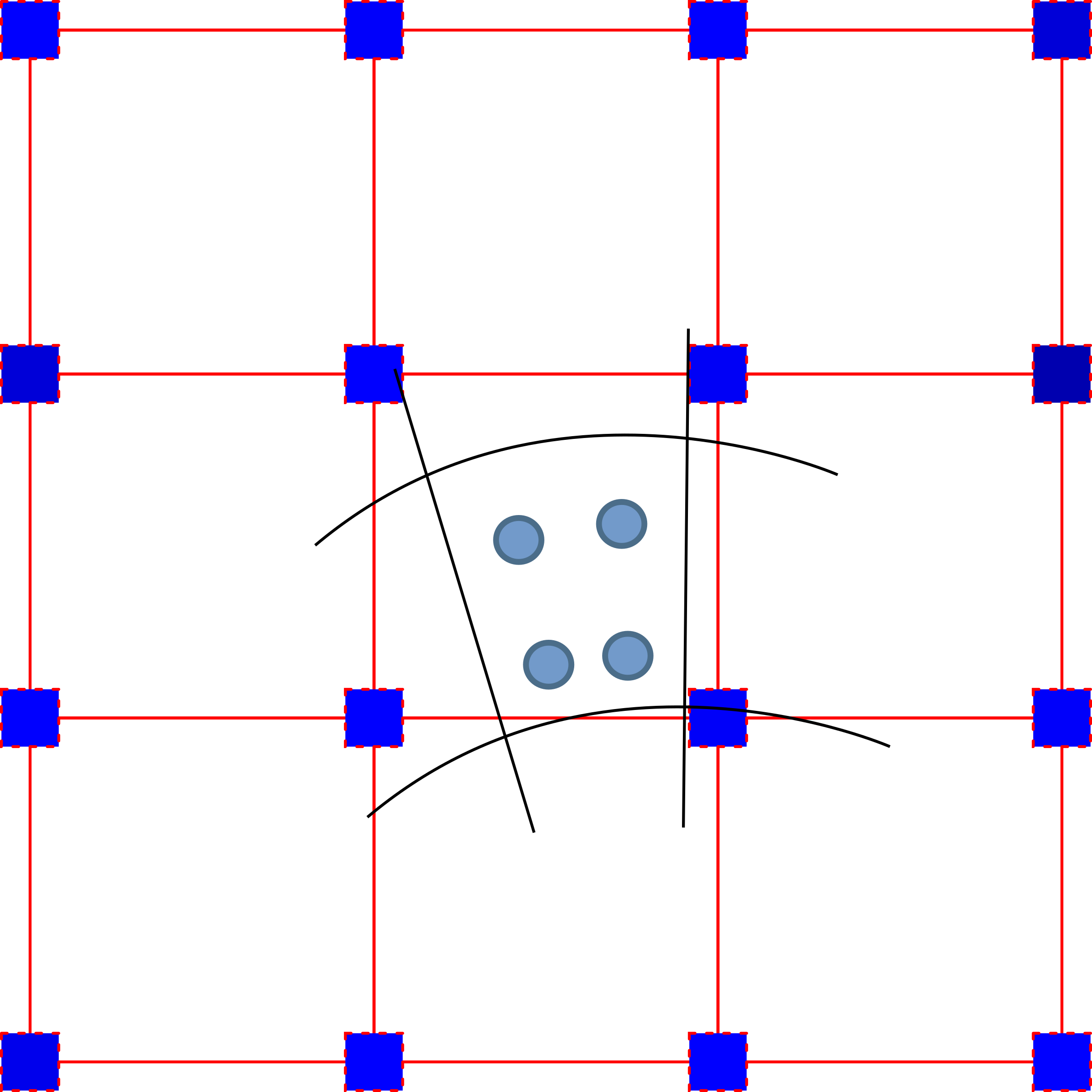}
	\end{minipage}
	\hfil
	\begin{minipage}[t]{.6\linewidth}
		\includegraphics[width=\textwidth]{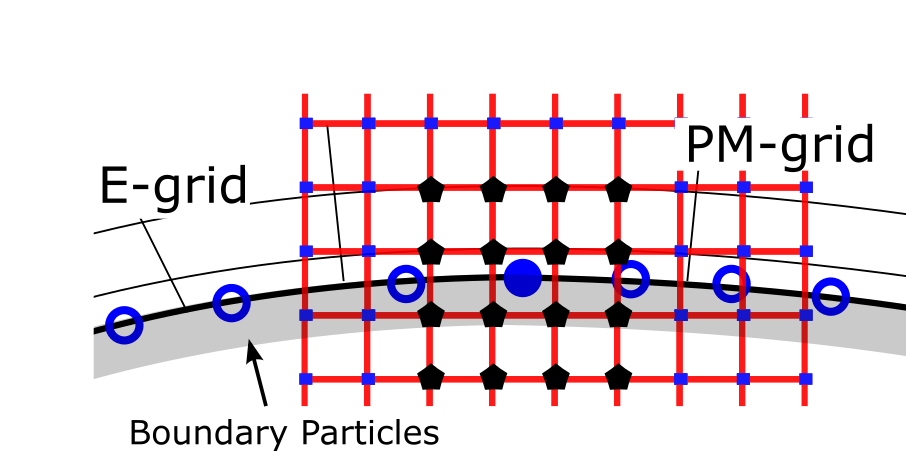}
	\end{minipage}
	\caption{\label{fig:cfdcreatedpar} Spatial distribution of the E- particles. 
		{ Left: One E-cell is shown within a 4x4 stencil of the PM grid corresponding to the support of the $M4'$ projection function. There are four $P_E$ particles in the E-cell marked as blue circles. Right: A close-up to the solid boundary is shown. On $S_B$, surface particles are shown as blue open circles. The middle surface particle is embedded in the $M4'$ stencil activated for its projection.} }
\end{figure}

\noindent The correction of  the L-solution is formulated by transforming the E-solution into particles that replace the existing (Lagrangian) ones within $D_E$. The particles $P_E$ that are generated from the E-solution, are placed at regular positions within every E-grid cell and carry $(\rho,\, \theta,\, \vec{\omega}, \,p)_{P_E}$ (Fig. \ref{fig:cfdcreatedpar}\,(left)). Since the E-solver is cell centered, $\theta$ and $\vec{\omega}$ are first calculated at the cell centers using the Green-Gauss formula and then, together with $\rho$ and $p$, they are interpolated at the $P_E$ particle positions. For this operation iso-parametric finite element approximations are used which also determine the associated volumes. The number of particles per cell depends on the cell size with respect to the PM spacing. In order to assure full space coverage and good particle density, more than one E-particle should be contained in every PM cell (Fig. \ref{fig:cfdcreatedpar}\,(left)). Similarly, surface particles that correspond to the surface terms in (\ref{eq:frep}) are also generated as shown in Fig. \ref{fig:cfdcreatedpar} right (for further details the reader is referred to \cite{Papadakis2014e,Papadakis2019}).

\subsection{Remeshing}

A well known problem in particle methods concerns the gradual loss of regularity in time. Sparsity of particles can result in loss of accuracy while high particle concentration can lead to numerical instabilities \cite{CottetKoumou}. This is analogous to grid irregularity or stretching in standard CFD solvers and is corrected with grid refinement. A similar procedure is also needed and applied in particle methods. Now the role of the grid is taken over by the particles themselves and grid refinement corresponds to the so called {\it re-meshing}. It consists of interpolating the known flow properties from the particle positions to regularly distributed ones. In the present implementation re-meshing is carried out at the end of every time step and makes use of the same interpolation function $W$ that has been associated to the projection and interpolation operators.

\subsection {The Fluid Structure Interaction (FSI) option}

\begin{figure}[H]
	\centering
	\begin{minipage}{0.5\linewidth}
		\includegraphics[width=\textwidth]{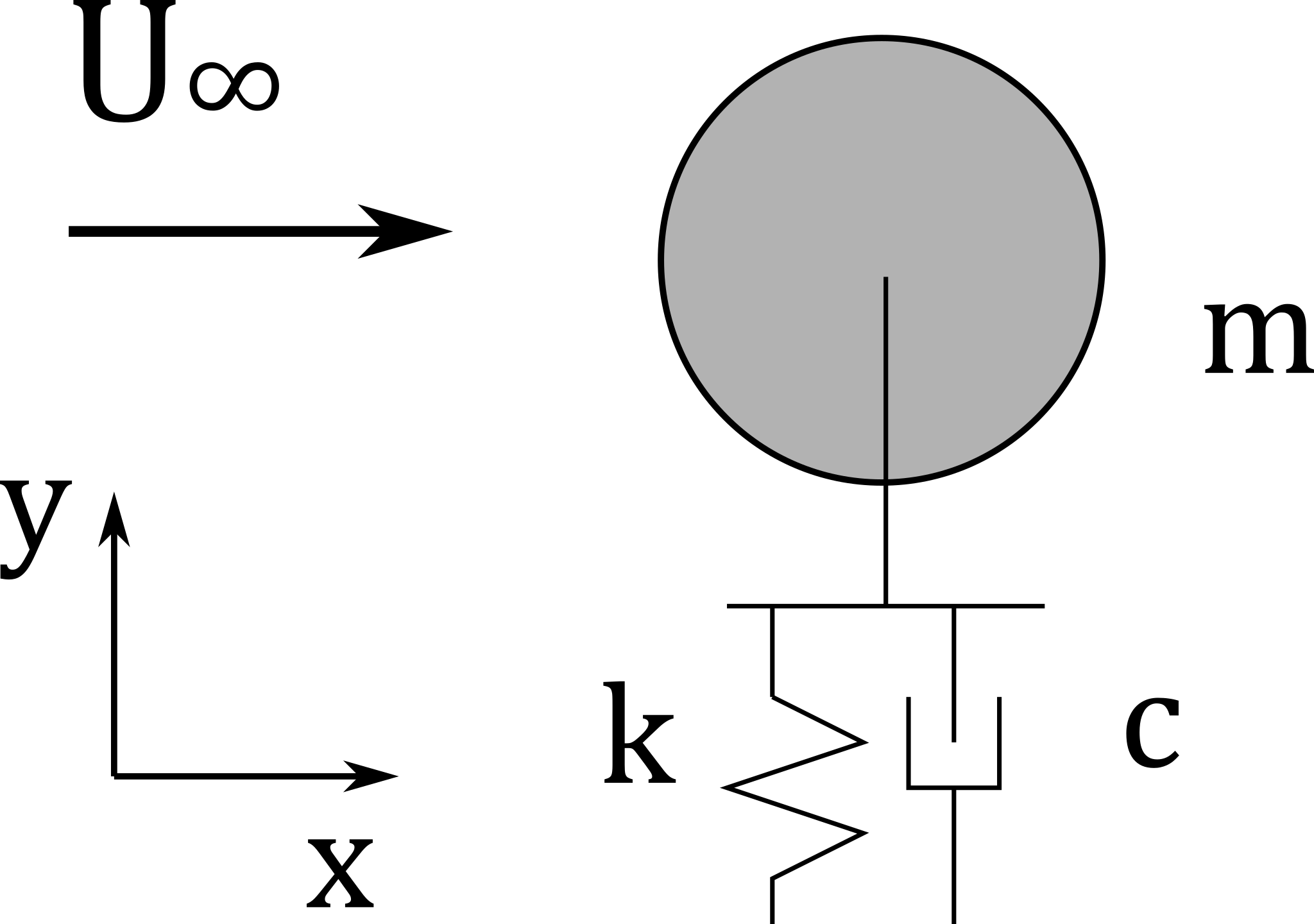}
	\end{minipage}
	\caption{The set-up of an elastically mounted cylinder. $m, c, k$ denote the mass, damping and stiffness of the system that are all concentrated at the center of the cylinder. 
		\label{fig:cyl1}}
\end{figure}

In case the solid bodies have flexible supports, they will vibrate in response to the loading due to the incoming flow. The dynamic (structural) equations are coupled with those of the flow through the boundary conditions. In this context, the flow solver provides the surface loading on $S_B$ while the structural one feeds back the surface velocity into the flow boundary condition. In the present work, the bodies are rigid cylinders that can only move in the direction perpendicular to the direction of the free stream velocity. This motion is either imposed, as in the rigid case, or is the result of elastic deflection due to a spring-damper support. In the latter case the structural equations also include inertia corresponding to a concentrated mass at the center of the cylinder (see Figure \ref{fig:cyl1}). 

Let $y, \dot{y}, \ddot{y}$ denote the displacement, the velocity and the acceleration respectively of the center of the cylinder. Then, 

\begin{equation}
m \ddot{y} + c \dot{y} + k y = {F}_y
\label{eq:rbdbasic}
\end{equation}

\noindent where $F_y$ denotes the y-component of the integral aerodynamic force due to pressure and shear stresses over the cylinder:
\begin{equation}
\vec{F}_{aero}= \begin{bmatrix} F_x \\F_y  \end{bmatrix}=\int_{cyl} (-p + \overleftrightarrow \sigma) \cdot \vec {n} dS 
\label{eq:faero}
\end{equation}

\begin{figure}[H]
	\centering
	\includegraphics[width=\textwidth]{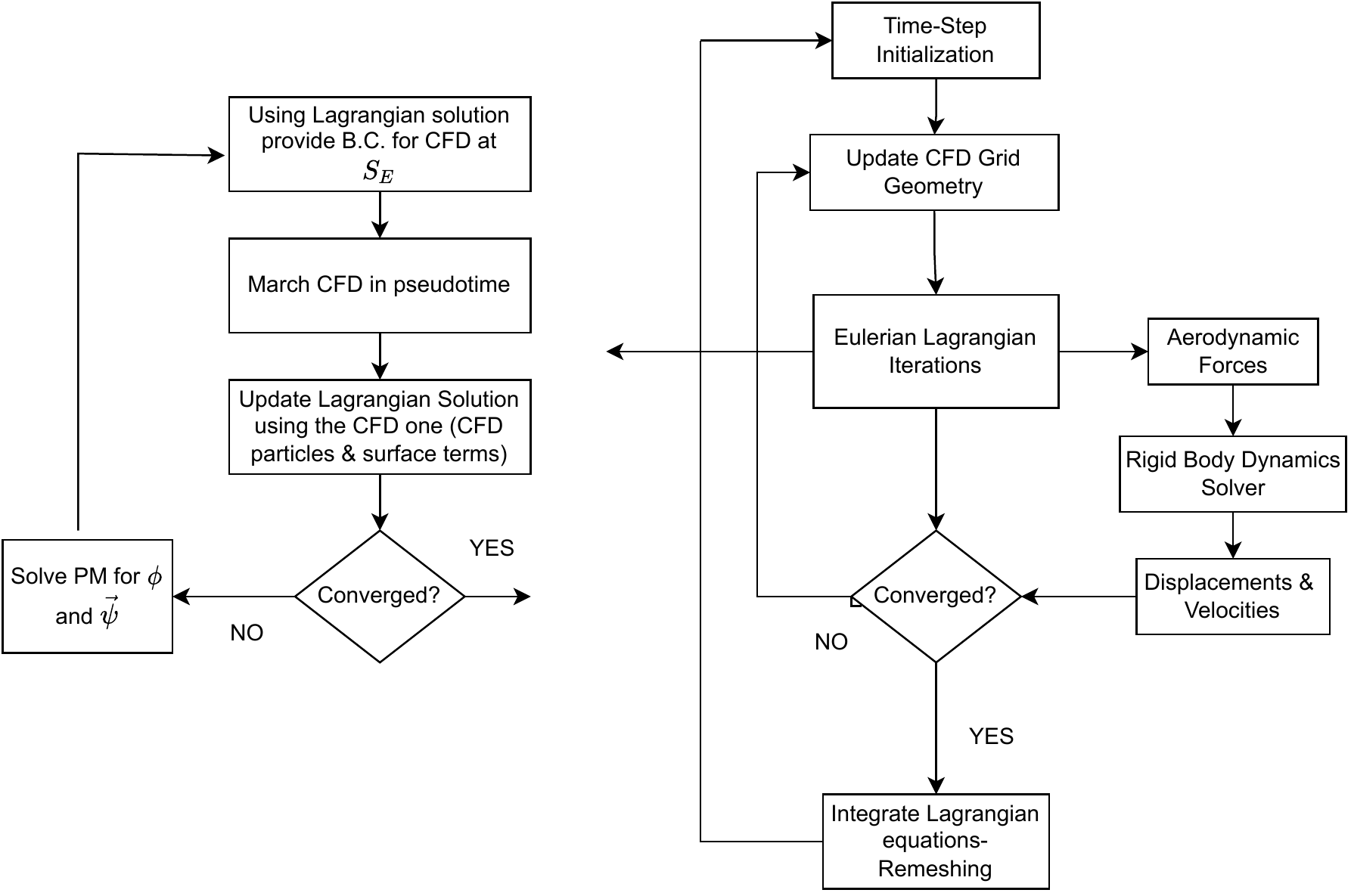}
	\caption{Flow chart of the strong coupling between the flow solver and the structural dynamics\label{fig:rbd}}
\end{figure}

The dynamic equations are integrated in time by means of the Newmark $\beta$ method \cite{Manolas2014}. However, due to the non-linear dependence of the driving force on the structural kinematics, in every time step the flow and dynamic equations must be solved iteratively. The algorithm converges when the difference in the calculated body acceleration between two successive iterations drops below $10e^{-09}$. This typically requires 4-5 iterations between the two solvers. A flow chart summarizing the aforementioned procedure is given in Figure \ref{fig:rbd}.  {\color{black} Regarding the Lagrangian-Eulerian iterations convergence is accomplished when the L2-norm residual for the CFD solution drops below 1e-09.} 

\section{Numerical Results} \label{sec:numres}
 The simulations that are next presented concern one and two circular cylinders of infinite length that are either still (rigid) or move in the direction perpendicular to the free stream velocity. The first case concerns an isolated cylinder at $Re=100$, which has been studied widely in the literature. Then, the case of an elastically supported cylinder is considered and predictions are compared to those obtained with a  spectral element method. The third case concerns two cylinders in tandem for which comparisons with predictions based on the immersed boundary condition method are provided. First the cylinders have fixed positions and then they are elastically mounted on separate supports. Finally, the convergence characteristics of the method in space and time  are presented in the appendix.



\subsection{Isolated Cylinder at $Re=100$}

\begin{figure}[H]
	\begin{minipage}{0.2\linewidth}
	    \includegraphics[width=\textwidth]{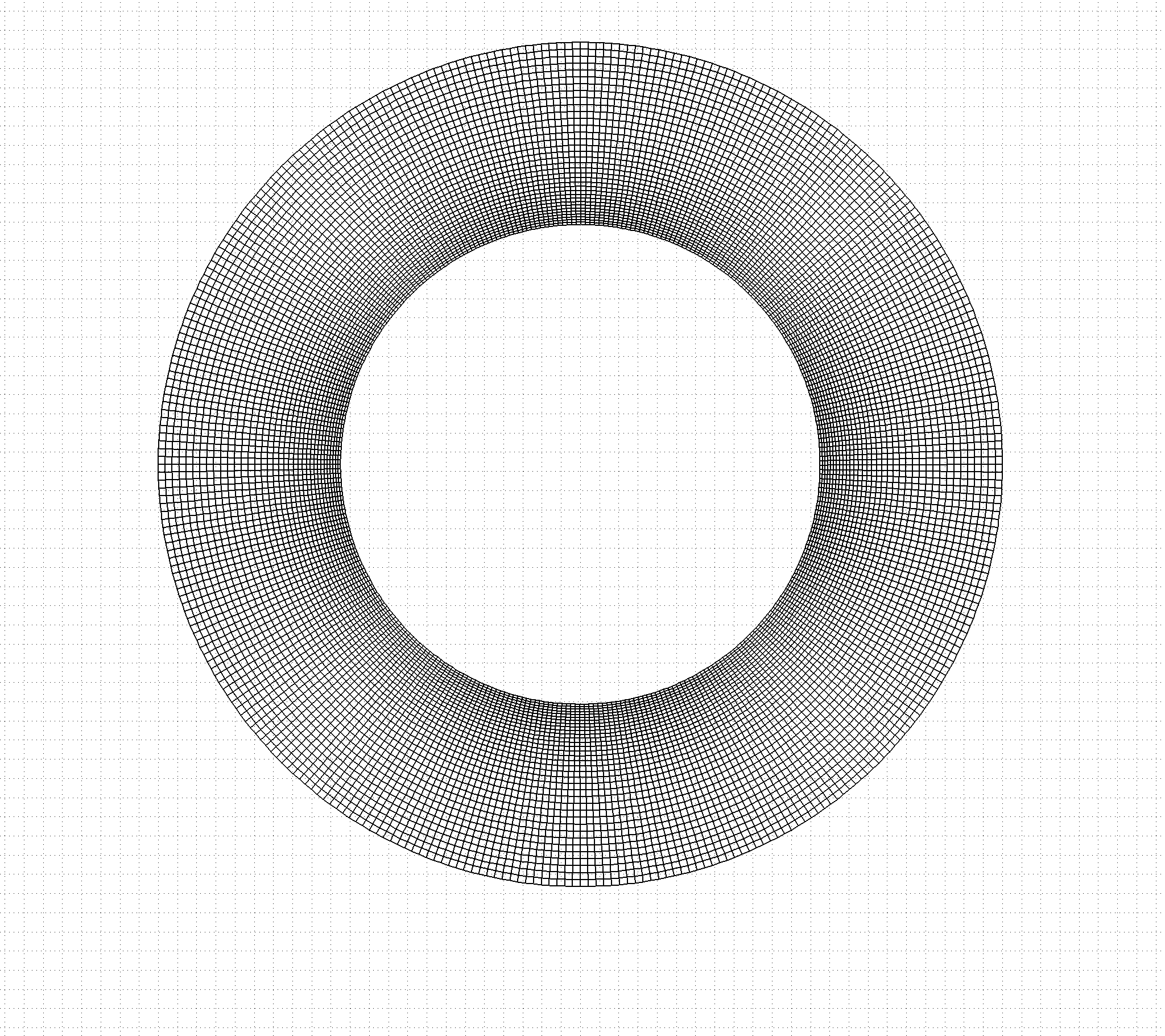}
       \end{minipage}
	\begin{minipage}{0.8\linewidth}
	    \includegraphics[width=\textwidth]{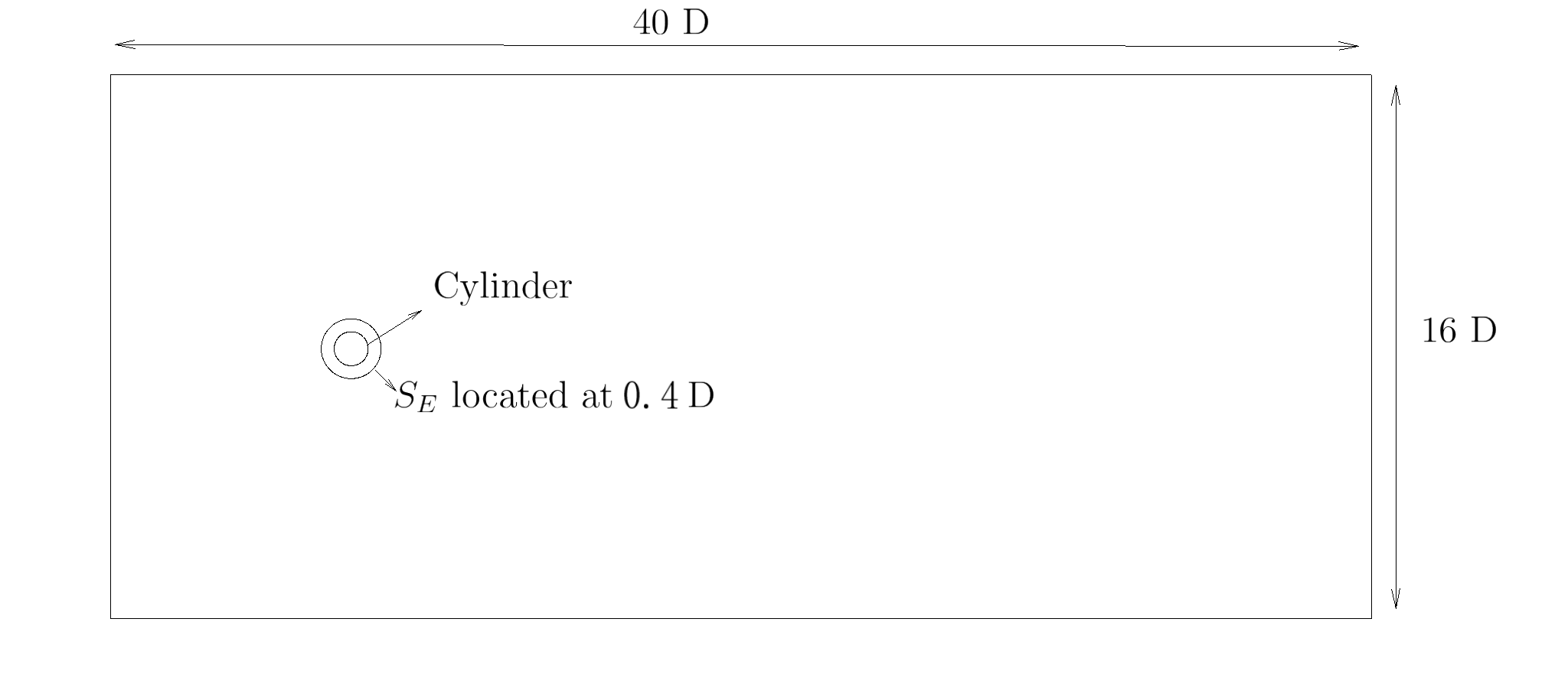}
       \end{minipage}
       \caption{\label{fig:cylsetup} The computational setup. The PM grid (left) extends 40 diameters in the streamwise direction and 16 diameters in the lateral one. The CFD grid (right) extends 0.4  diameters.}
\end{figure}

This case has been extensively studied both numerically \cite{Sharman2005,Park1998,Stalberg2006,Qu2013,Posdziech2007} and  
experimentally \cite{Williamson1996}. At $Re=100$, the flow is dominated by laminar vortex shedding which renders the flow unsteady.  Spectral element \cite{Posdziech2007}, as well as high order schemes \cite{Stalberg2006} have been applied in order to obtain reference predictions. In this respect, the resolution of the grid as well as the extent of the computational domain are important as shown in \cite{Posdziech2007} and \cite{Qu2013}. For the hybrid method, $D_E$ is chosen to cover a span of $\approx$ 0.4 diameters around $S_B$ while the PM mesh extends up to 40D (Figure \ref{fig:cylsetup}).

The CFD grid in the near-cylinder region consists $\approx 19000$ elements with 320 nodes around the cylinder while the centers of the first row of cells is located at 0.01 from the cylinder boundary. The CFD grid is considered adequate for this Reynolds number and thus grid independence is carried out with respect to PM spacing ($h$). Three different resolutions are tested $h=0.02,0.04 \text{ and } 0.06$ with a step of $0.02$. In all computations  a time step of $\Delta t=0.004$ is used. 

\begin{table}[H]
    \centering
    \small{
    \caption{Comparison of the present methodology using three PM spacing with various method from the literature. FV : Finite Volume, FD: Finite Difference, SE: Spectral Element, EXP: Measurements}
    \label{tab:steadycyl}
    \begin{tabular}{l|l|c|c|c|c|c|c|c|c}
	&Method &Strouhal  & $\overline{Cd}$ & $\overline{Cdp}$ & $\overline{Cdv}$ & $Cd_{std}$ & $Cdp_{std}$ & $Cdv_{std}$ & $Cl_{std}$ \\
    	\hline		
    	Current  h=0.06 & &0.1628 &1.3107 &0.9730 & 0.337 &0.0074 &0.0067 &0.0007 &0.241 \\
    	Current h=0.04 & &0.1648 &1.3163 &0.9775 & 0.338 &0.0074 &0.0067 &0.0007 &0.242 \\
        Current	h=0.02 & &0.1652 &1.3137 &0.9754 & 0.338 &0.0080 &0.0072 &0.0008 &0.239 \\
        \hline	
    	Park et al \cite{Park1998}           & FV & 0.165  & 1.33  &0.99  & 0.34  & 0.0064 & 0.0058& 0.0007    &0.23 \\
		Sharman et al\cite{Sharman2005}      & FV & 0.164  & 1.33  &0.99  & 0.34  & 0.0064 & 0.0058& 0.0007    &0.23 \\
		Posdziech et al \cite{Posdziech2007} & SE & 0.1633 & 1.312 & -    &  -    &   -    &  -    &  -       & -\\
   	    Stalberg et al \cite{Stalberg2006}   & FD & 0.166  & 1.32  &0.972 & 0.348 &   -    &  -    &  -       &0.23 \\
		Qu et al \cite{Qu2013}               & FV & 0.1648 & 1.319 &0.984 & 0.335 &   -    &  -    &  -       &0.225\\
        Williamson  \cite{Williamson1996}    & EXP& 0.164   & - &- & - &   -    &  -    &  -       &-
    \end{tabular} 
}
\end{table}

In Table \ref{tab:steadycyl} results from various simulations are compared in terms of lift and drag mean values and standard deviations while the experimental Strouhal number is also provided. Even though essentially different methods are compared, there is fair agreement. It also follows from this comparison, that all three PM grids provide consistent results although the coarser one ($h=0.06$) predicts lower  Strouhal number. It is noted that the difference between $h=0.04$ and $h=0.02$ in the prediction of the Strouhal number is less than 0.25\%. 

\subsection{Single cylinder flow-induced vibration}
The second case concerns, the flow induced vibration of an elastically mounted cylinder. Results of the hybrid method are compared to those of the spectral element  method, published in \cite{Leontini2006}. Prior to that grid and time independence is examined with respect to the PM grid spacing $h$ and the time-step $dt$. Three different spacings are considered: $h=0.02,0.04,0.06$, and three time-steps: $dt=0.002,0.004,0.008$.

For the sake of the comparisons made in the next section, the non-dimensionalisation of $U_\infty$, $m,c,k$ used in \cite{Leontini2006,Griffith2017b}, is adopted. The reduced velocity $U^*$ and the mass ratio $m^*$ are defined as follows 

\begin{equation}
	U^* = \frac{U_\infty}{f_n D}, \hspace{0.5cm} m^* = \frac{m}{m_f} , 
	\label{eq:reduced}
\end{equation}

\noindent where $m_f=4\rho \pi D ^2$ is the mass of equivalent volume of fluid and $f_n= \frac{1}{2\pi}\sqrt{\frac{k}{m+m_f}}$ denotes the natural frequency including the effect due to added mass (see \cite{Griffith2017b},\cite{khalak1999motions}. Finally, by also introducing the critical damping ratio $\xi$, the dynamic equation takes the form: 

\begin{equation}
	\ddot{y}  + 4 \pi f_n \xi \dot{y} + \left( 1 + \frac{1}{m^*} \right) (2 \pi  f_n)^2 y = \frac{F_y}{m}
	\label{eq:1cylrbdf}
\end{equation}



\noindent By changing the reduced velocity $U^*$, the natural frequency and the spring constant  also change. For the results presented here the mass ratio was set to $m^*=1$. 

\begin{figure}[H]
\begin{minipage}{0.5\textwidth}
	\includegraphics[width=\linewidth]{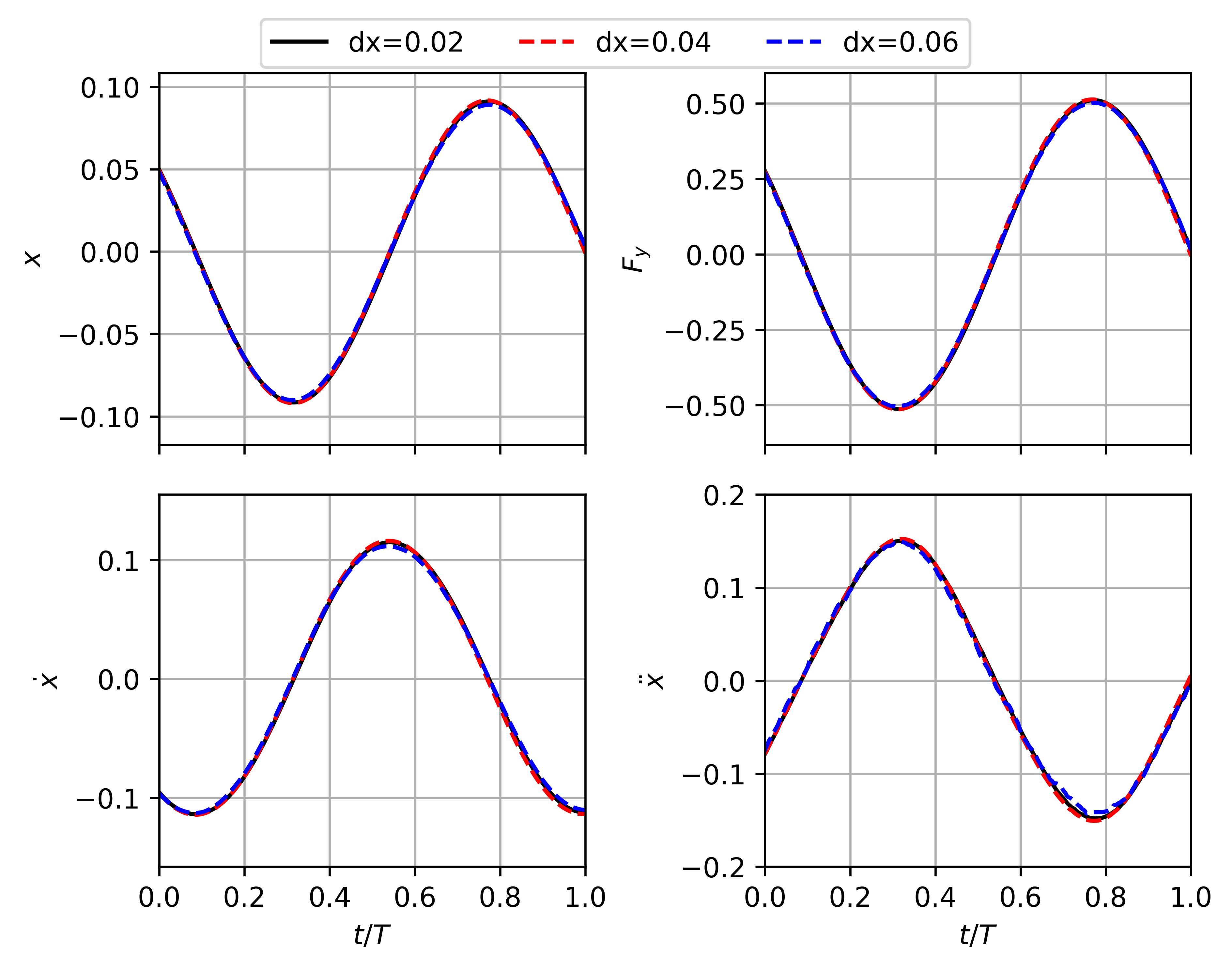}
\end{minipage}
\hfil
\begin{minipage}{0.5\textwidth}
	\includegraphics[width=\linewidth]{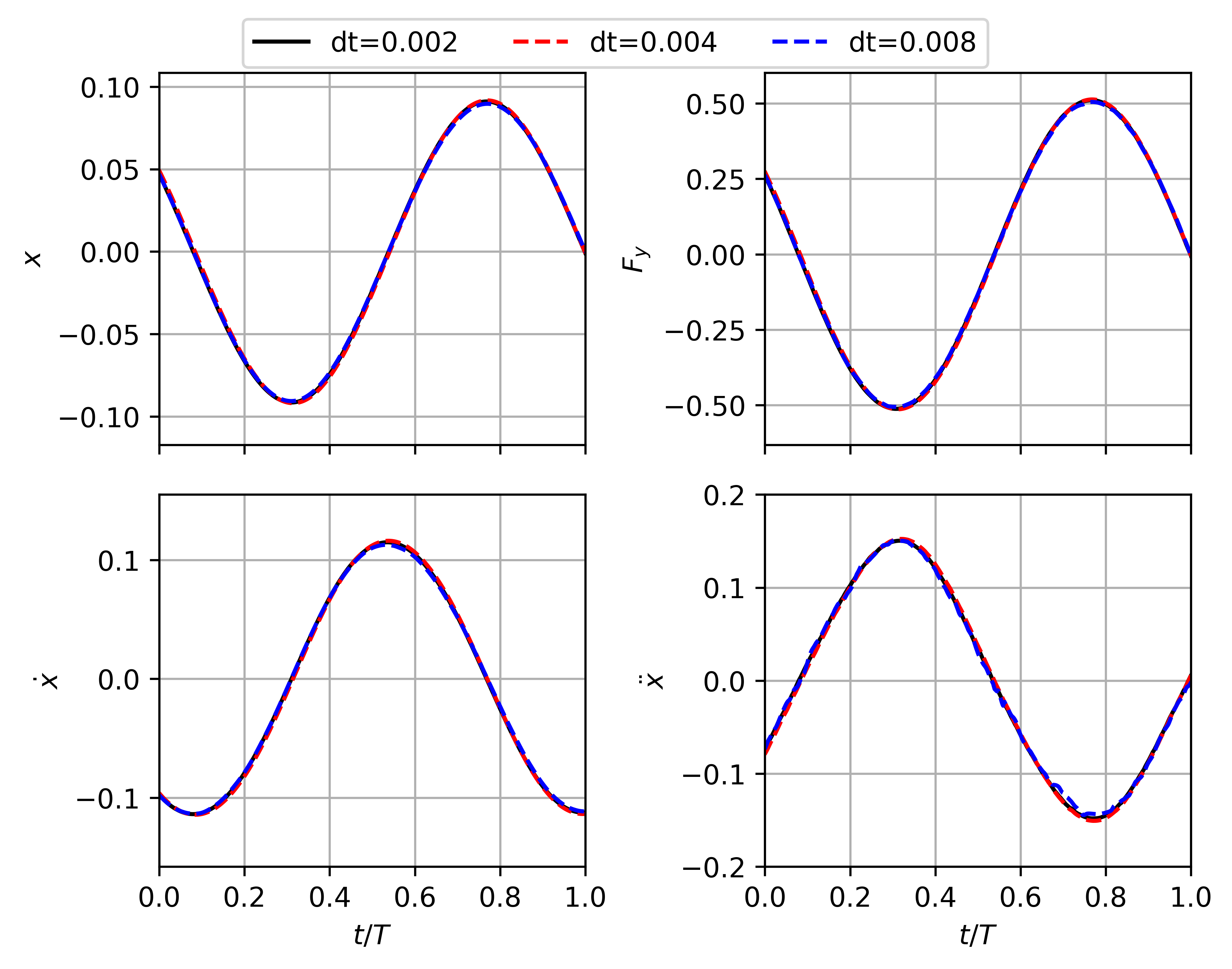}
\end{minipage}

\caption{Grid(left) and Time (right) independence  study for $U^*=3$. Displacement ($x$) (top left), force (top right), velocity (bottom left) and acceleration (bottom right) are presented for each one. The agreement is very good even for the coarser resolution. Minor discrepancies can be only seen in the predicted acceleration.\label{fig:1cyl-dx}}
\end{figure}

In Figure \ref{fig:1cyl-dx} (left), one period of the response of the cylinder is shown after periodic conditions have been reached. The responses with the three different spatial resolutions and $dt=0.004$ are compared, while in Figure \ref{fig:1cyl-dx} (right) the same is done for responses with the three  different time resolutions and $h=0.04$. In both figures the signals almost coincide. Slight deviations are only noted in the acceleration signals at around t/T=0.75. So for the simulations that follow, $dt=0.004$  and $h=0.04$ are used as reference values.
 
\begin{figure}[H]
	   \includegraphics[width=\linewidth]{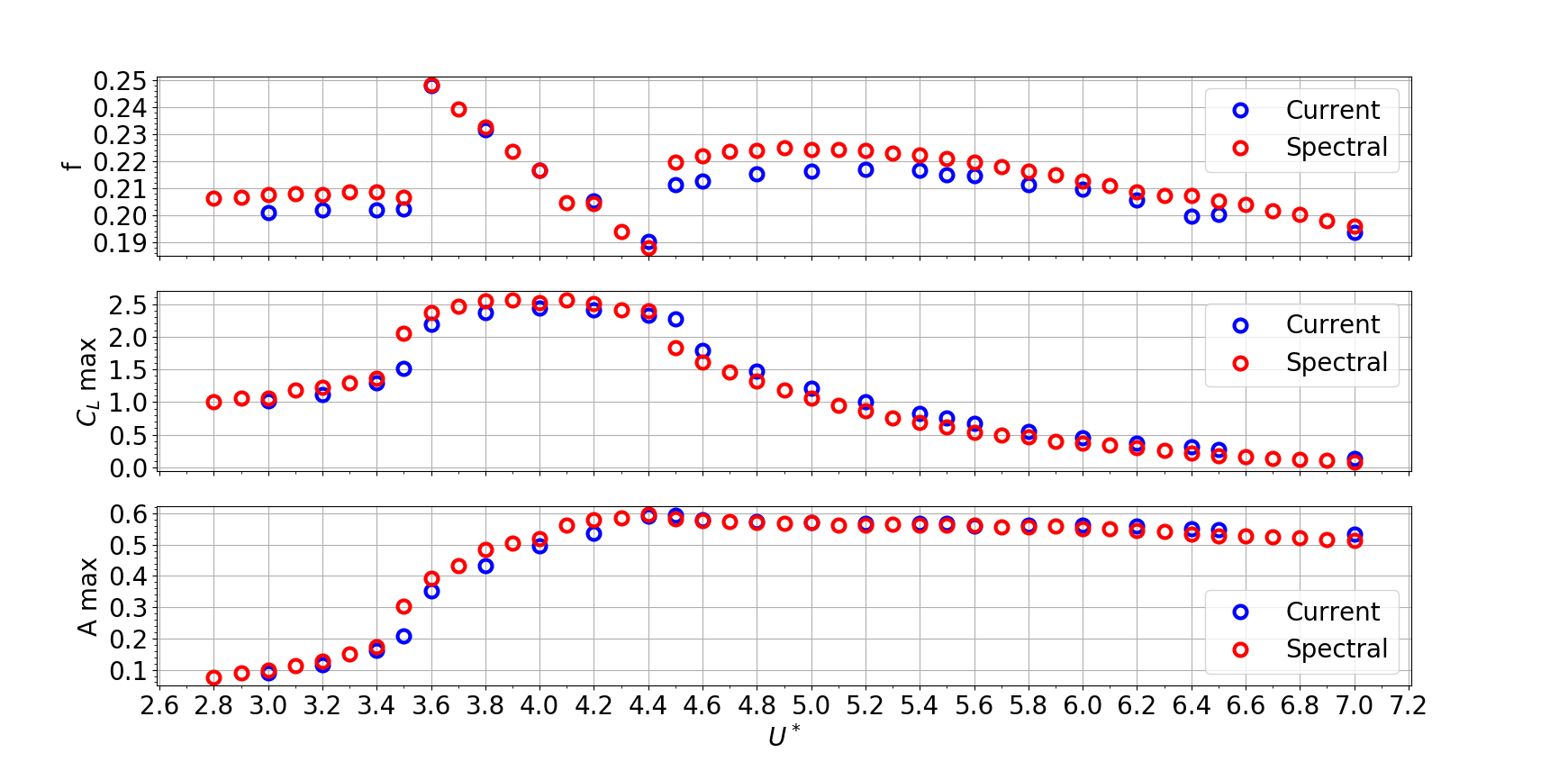}
       \caption{Comparison of the predicted main frequency (top), maximum lift coefficient (middle) and maximum amplitude (bottom) vs reduced velocity ($U^*$) for the single cylinder. Hybrid results are compared with the  spectral element method  predictions from \cite{Griffith2017a}. The overall agreement is good. The hybrid solver predicts smaller dominant frequencies for  $U^*<3.6$ and $U^*>4.4$.
       	\label{fig:1cyl-moving}}
\end{figure}

In Figure \ref{fig:1cyl-moving}, predicted responses of the present model are compared against the  prediction of the spectral element method presented in \cite{Leontini2006}. The response frequency,  the maximum amplitude and the maximum $C_L$ are recorded and compared over the range: $U^*=2.6 - 7$. In order to exclude transient effects, the last 50000 steps (or 200 non-dimensional time units) out of the total of 150000 times-steps (or 600 non-dimensional time) are processed. Good agreement between the two sets is noted despite the very different numerical approach they use. Outside the lock-in region ($U*<3.6$ and $U*>4.4$) the present method predicts slightly lower frequency of the cylinder response. A perfect match is noted within the lock-in region where the response is dominated by the natural frequency of the system. Outside the lock-in region the response is dominated by the shedding frequency which is predicted ~5\% lower by the present method. 

Regarding the  maximum lift coefficient($C_{LMAX}$) and maximum amplitude ($A_{MAX}$) the predictions compare well. In between $3.6< U^* < 4.4$ the agreement is very good with the exception of $U^*=3.5$ where differences in $C_{LMAX}$ and $A_{MAX}$ are evident.  

\subsection{Cylinders in tandem arrangement}
\begin{figure}[H]
	\begin{minipage}{0.5\linewidth}
		\includegraphics[width=\textwidth]{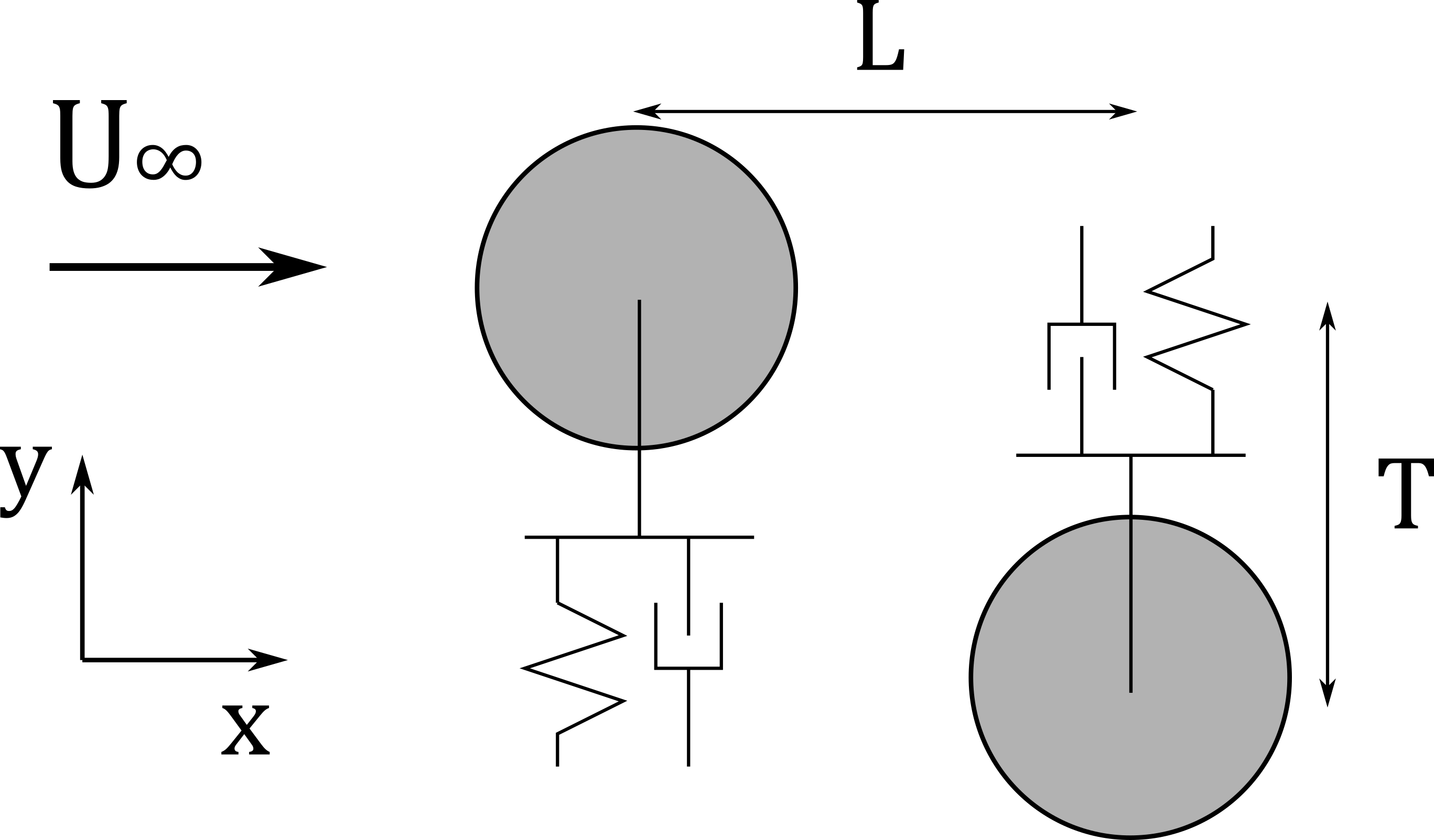}
	\end{minipage}
	\hfil
	\begin{minipage}{0.5\linewidth}
		\includegraphics[width=\textwidth]{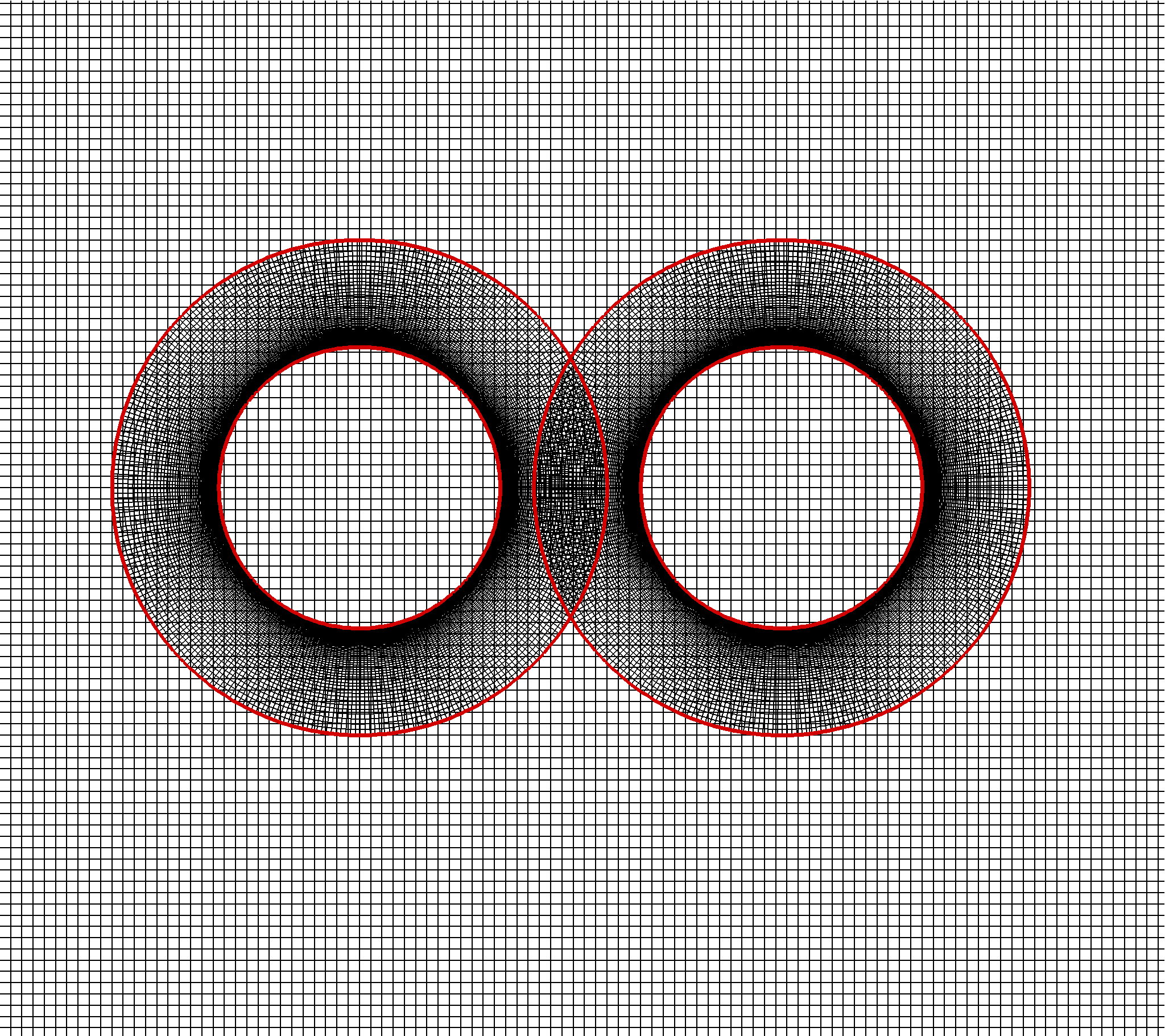}
	\end{minipage}
	
	\caption{Numerical Set-up of the cylinders in tandem arrangement (left). The Eulerian grids and the underlying particle mesh grid are shown on the right.The far-field boundary of every E-grid is at 0.87D (0.37D from the wall) from the cylinder center, so the two grids overlap.
		\label{fig:cyl2}}
\end{figure}

In this section the case of two cylinders in tandem arrangement is considered. The overall set-up is illustrated in Figure \ref{fig:cyl2}-left. Following \cite{Griffith2017a} and \cite{Borazjani2009} the streamwise distance between the cylinders is set equal to $L=1.5 D$. As indicated in Figure \ref{fig:cyl2} (right) in this set up the Eulerian grids of the two cylinders, overlap. However since all flow communication for every E-grid is done through the PM solution, there in no need for any special treatment.

\subsubsection{Stationary Cylinders}
\begin{figure}[H]
   \begin{subfigure}{0.9\linewidth}
	\begin{minipage}{0.3\linewidth}
		\includegraphics[width=\textwidth]{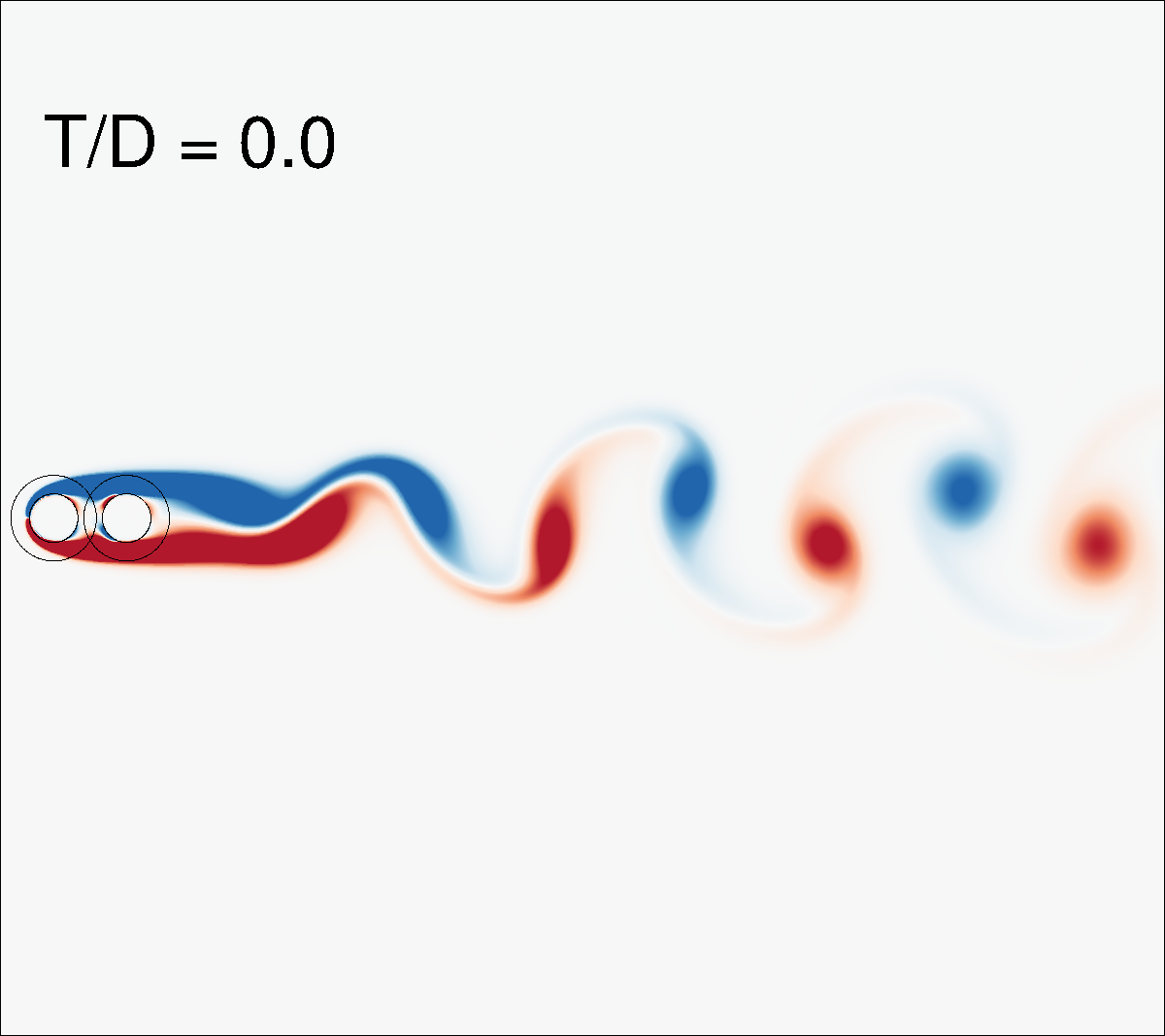}
	\end{minipage}
	\hfil
	\begin{minipage}{0.75\linewidth}
		\includegraphics[width=\textwidth]{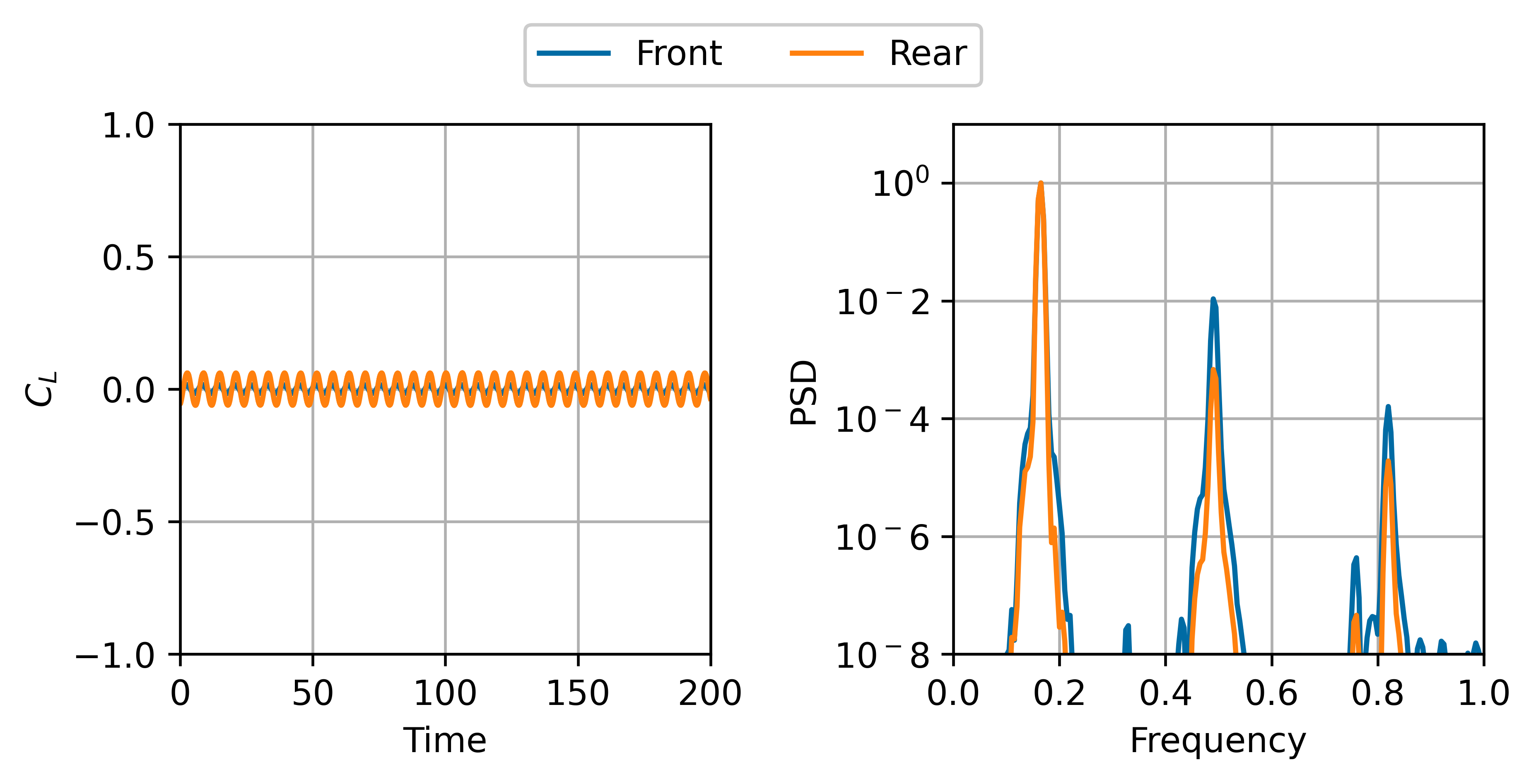}
	\end{minipage}
  \end{subfigure}

  \begin{subfigure}{0.9\linewidth}
	\begin{minipage}{0.3\linewidth}
		\includegraphics[width=\textwidth]{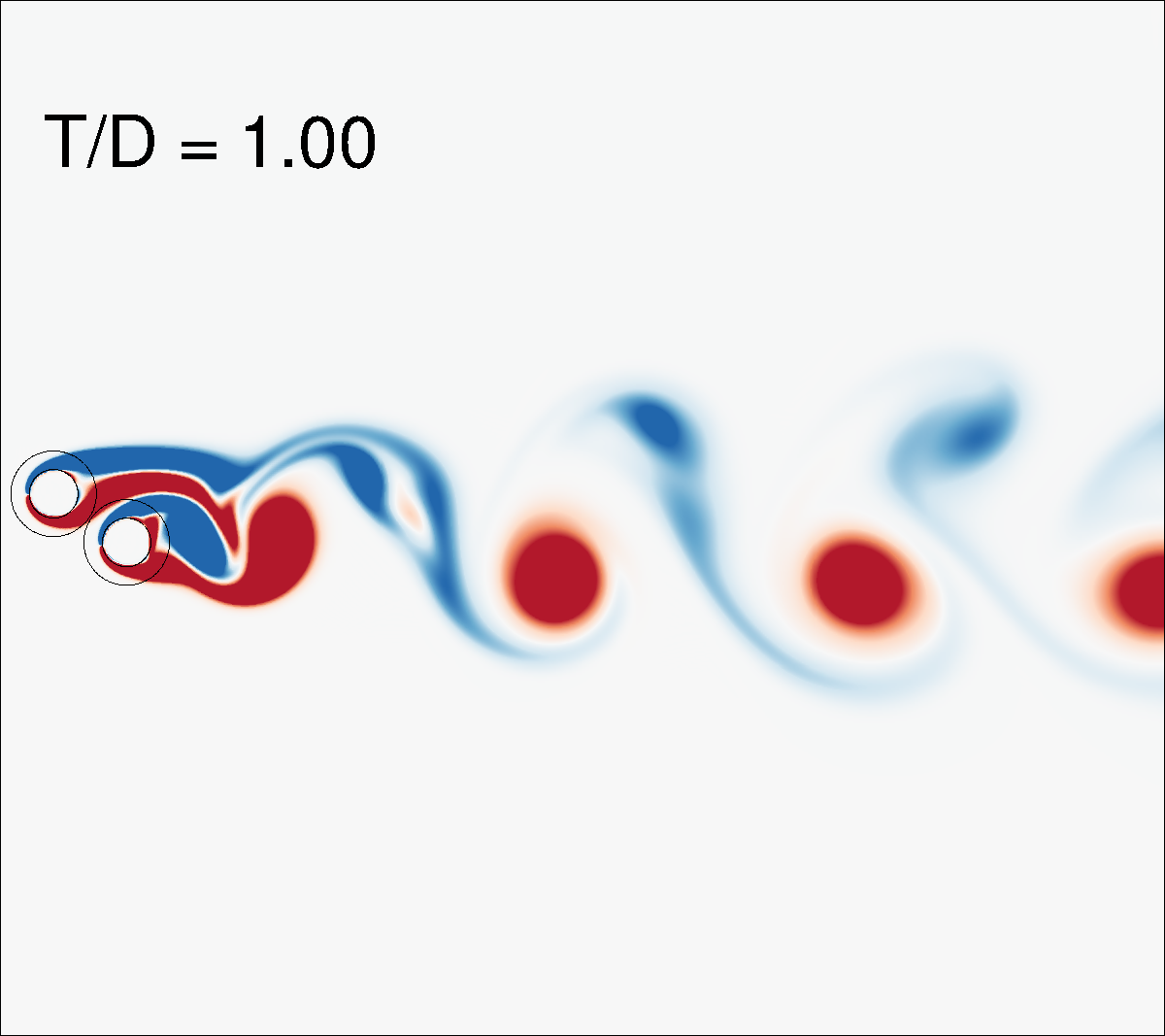}
	\end{minipage}
	\hfil
	\begin{minipage}{0.75\linewidth}
		\includegraphics[width=\textwidth]{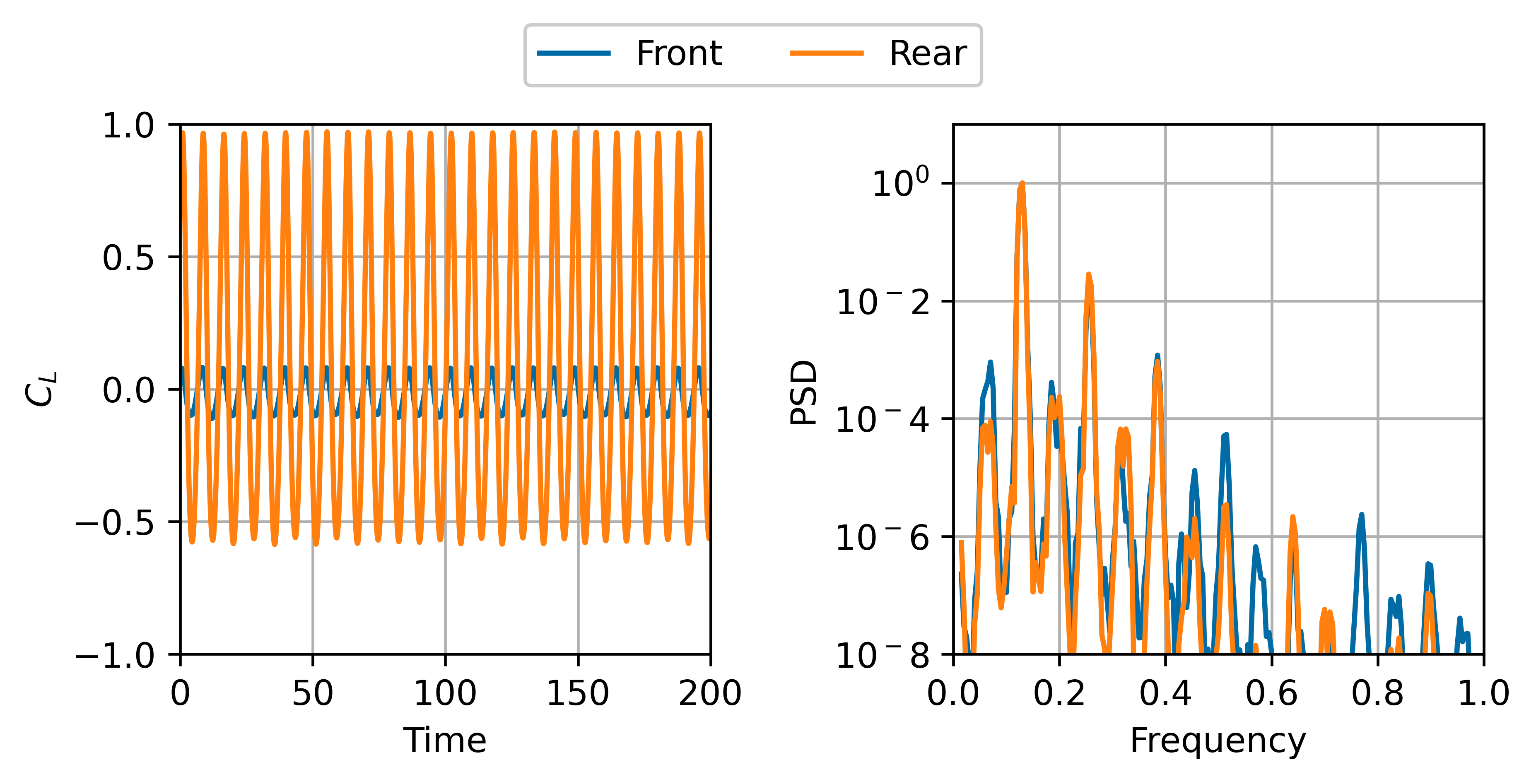}
	\end{minipage}
  \end{subfigure}

 \begin{subfigure}{0.9\linewidth}
	\begin{minipage}{0.3\linewidth}
		\includegraphics[width=\textwidth]{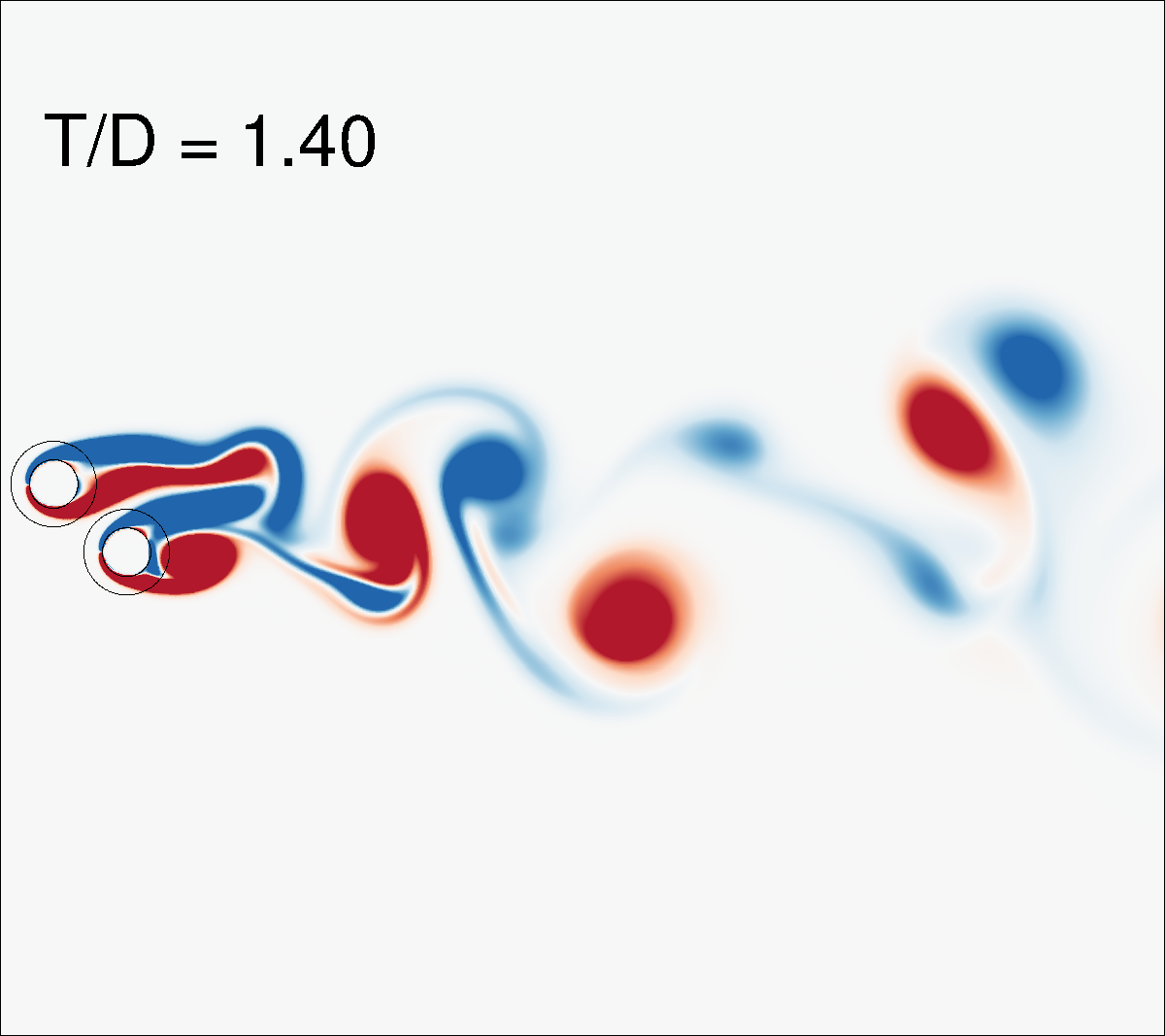}
	\end{minipage}
	\hfil
	\begin{minipage}{0.75\linewidth}
		\includegraphics[width=\textwidth]{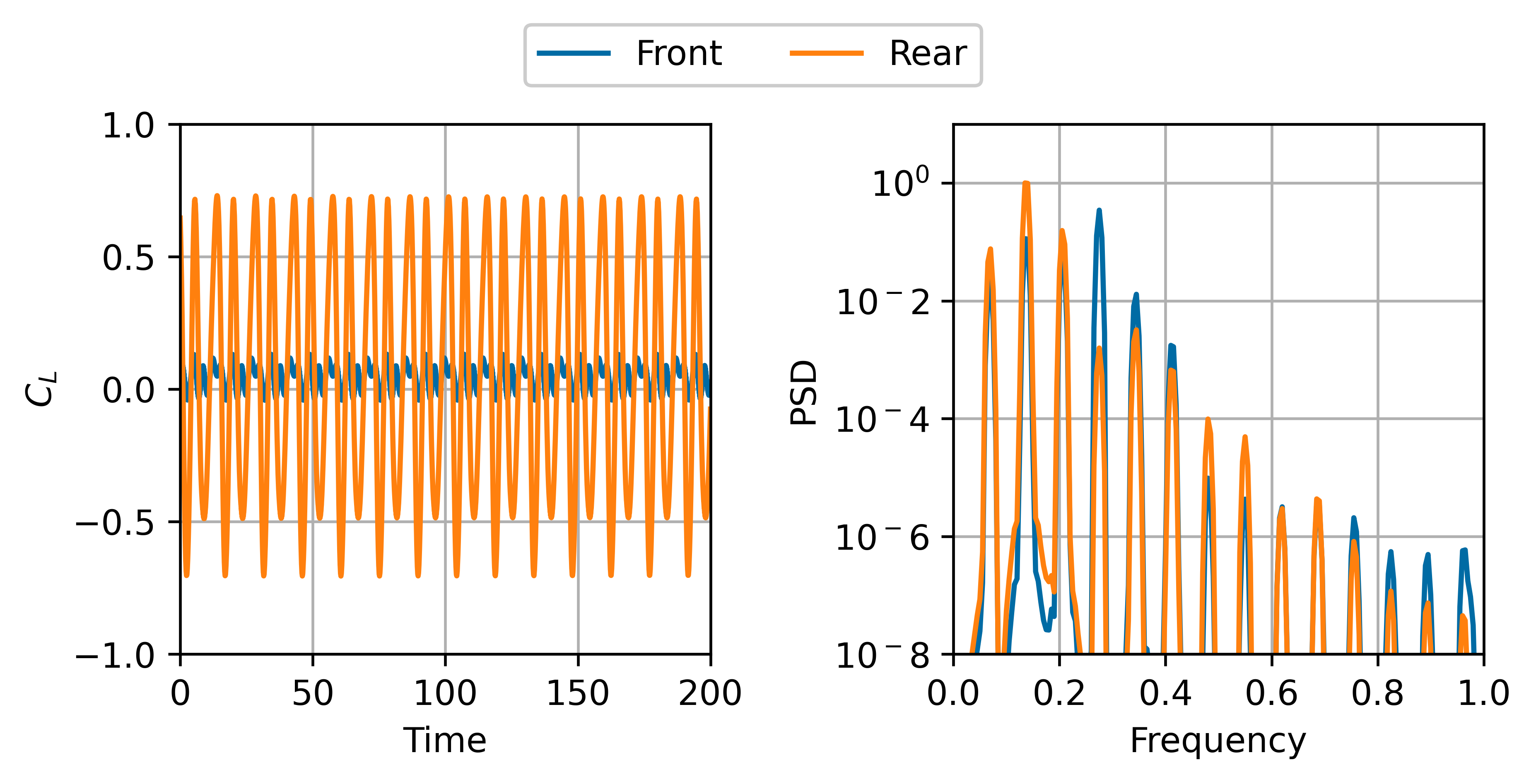}
	\end{minipage}
 \end{subfigure}

 \begin{subfigure}{0.9\linewidth}
	\begin{minipage}{0.3\linewidth}
		\includegraphics[width=\textwidth]{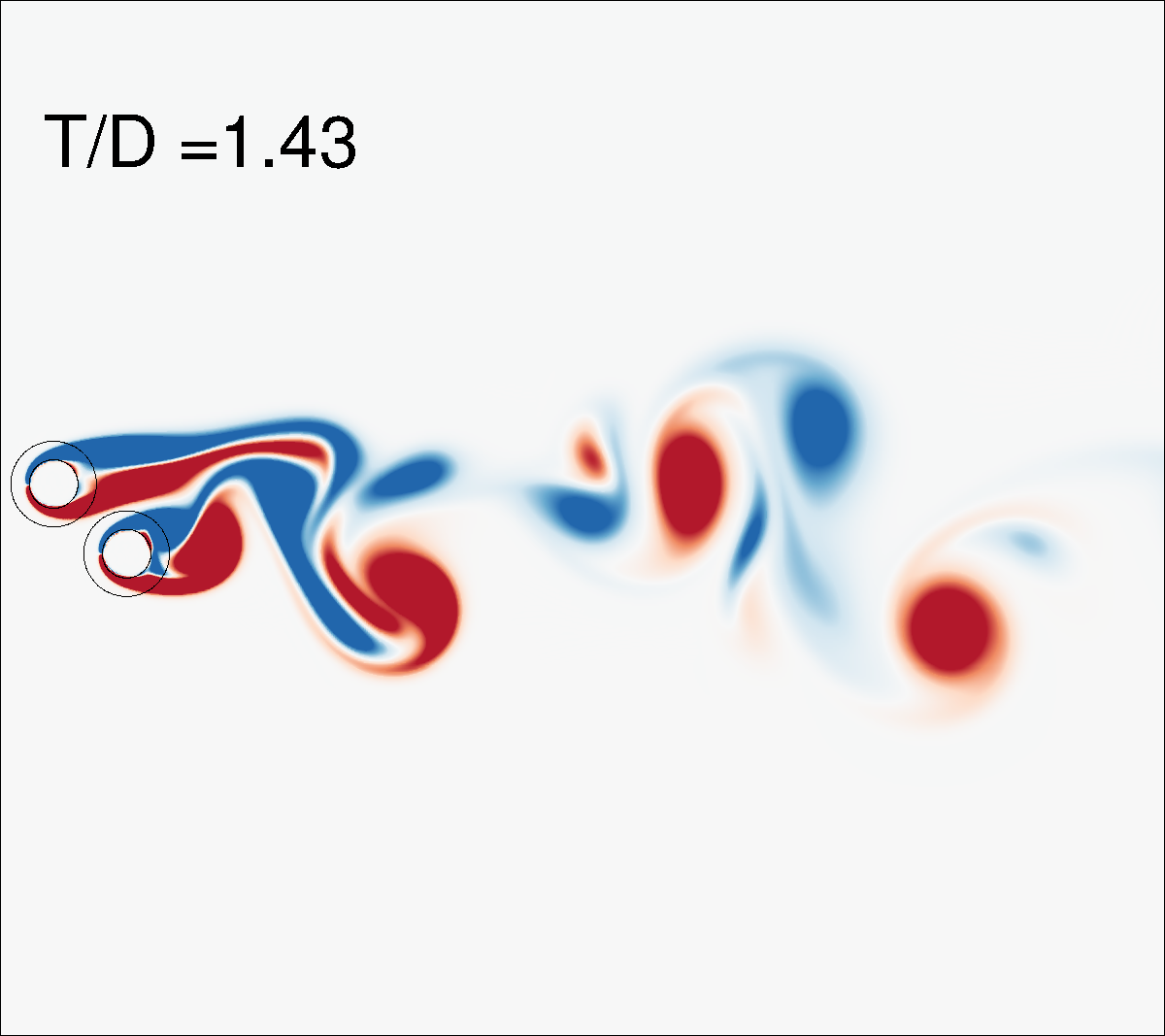}
	\end{minipage}
	\hfil
	\begin{minipage}{0.75\linewidth}
		\includegraphics[width=\textwidth]{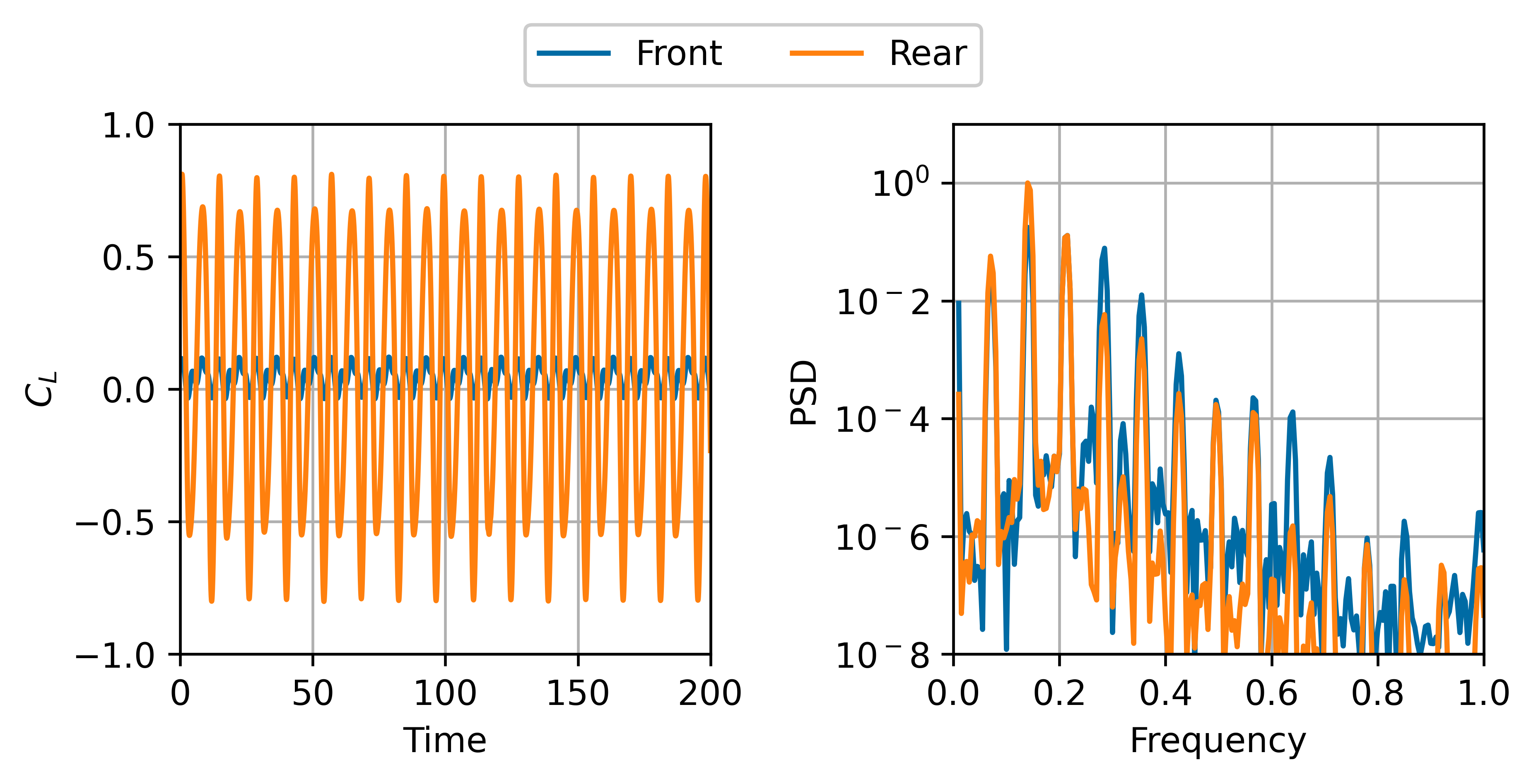}
	\end{minipage}
 \end{subfigure}
\caption{Vorticity contours ranging from -1 to 1 (left), Lift Coefficient($C_L$) time signals (center) and the corresponding spectra (right)  for the case of two stationary cylinders. The offset between the cylinders varies from $T/D=0 : 1.43$.\label{fig:static1}}
\end{figure}

\begin{figure}[H]
	\begin{subfigure}{0.9\linewidth}
		\begin{minipage}{0.3\linewidth}
			\includegraphics[width=\textwidth]{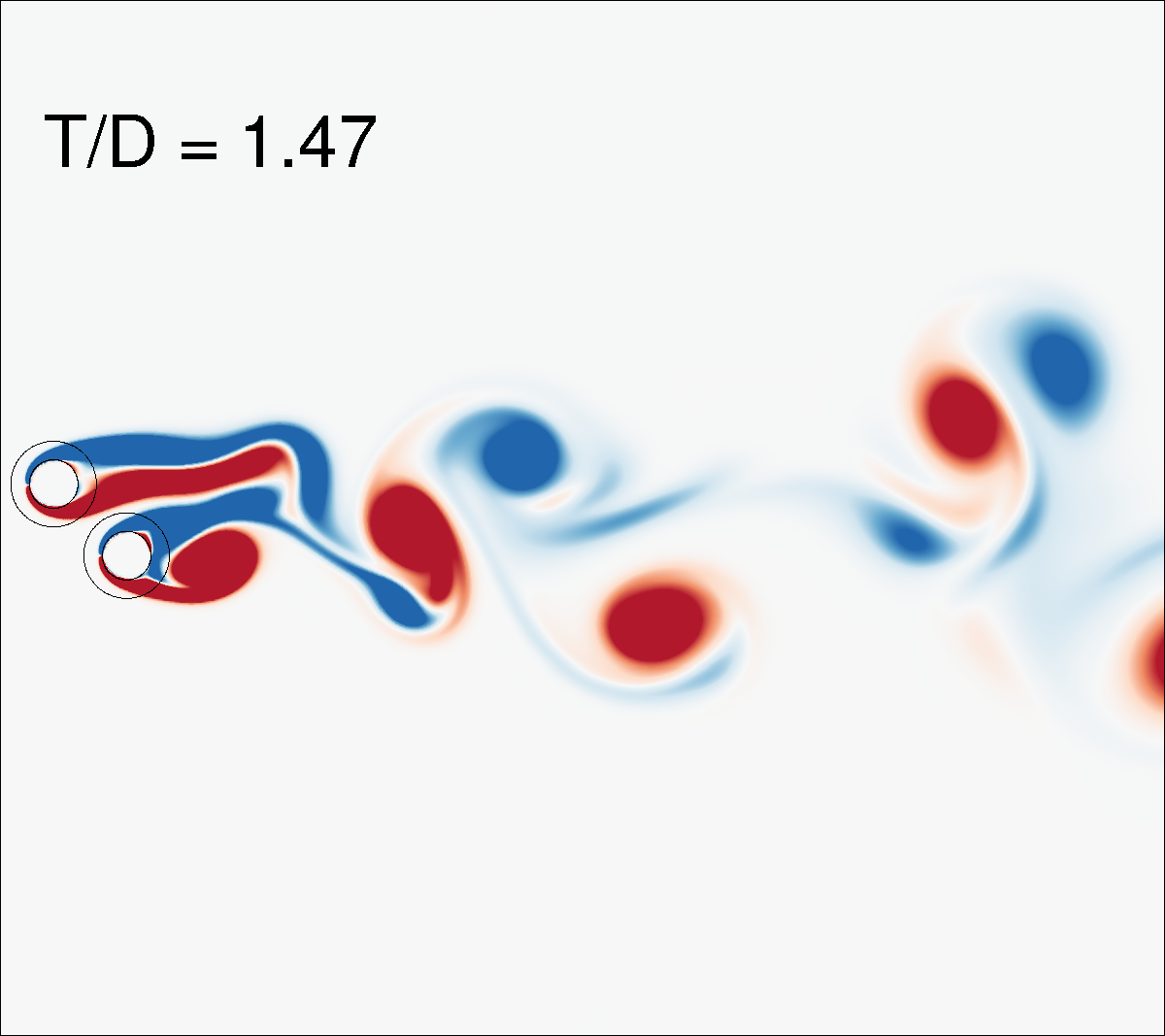}
		\end{minipage}
		\hfil
		\begin{minipage}{0.75\linewidth}
			\includegraphics[width=\textwidth]{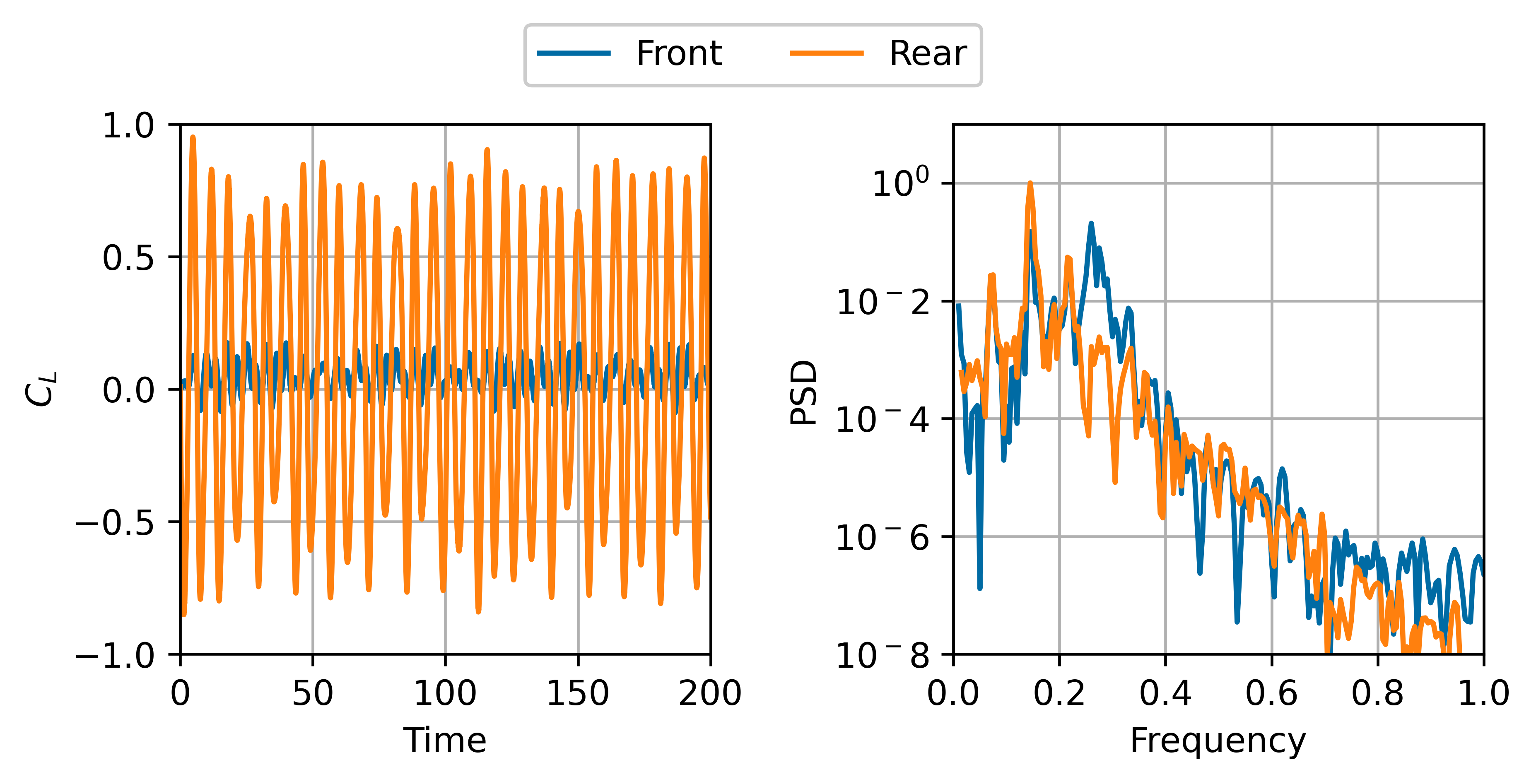}
		\end{minipage}
	\end{subfigure}
	
	\begin{subfigure}{0.9\linewidth}
		\begin{minipage}{0.3\linewidth}
			\includegraphics[width=\textwidth]{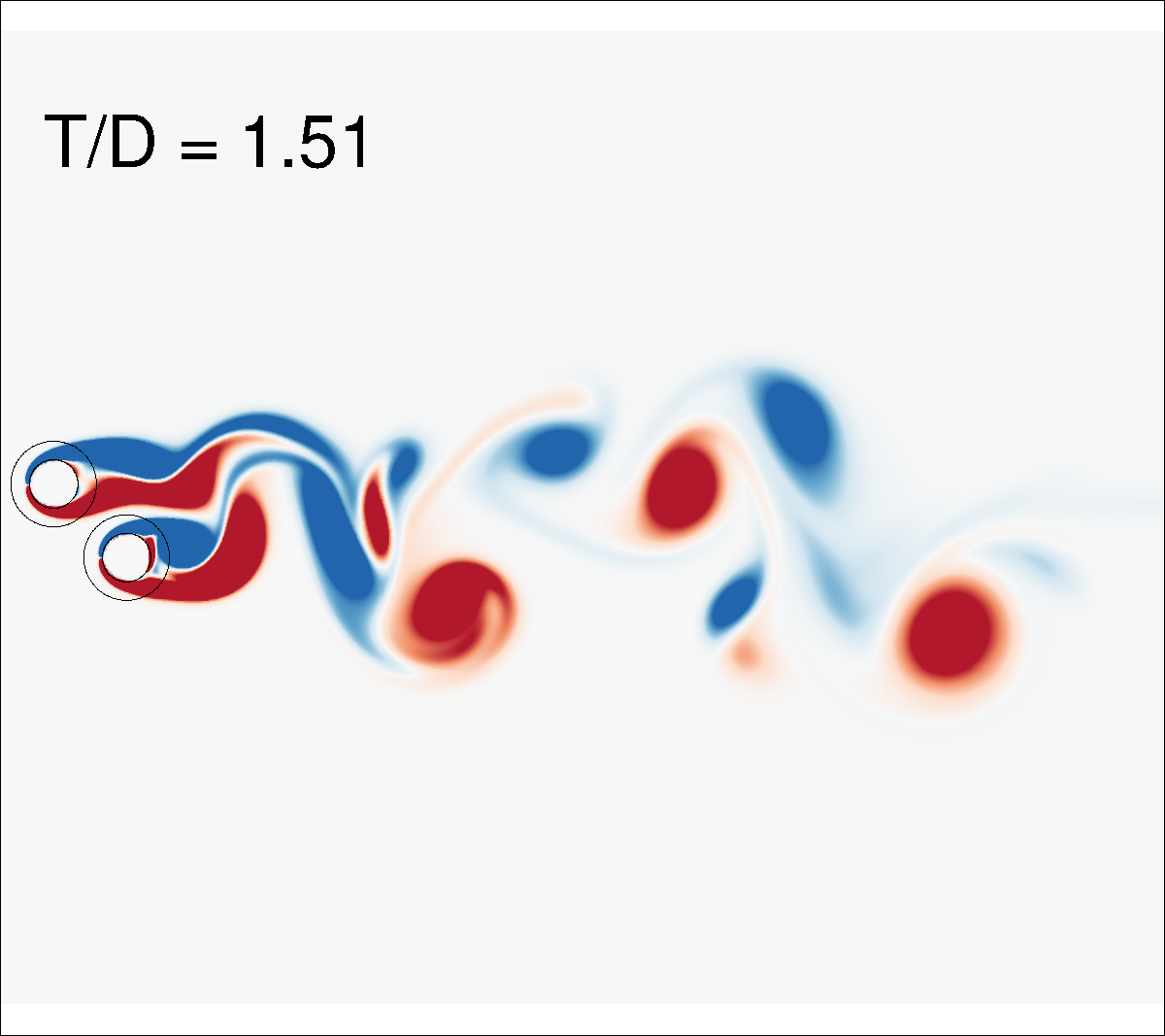}
		\end{minipage}
		\hfil
		\begin{minipage}{0.75\linewidth}
			\includegraphics[width=\textwidth]{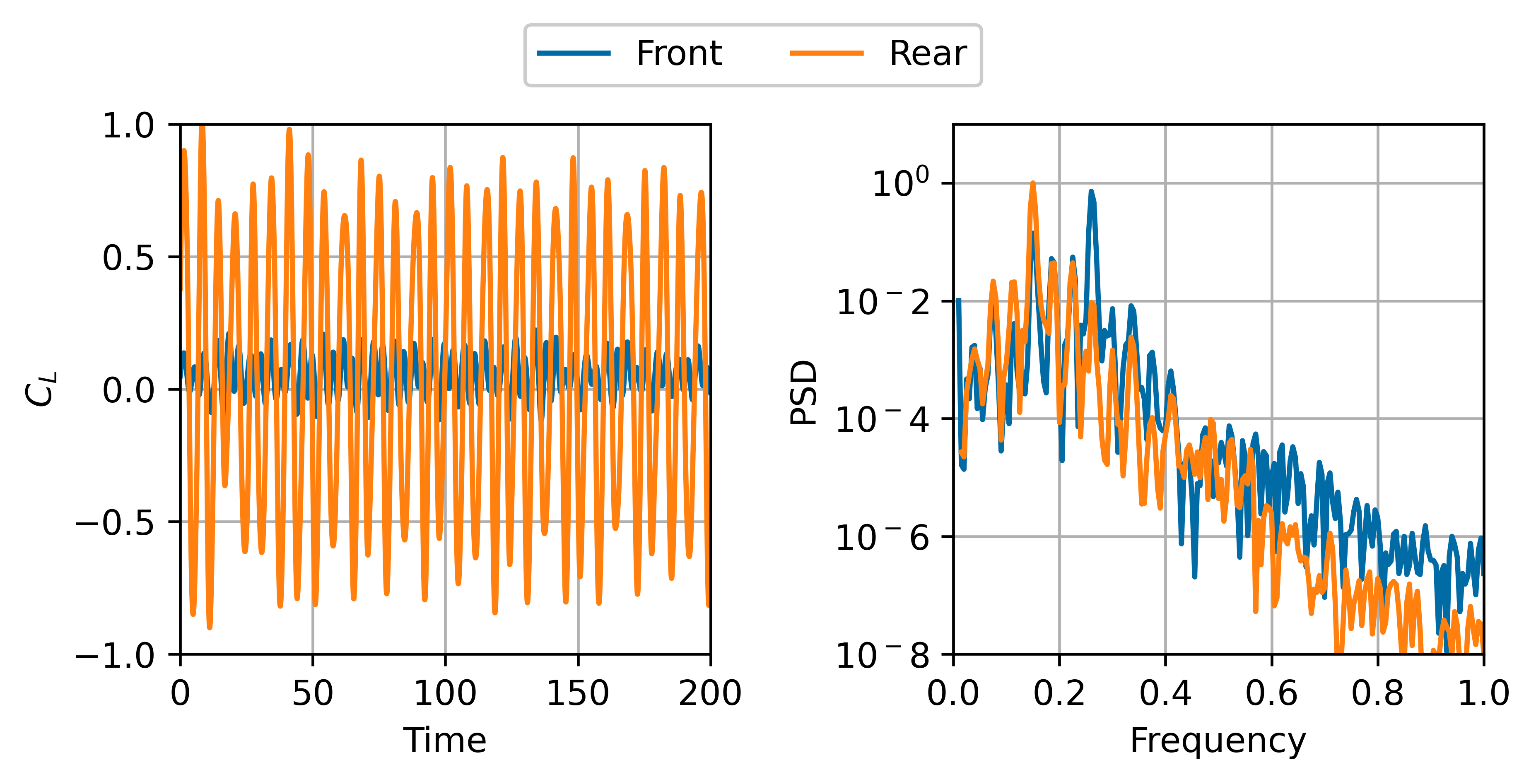}
		\end{minipage}
	\end{subfigure}
	
	\begin{subfigure}{0.9\linewidth}
		\begin{minipage}{0.3\linewidth}
			\includegraphics[width=\textwidth]{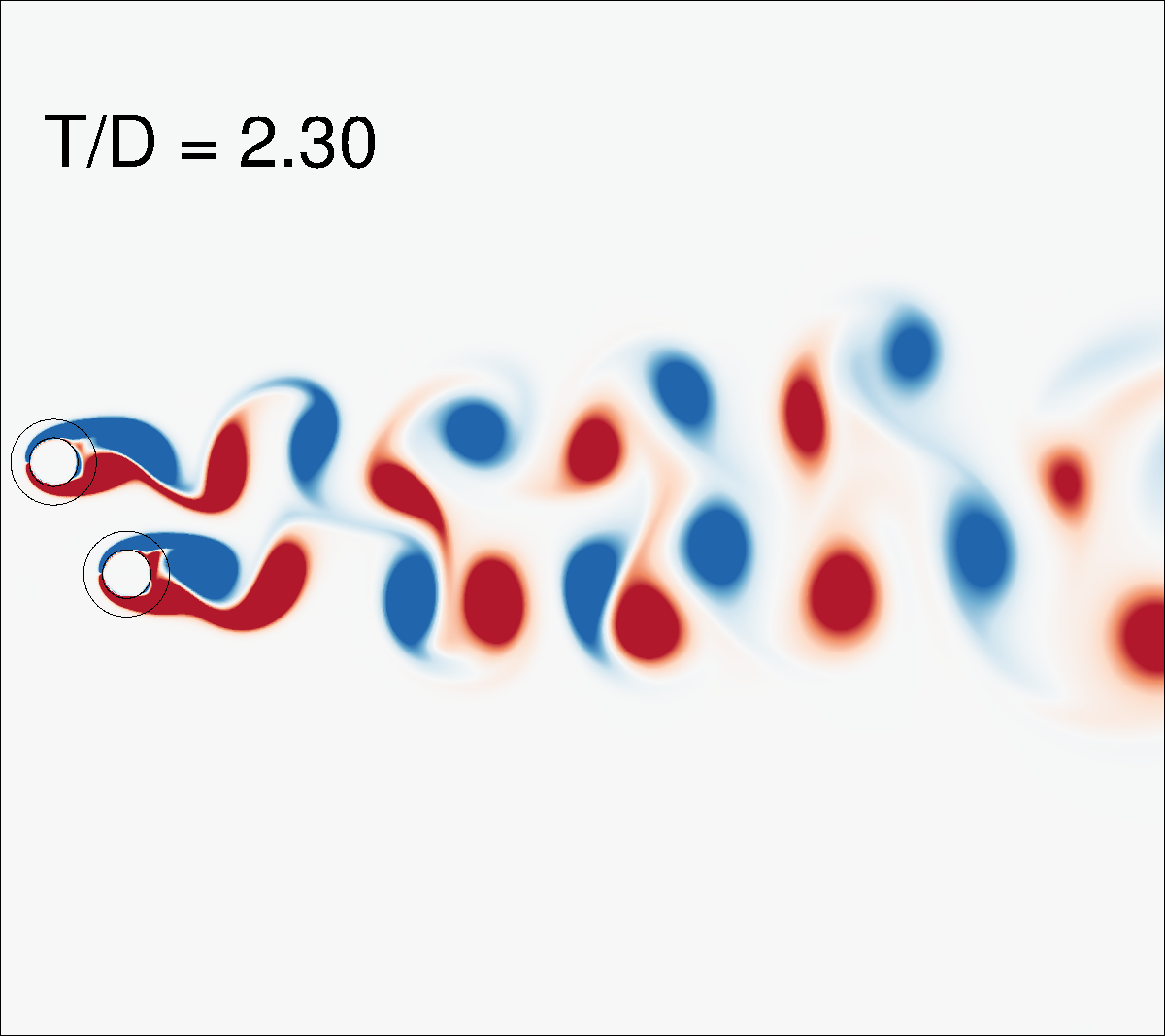}
		\end{minipage}
		\hfil
		\begin{minipage}{0.75\linewidth}
			\includegraphics[width=\textwidth]{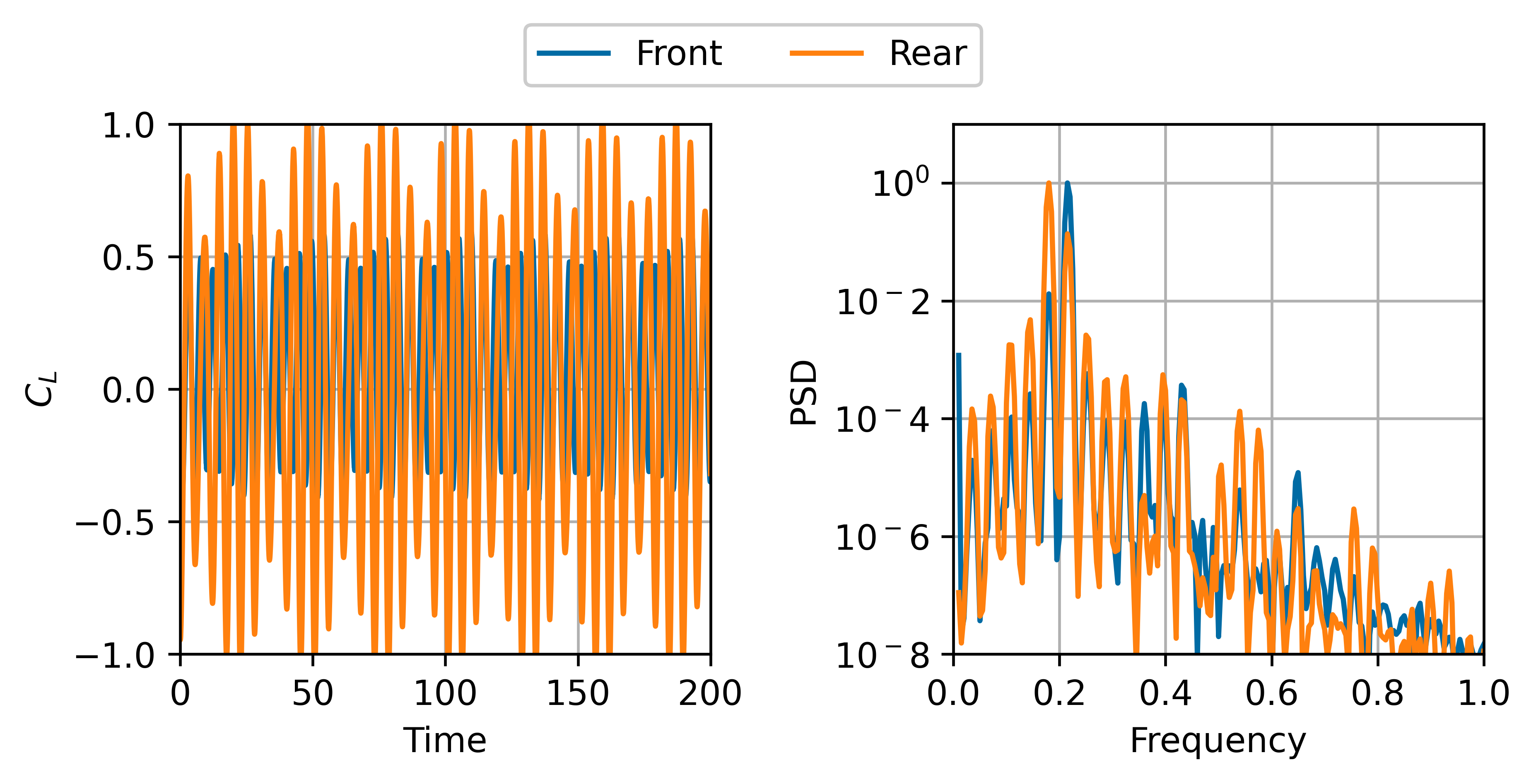}
		\end{minipage}
	\end{subfigure}
	
	\begin{subfigure}{0.9\linewidth}
		\begin{minipage}{0.3\linewidth}
			\includegraphics[width=\textwidth]{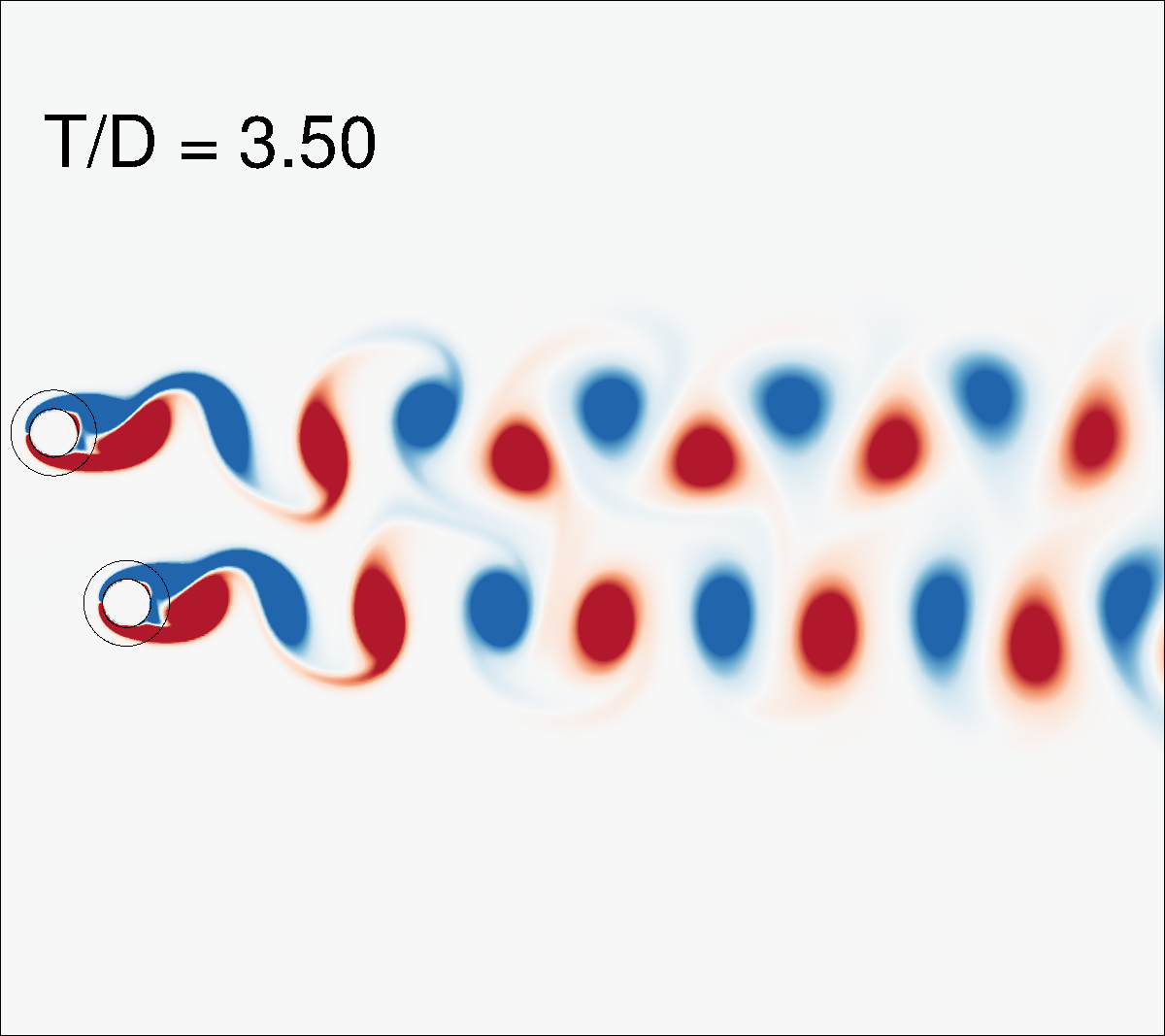}
		\end{minipage}
		\hfil
		\begin{minipage}{0.75\linewidth}
			\includegraphics[width=\textwidth]{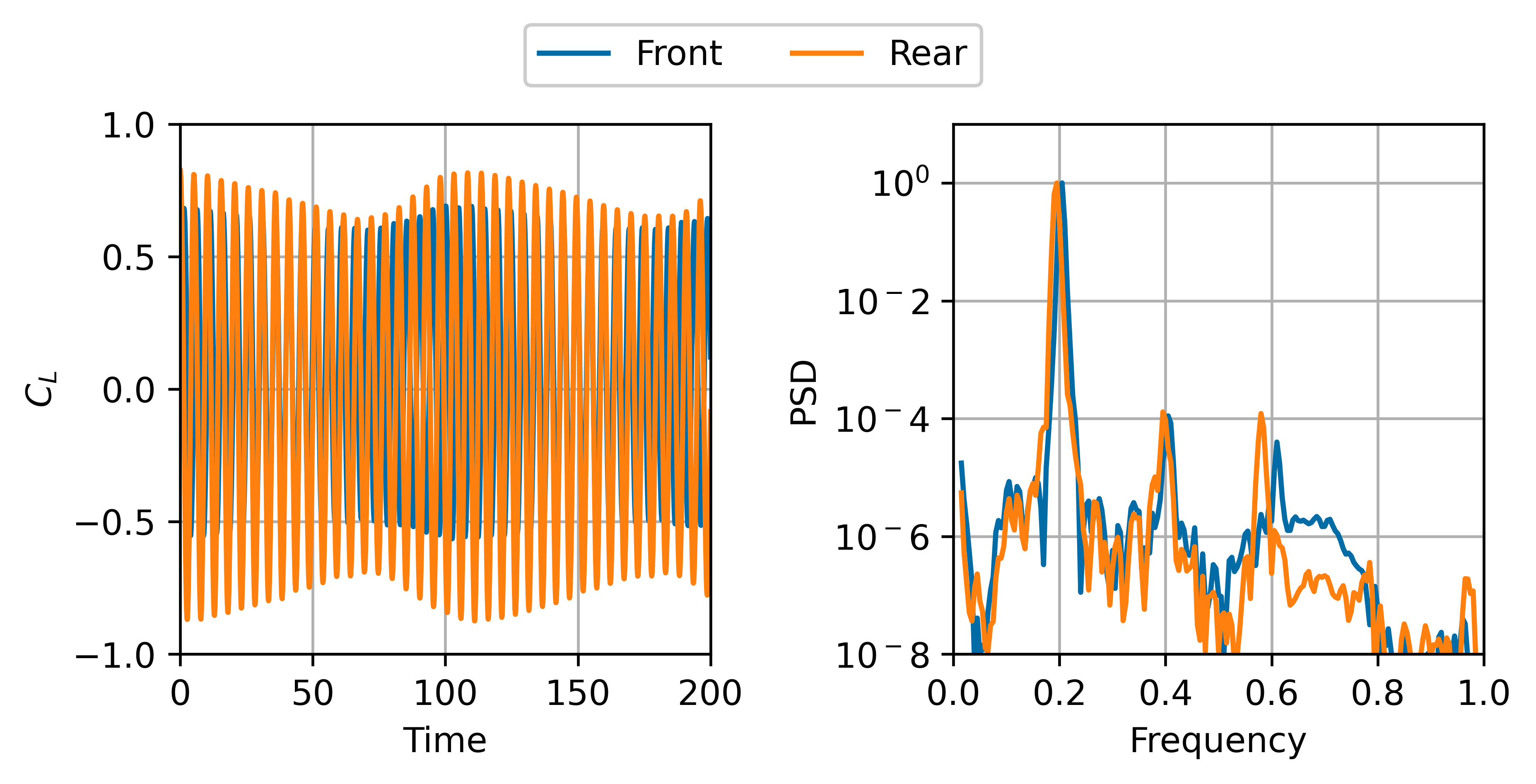}
		\end{minipage}
	\end{subfigure}
	\caption{Vorticity contours ranging from -1 to 1 (left), Lift Coefficient($C_L$) time signal (center) and the corresponding spectra(right) for the case of two stationary cylinders. The offset between the cylinders varies from $T/D=1.47 : 3.50$. \label{fig:static2}}
\end{figure}

First, the case of stationary cylinders ($U^*=0$) is examined for different transverse distances  in the range $T/D=0$ to $T/D=3.5$. Vorticity contour plots together with $C_L$ signals and spectra are presented in Figures \ref{fig:static1}, \ref{fig:static2}. 

As contour plots show, the $T/D$ offset modulates the interaction between the cylinders and their wakes. At $T/D=0$ a single wake system is formed. The mean $C_L$ of both cylinders is zero while 3 distinct frequencies are excited. By increasing the offset to $T/D=1$, the mean $C_L$ value is no longer zero, the amplitude of the rear cylinder increases while more frequencies are excited. This offset allows  interaction of the two wakes leading to the excitation of additional harmonics while the main frequency is close to the one in the $T/D=0$ case. By further increasing the offset, a more populated spectrum is obtained,  while the main frequency remains the same. 

At $T/D=1.47$ and $1.51$ (see Figure \ref{fig:static2}) the $C_L$ signals do not converge to a true periodic state which renders the spectrum broadband. By further increasing the gap to $T/D=2.3$ and $ 3.50$, coherent structures start to form again in the wake and distinct harmonics re-emerge. This is more pronounced at $T/D=3.50$ where the typical frequency of a single cylinder is recovered.

In the majority of the cases, the load on the rear cylinder is much higher compared to that of the front one. This is due to the impingement of the front wake on the rear cylinder. As the gap increases, the two wakes gradually decouple resulting in a drop of the $C_L$ of the rear cylinder and an increase in the force on the front one. 

In comparison to the results by Griffith et al \cite{Griffith2017a}, visual inspection suggests good agreement. At $T/D=0$ the hybrid solver gives slightly higher $C_L$ amplitudes and there is excitation to more than one frequencies. At $T/D=1$ the amplitudes are close and so are the peaks in the spectrum. In the hybrid results there is also excitation to intermediate frequencies while the $C_L$ signals are not 100\% periodic. In this respect, a possible reason of these differences, is that the present simulations are 30\% shorter than those in \cite{Griffith2017a}. At $T/D=1.4$ there is a slight frequency shift in the power spectral density (PSD)  plot which is also depicted in the lift signal. At $T/D=1.43$ the PSD in \cite{Griffith2017a} is broadband while a similar change is here found starting from $T/D=1.47$. At higher $T/D$ the agreement is in all respects good. Let us note in all the present spectra the blackman filter was applied in order to sharpen the dominating frequencies.
 
\subsubsection{The two-cylinder oscillating system}
In this section the two cylinders are no longer rigid but they are allowed to  oscillate in the y-direction, each having a separate flexible support. The mass ratio is $m^*=1$ for both cylinders and the damping is set to zero ($\xi=0$). Structurally the motions of the cylinders are uncoupled, but being submerged in the same flow, coupling is established through the flow-induced forcing. Hybrid simulations with $h=0.04$ and $\Delta_t=0.004$ are carried out over the range $3.0\le U^* \le 14.0$ and predictions are compared to those  in \cite{Borazjani2009} and \cite{Griffith2017a}. The mass and the spring constant are defined by:
\begin{equation}
	m =4\pi D m^* , k= (2\pi f_n)^2 m 
\end{equation}
where it is noted that the added mass contribution is not longer included (as opposed to equation \ref{eq:1cylrbdf}).

\begin{figure}[H]
	\includegraphics[width=\linewidth]{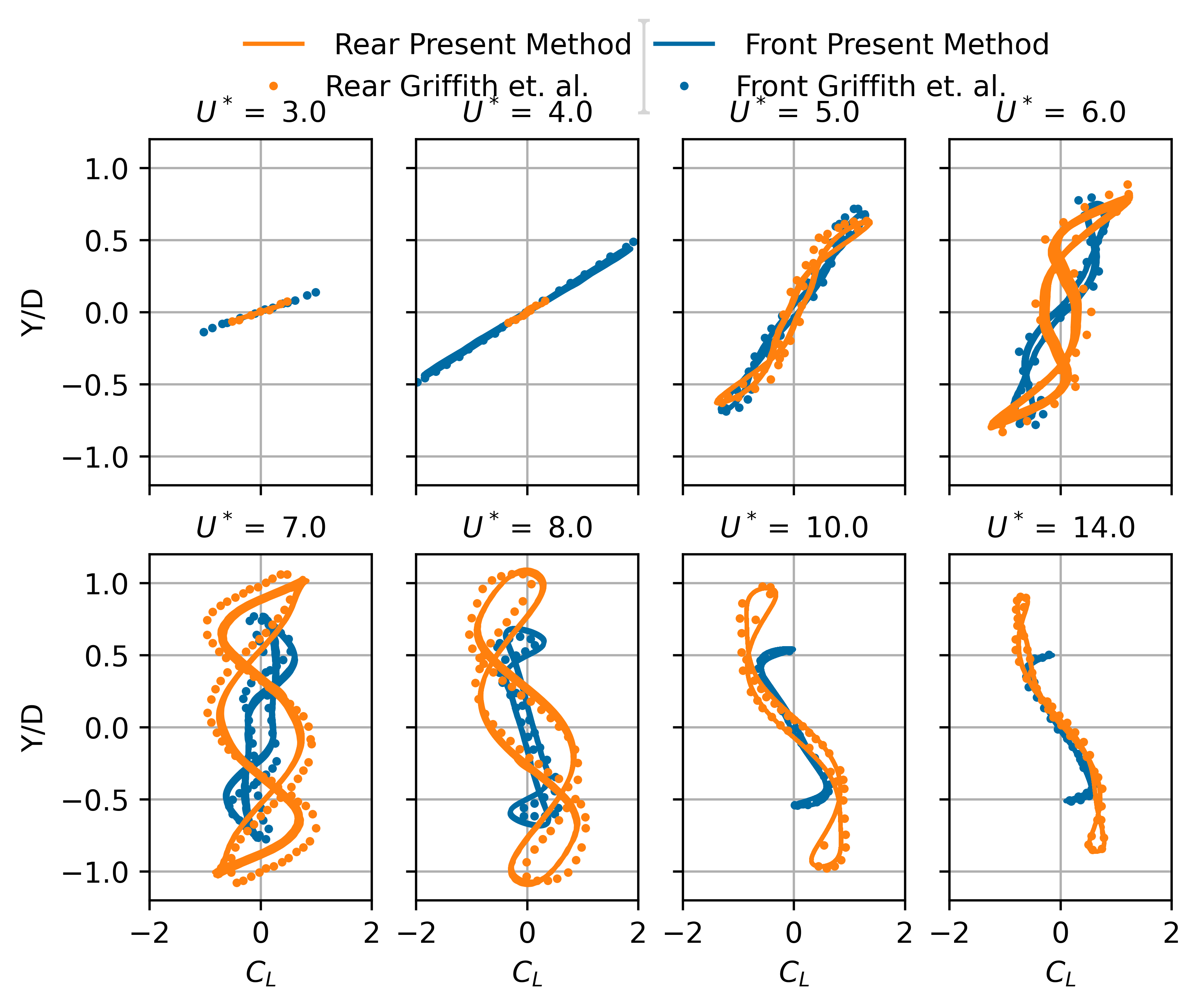}
	\caption{Lissajous curves of the front  and rear  cylinder for the various reduced velocities ($U^*$). Hybrid predictions (continuous lines) are compared to those by Griffith et al \cite{Griffith2017a} (dots).  
		\label{fig:2cyl-liz}}
\end{figure}

In Figure \ref{fig:2cyl-liz} the converged Lissajous curves (lift coefficient vs displacement) are shown   in comparison to those by Griffith et al  \cite{Griffith2017a} over the full range of $U^*$. A good agreement is noted in spite of the very different numerical method employed in the two simulations (hybrid vs. immersed boundary). As $U^*$ increases, a three branch hysteresis loop gradually develops. Besides a small swift in $U^*$, that diminishes at high $U^*$ values, the two sets follow the same forming stages. Higher differences are noted at  $U^*=3.0$ where the hybrid solver predicts smaller amplitudes. 

\begin{figure}[H]
	\includegraphics[width=\linewidth]{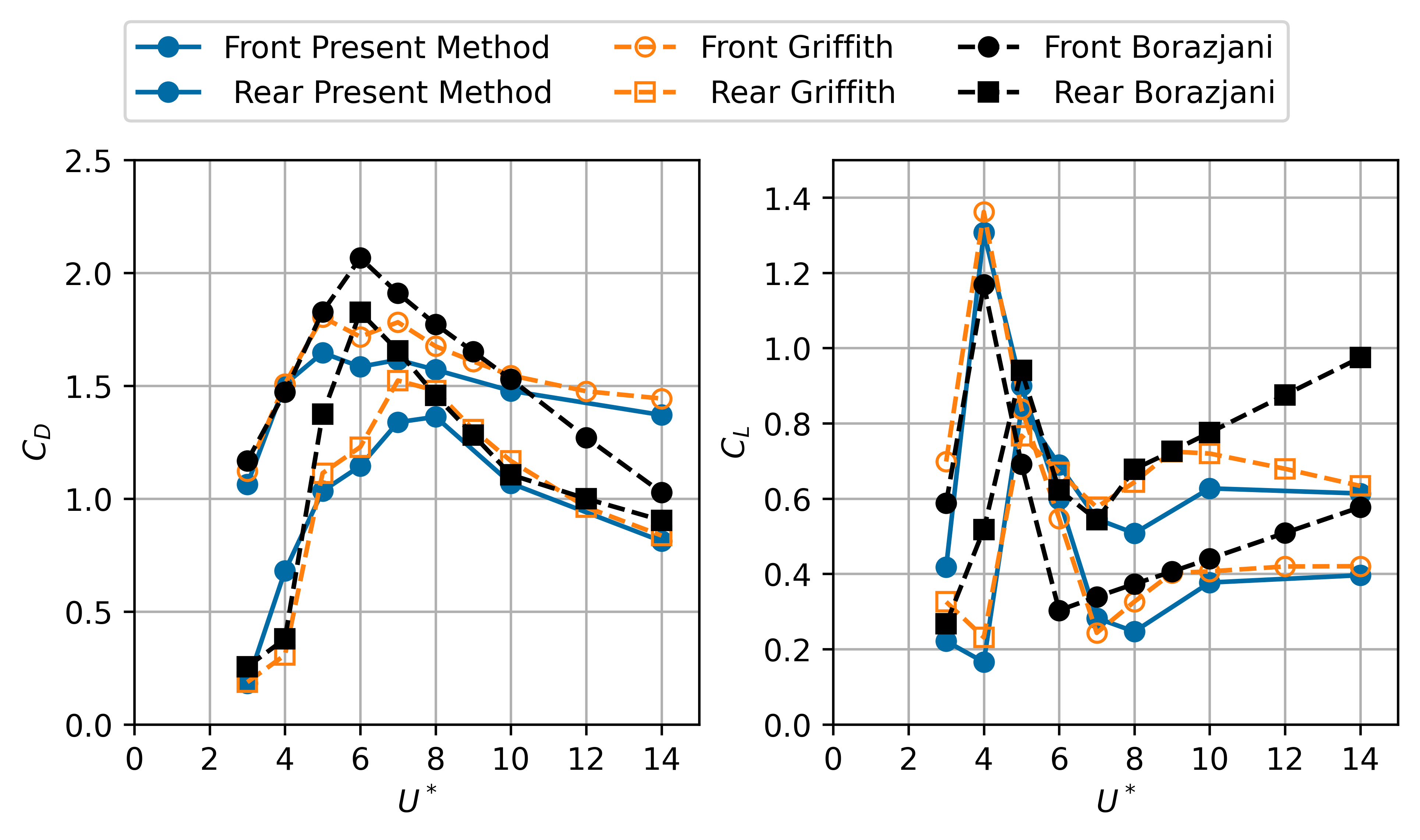}
	\caption{Variation of drag  ($C_D$-left)  and lift ($C_L$-right) coefficient for different reduced velocities ($U^*$). Hybrid predictions are compared with those from \cite{Griffith2017a}, \cite{Borazjani2009}\label{fig:2cyl-clcd}}
\end{figure}

Next in Figure \ref{fig:2cyl-clcd} loads are compared in terms of mean lift ($C_L$)  and drag coefficients ($C_D$) of the two cylinders for different $U^*$ values. Results from \cite{Griffith2017a} and \cite{Borazjani2009} are included in this comparison. The hybrid results are in better agreement with those of \cite{Griffith2017a}. Between the two sets, Griffith et al \cite{Griffith2017a} predict higher drag for intermediate $U^*$ values while in \cite{Borazjani2009} lift is significantly higher and the $C_D$ variation has rather different shape. In terms of shape there is better agreement in the lift plots up to $U^*=6$. At higher $U^*$ values, Borazjani et al \cite{Borazjani2009} give an almost linear variation while the other two sets have similar trends. However the $C_L$ local minimum in \cite{Griffith2017a} and the present results is not the same. The hybrid method predicts the minimum at $U^*=8$ while \cite{Griffith2017a} at $U^*=7.0$. 

\begin{figure}[H]
	\includegraphics[width=\linewidth]{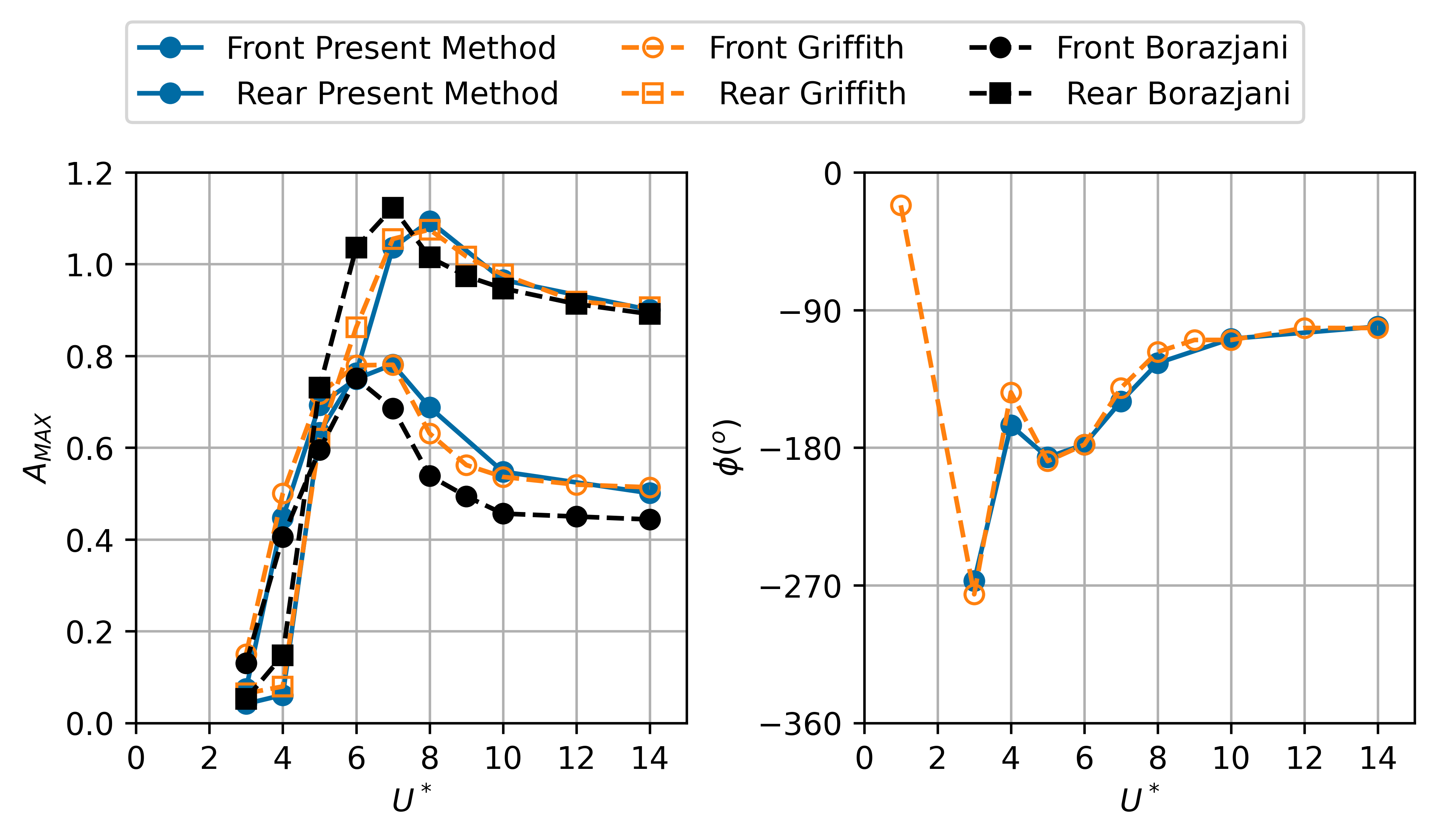}
	\caption{The predicted maximum amplitude ($A_{MAX}$) and phase difference between the front and the rear cylinder. Results using the present method are compared with the ones from \cite{Griffith2017a},\cite{Borazjani2009}. 
		\label{fig:2cyl-amax}}
\end{figure}

Next the comparison proceeds with the variation of the maximum amplitude and the phase difference in Figure \ref{fig:2cyl-amax}. All three solvers produce similar variations over the whole range of $U^*$ considered where Griffith's results and the present ones are in good quantitative agreement which is also seen in the  phase difference plots.

\begin{figure}[H]
	\begin{minipage}{\linewidth}
	\begin{subfigure}{0.6\linewidth}
		\includegraphics[width=\textwidth]{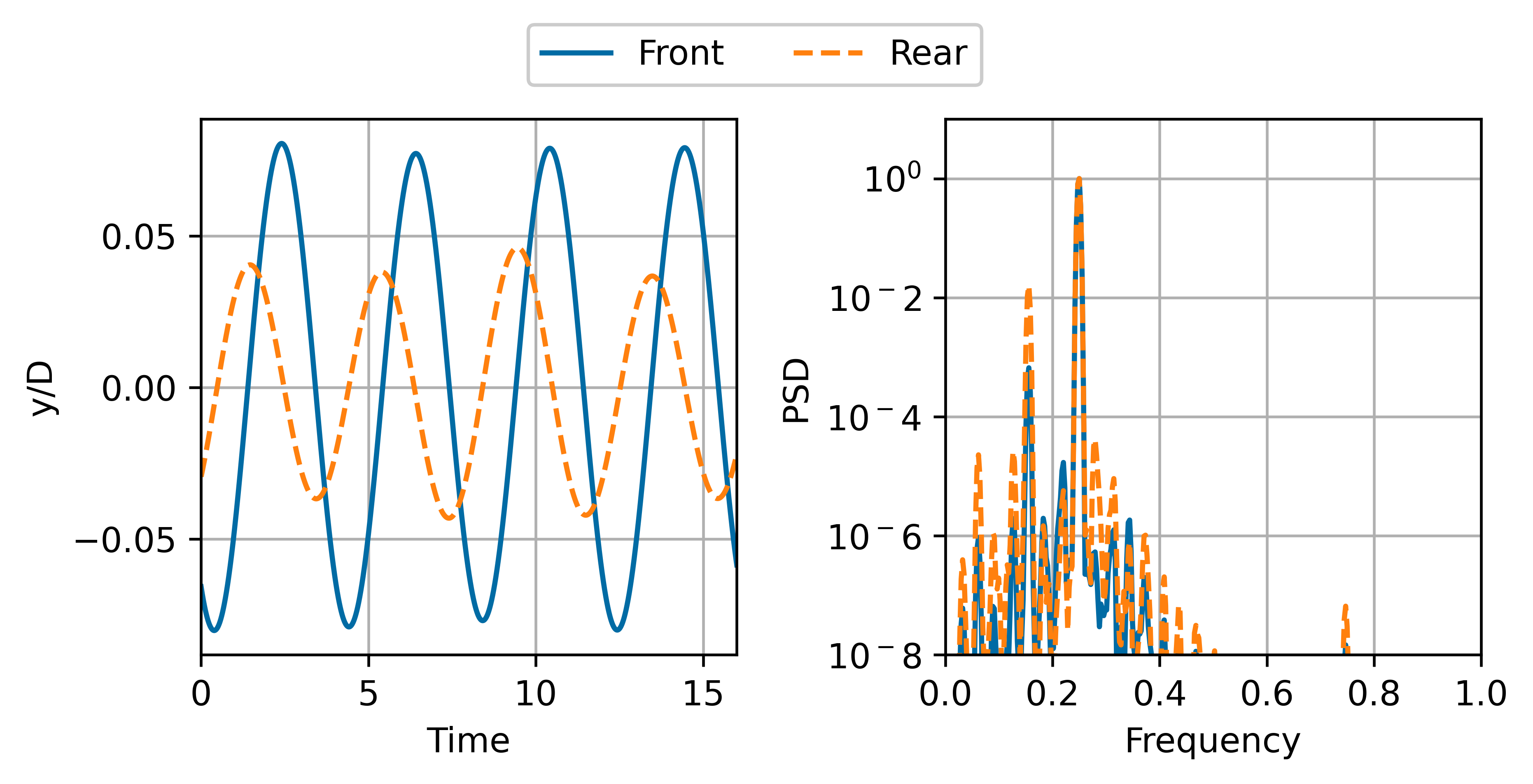}
		\caption{$U^*$=3.0\label{fig:2cyl-fft3}}
	\end{subfigure}
	\begin{subfigure}{0.6\linewidth}
		\includegraphics[width=\textwidth]{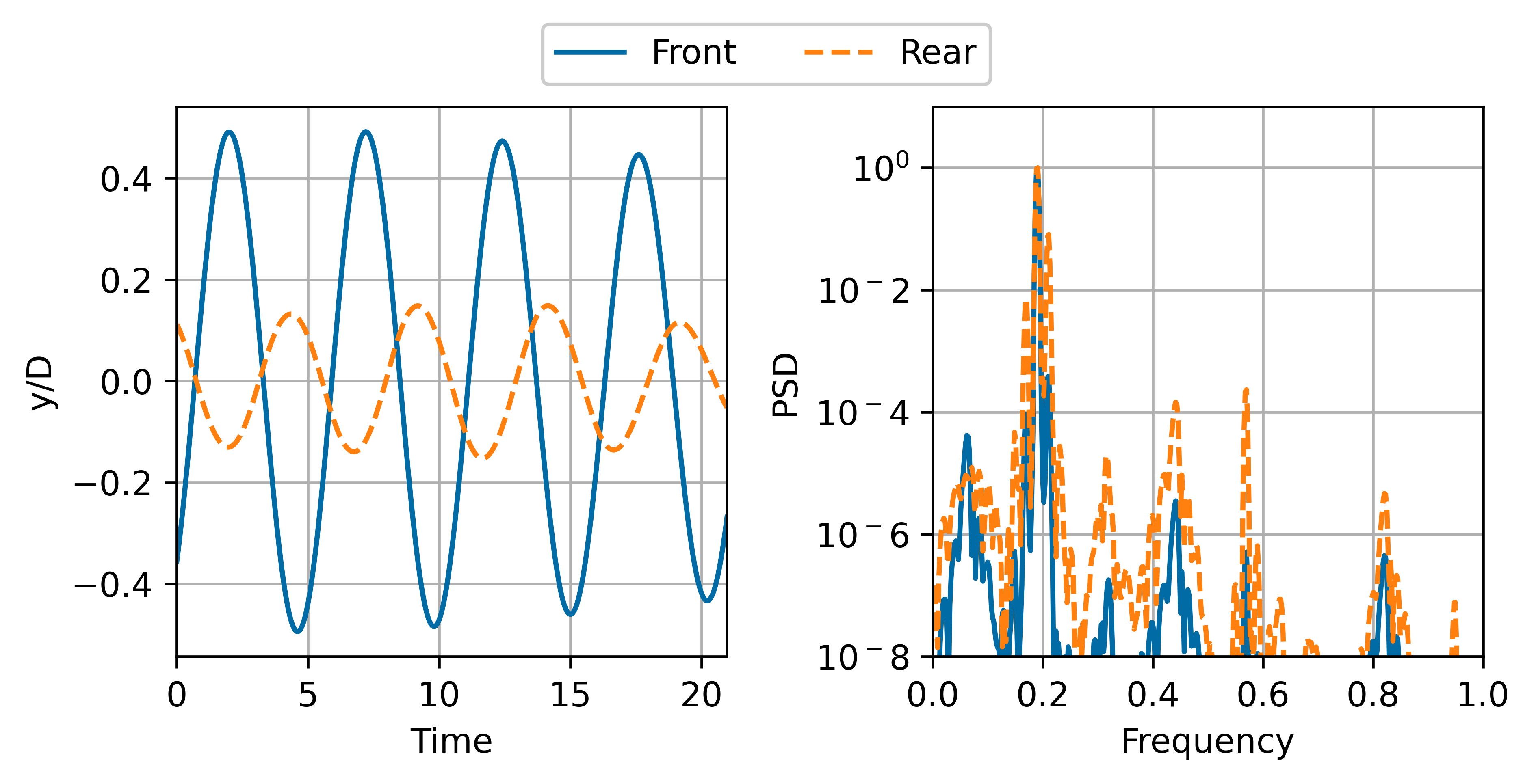}
		\caption{$U^*$=4.0 \label{fig:2cyl-fft4}}
	\end{subfigure}
	\begin{subfigure}{0.6\linewidth}
		\includegraphics[width=\textwidth]{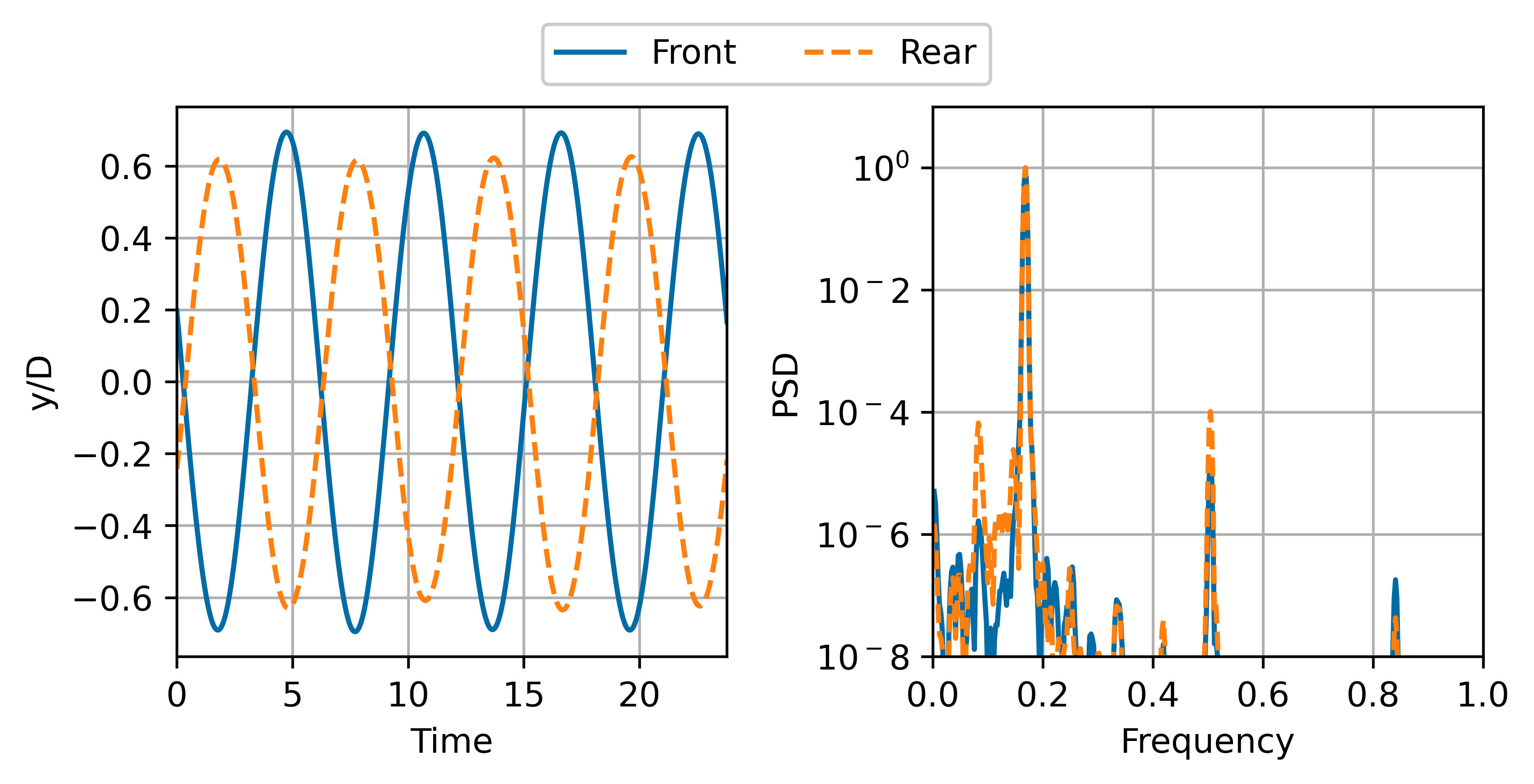}
		\caption{$U^*$=5.0 \label{fig:2cyl-fft5}}
	\end{subfigure}
	\begin{subfigure}{0.6 \linewidth}
		\includegraphics[width=\textwidth]{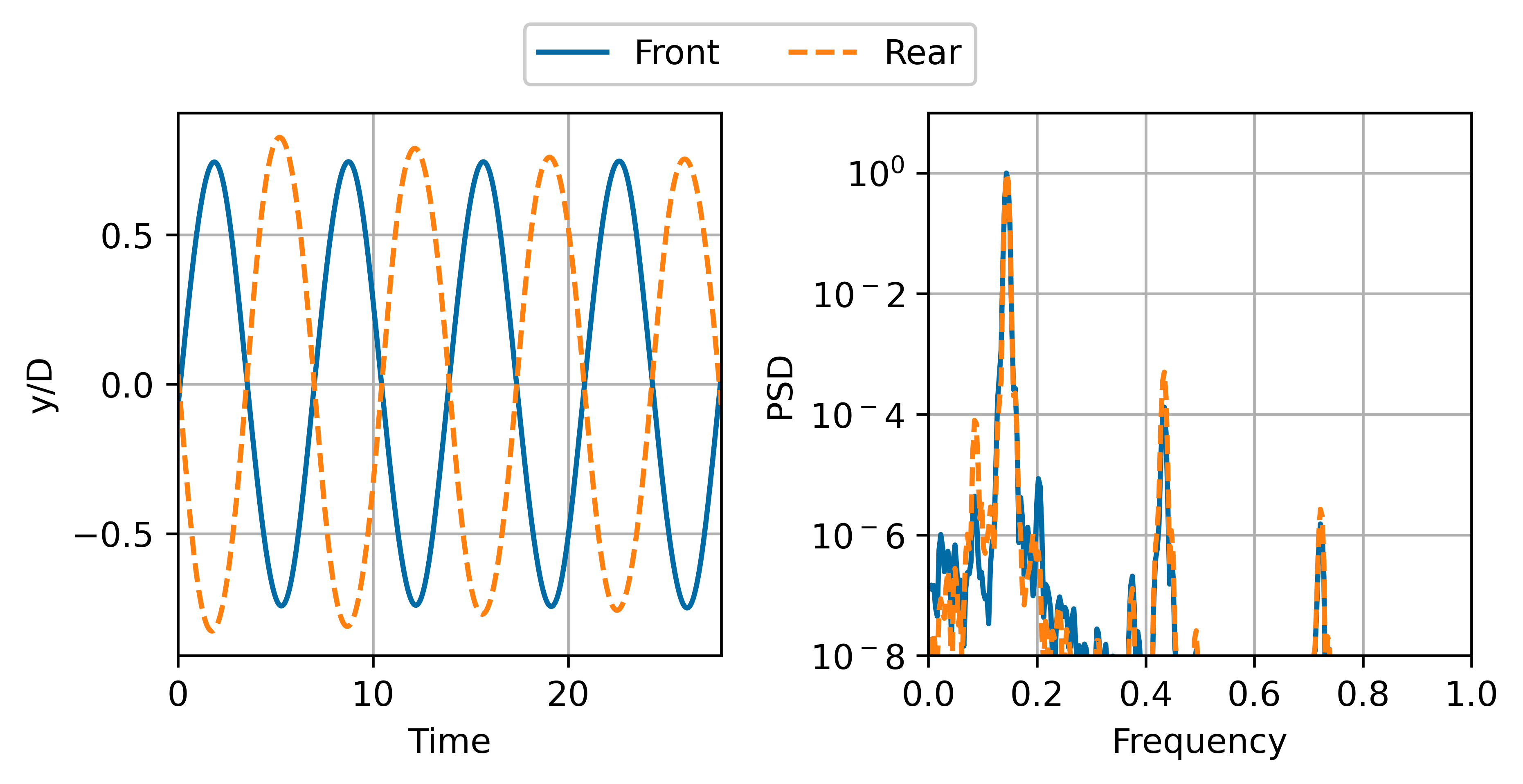}
		\caption{$U^*$=6.0 \label{fig:2cyl-fft6}}
	\end{subfigure}
   \end{minipage}
	\begin{minipage}{\linewidth}
	\begin{subfigure}{0.6\linewidth}
		\includegraphics[width=\textwidth]{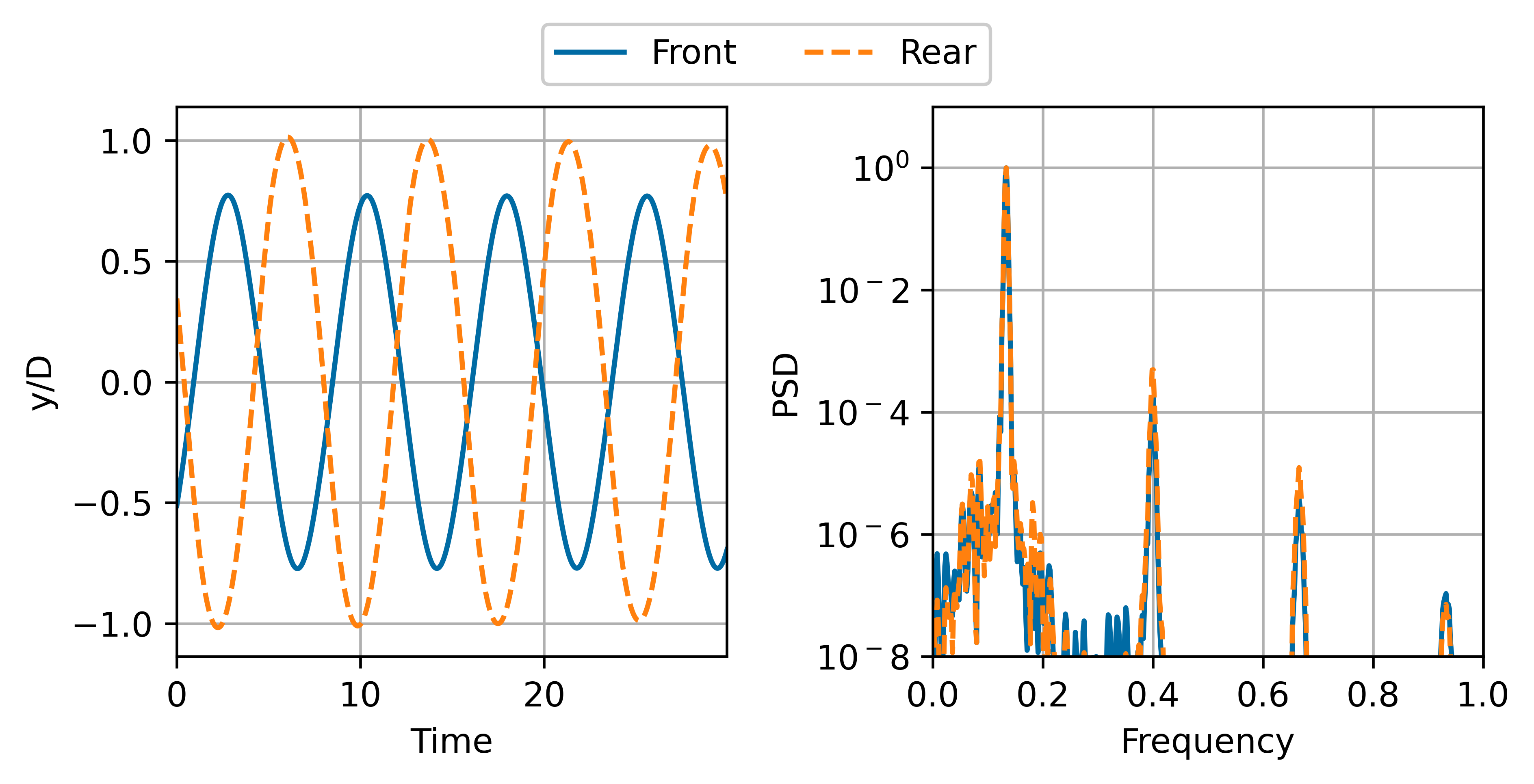}
		\caption{$U^*$=7.0 \label{fig:2cyl-fft7}}
	\end{subfigure}
	\begin{subfigure}{0.6\linewidth}
		\includegraphics[width=\textwidth]{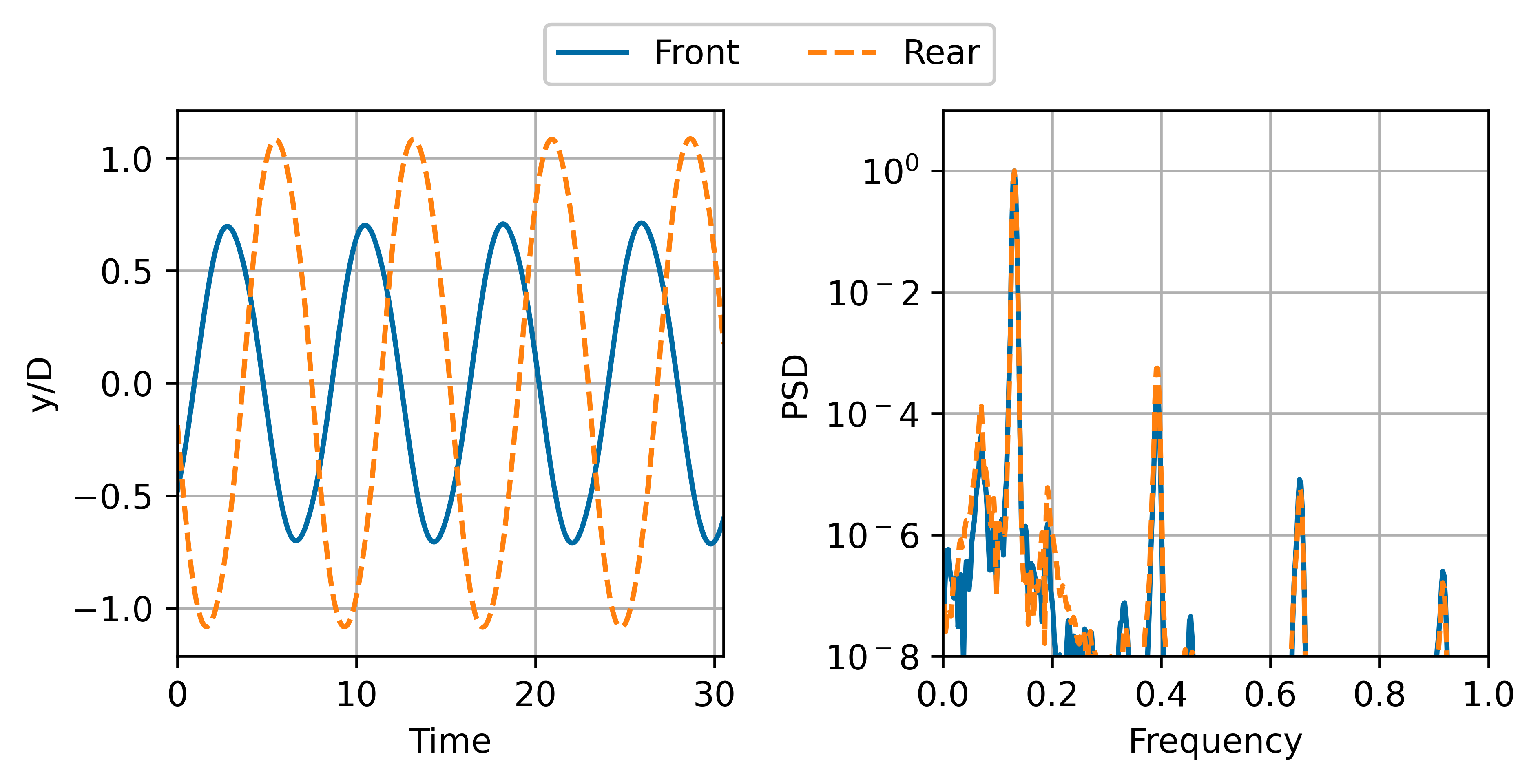}
		\caption{$U^*$=8.0 \label{fig:2cyl-fft8}}
	\end{subfigure}
	\begin{subfigure}{0.6\linewidth}
		\includegraphics[width=\textwidth]{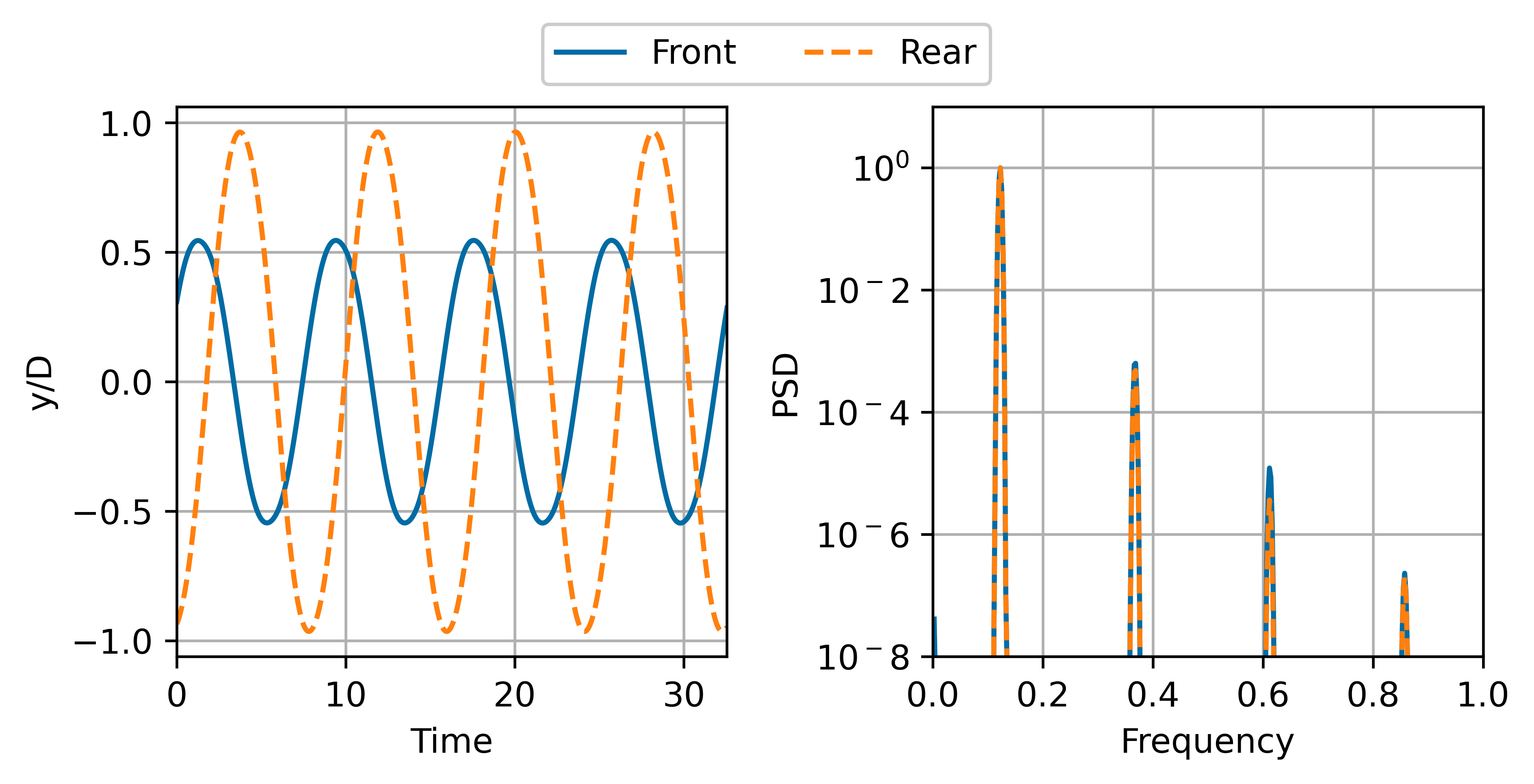}
		\caption{$U^*$=10.0\label{fig:2cyl-fft10}}
	\end{subfigure}
	\begin{subfigure}{0.6\linewidth}
		\includegraphics[width=\textwidth]{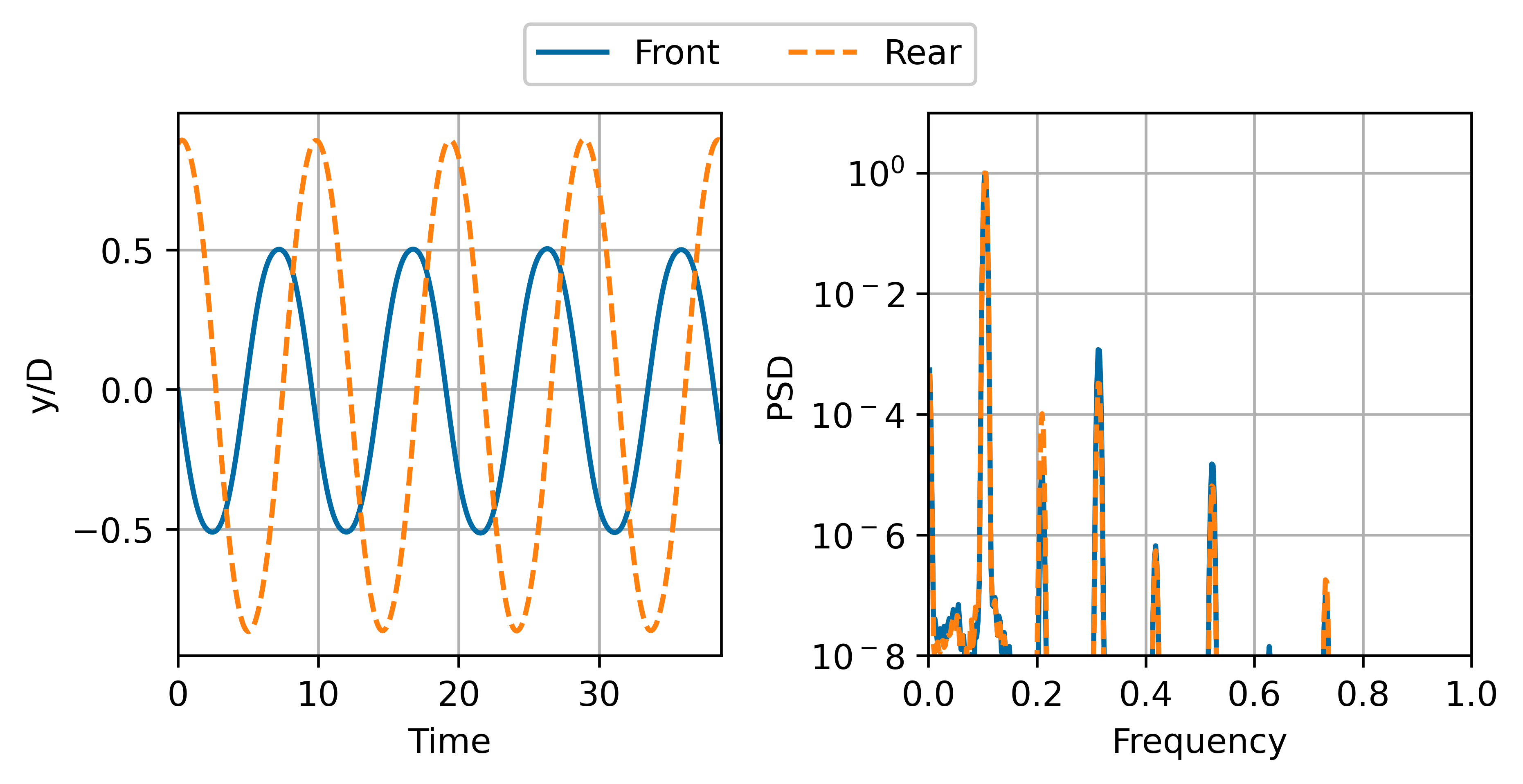}
		\caption{$U^*$=14.0 \label{fig:2cyl-fft14}}
	\end{subfigure}
\end{minipage}
	\caption{Displacements and spectra of the two cylinders for the various reduced velocities \label{fig:2cyl-fft}. For the PSD plots the blackman filter was used to avoid spectral-leakage related errors.}
\end{figure}

Finally Figure \ref{fig:2cyl-fft} presents the displacement of the two cylinders as well as the corresponding spectra. The displacement is plotted for 4 periods based on the dominant frequency. Initially for $U^*=3-4$ the displacement of the front cylinder is greater that the one of the rear. Gradually, as the reduced velocity increases the displacement of the rear cylinder becomes larger. For $U^*\ge7.0$ the amplitude of the oscillation of the rear cylinder dominates. This is in agreement with the observations made in \cite{Borazjani2009}. As \cite{Borazjani2009} and \cite{Griffith2017a} suggest  two distinct flow states can be defined. For $U^*\le 4.0$  where the larger amplitude  of the front cylinder state 1 is defined. State 2 appears when the amplitude of rear cylinder dominates ($U^*\ge 7.0$)  while for $4.0< U^* < 7.0$ a transition state occurs.
 
%
%

\section{Discussion and concluding remarks}\label{sec:conc}

The hybrid Eulerian-Lagrangian flow solver that was first introduced in \cite{Papadakis2019}  for compressible inviscid flows, has been extended to low speed laminar viscous external flows. The cases considered involve one or more cylinders in close interaction. In all cases the flow is dominated by massive separation and strong vortex shedding. A further complication was added by having two cylinders on independent elastic supports and simulating the corresponding fluid-structure interaction problem.  

The hybrid predictions were found in good agreement with test data in the case of a standing cylinder and with other predictions in all other cases. In the more complicated fluid-structure interaction cases, good agreement adds confidence in the results since the corresponding solvers are completely different. The present method uses a body-fitted grid close to solid boundaries instead of the immersed boundary technique that is used in \cite{Griffith2017b}; the hybrid solver solves the compressible equations with low Mach preconditioning instead of the pressure correction approach that the immersed boundary solvers apply; over most of the flow-field the particle method is here applied, an approach known to be less diffusive than conventional CFD \cite{CottetKoumou}. Another important finding of the present work is that the hybrid method can handle bodies in close proximity and that had no difficulty even when the separate E-grids overlap. 

\begin{figure}[H]
	\begin{minipage}{\linewidth}
		\begin{subfigure}{0.4\linewidth}
			\includegraphics[width=\textwidth]{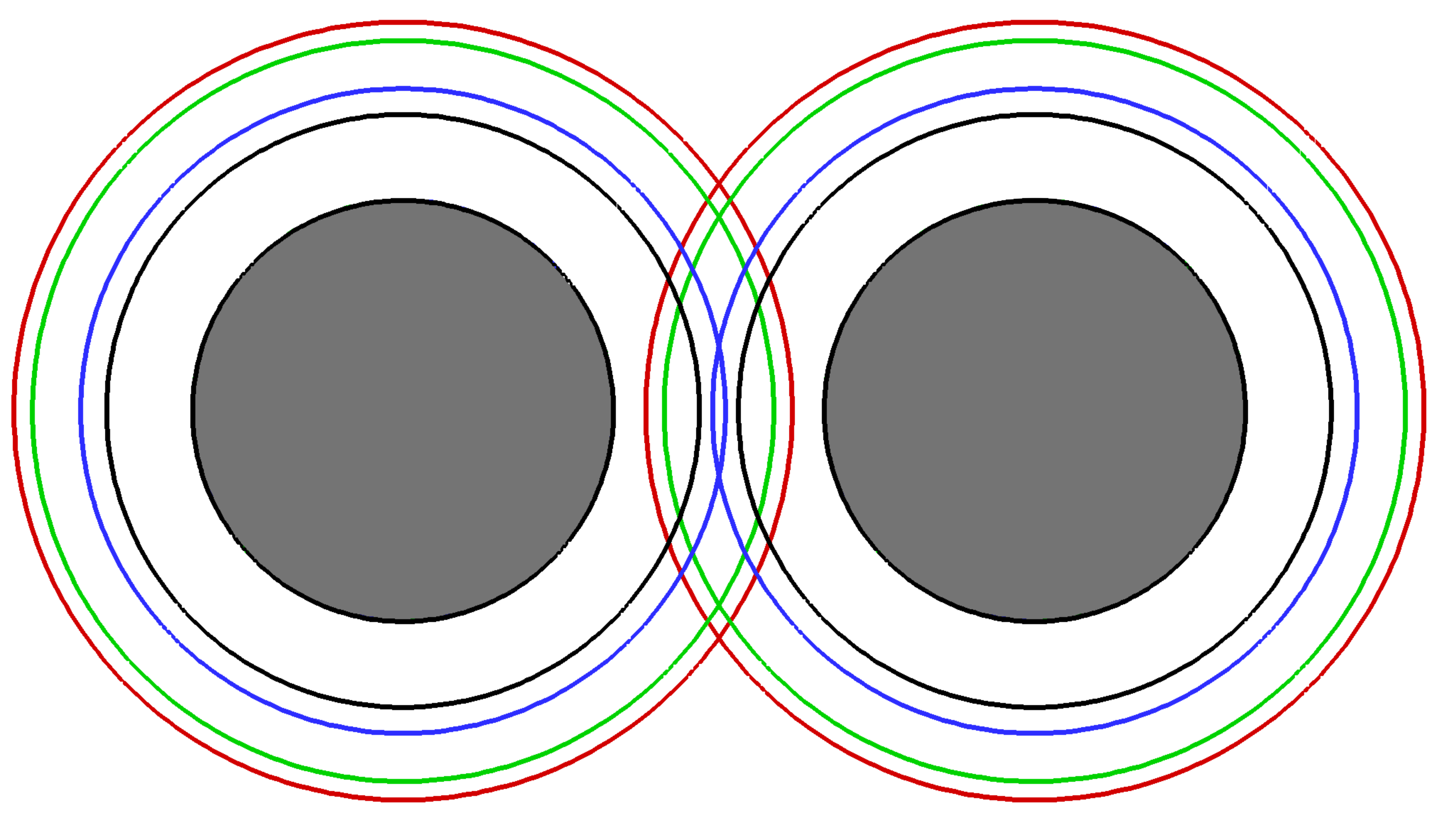}
			\caption{\label{fig:2cyl100f}}
		\end{subfigure}
		\begin{subfigure}{0.6\linewidth}
			\includegraphics[width=\textwidth]{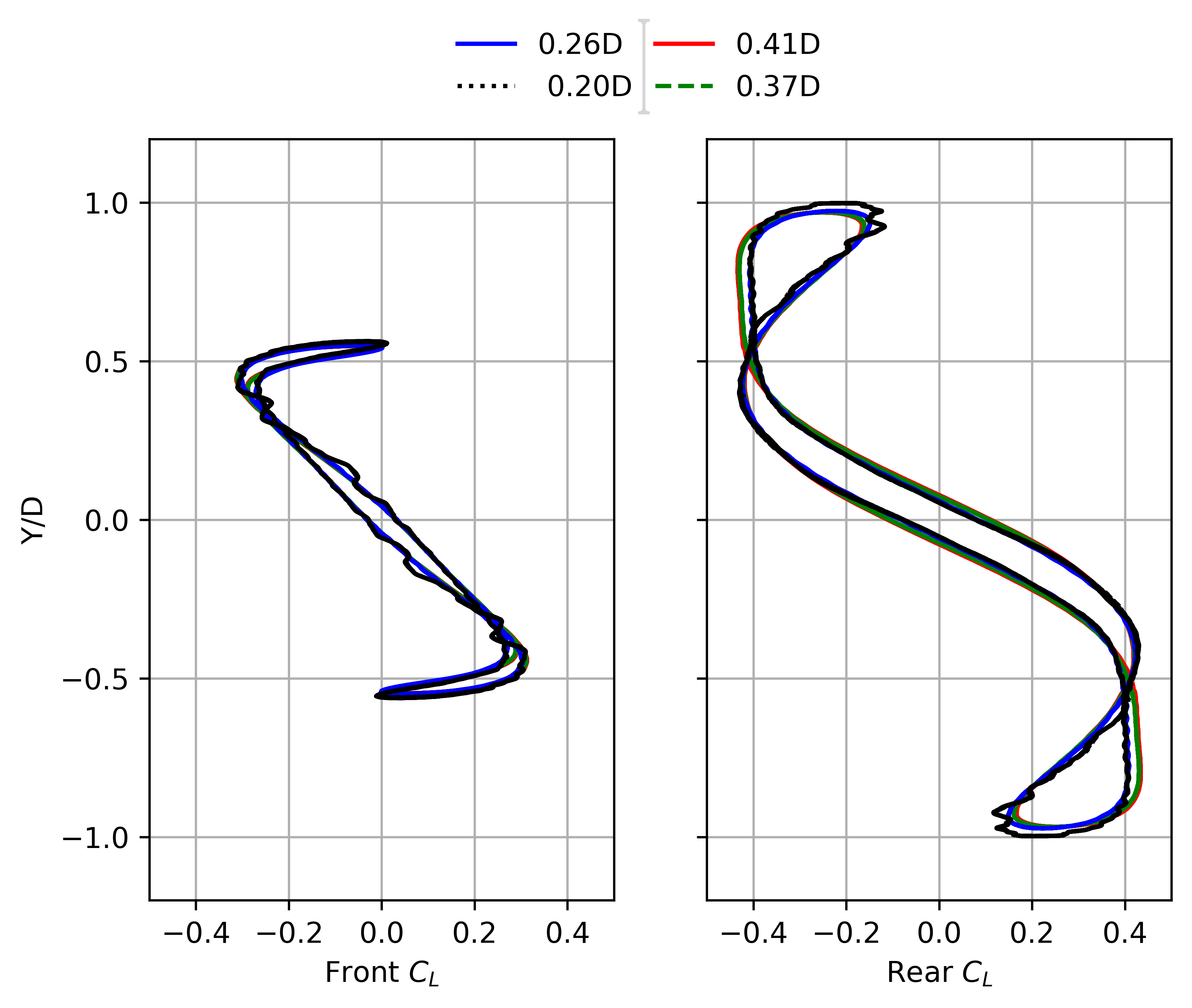}
			\caption{\label{fig:2cyl100liz}}
		\end{subfigure}
	    \caption{(Left) Four different grids are generated with varying position of the external boundary $S_E$. The furthest one (red) is located at $0.41D$ while the nearest (black) one is located at 0.20R. The green one corresponds to the grid used in the previous analysis. (Right)The Lissajous curves for $U^*=10$ for the various grids. Results using the larger grids ($0.37-0.41D$) are identical. When bringing the farfield  boundary closer ($0.20-0.25D$) numerical noise starts to appear, nevertheless, the overall comparison is considered fair.}
	\end{minipage}
\end{figure}

In order to clarify this point, four E-grids of different width were tested in the case of two oscillating cylinders at $U^*=10$. The specific reduced velocity was chosen since in this case the amplitude of the motion for both cylinders is relatively large while the solution locks in a periodic state. Furthermore  the comparison indicated negligible differences as shown in Figure \ref{fig:2cyl-liz}). The reference grid with which all previous simulations were carried out had a width of $0.37D$ around the cylinder (green line in Figure \ref{fig:2cyl100f}). Then widths of $0.41D$ (red line), $0.26D$ (blue line) and $0.20D$ (black line) were added. It's worth noticing that all the CFD domains overlap except for the smaller one. 

Results are compared in Figure \ref{fig:2cyl100liz} in terms of the amplitudes with respect to the lift coefficient ($C_L$) for the front (left) and the rear (right) cylinder. Although visually the results are in agreement, a closer look reveals some discrepancies. Starting from the smaller domain at $0.20D$ (black line) there is indication that the curve is  "polluted" by  numerical errors. As the CFD domain increases in size the (blue line) the L-solution becomes more accurate (since the distance from $S_B$ increases), the numerical errors disappear and the curve becomes smoother.  Still, there is some minor deviation from the reference solution (green line) especially for the rear cylinder in the peak amplitude area. The next two solutions, the green and red lines, that correspond to the wider grids are almost identical. 

Although the aim of this analysis was to check the behavior of the solver when there is overlapping, a point can be made regarding the lower limit of the E-grid. This is constrained by the way the method treats the boundary terms. Replacing their contribution with that of point singularities has an error inversely proportional to the distance. In \cite{Papadakis2019}, the limit was placed at $0.25$ of the airfoil chord which is close  to the $0.41D$ here chosen. 



In summary, the hybrid solver has proven to provide accurate results. This is evident on the study of the isolated cylinder at $Re=100$ where comparison is made between measurements an other computational results. Additionally for the single vibrating cylinder the current method produces similar results to the spectral element  method presented in \cite{Leontini2006}. 

The most challenging case, is the two vibrating cylinder in tandem arrangement. The relative movement of the two cylinders in very close proximity poses an additional challenge for traditional computational methods. Using they hybrid method this is handled implicitly, without any special treatment. The results suggest a qualitative and quantitative  agreement with previous work for \cite{Borazjani2008} and \cite{Griffith2017a}.

Regarding the efficiency of the hybrid solver two remarks can be made. {\color{black} The computational performance} of the method mainly depends on the solution of the Poisson equations for the scalar and vector potential. Employing an FFT-based Poisson solver can greatly reduce the computational time and indeed such techniques can render such solvers scalable \cite{caprace2021flups},\cite{McCorquodale2005}. {\color{black} Nevertheless, FFT-based Poisson solvers require constant spacing  in each  spatial dimension. This can potentially lead to a greater number of nodes as the domain size increases. On the contrary, it is common practice for  E-solvers to increase the cell sizing away from the solid boundaries in order to  reduce the overall computational cost. A way to alleviate the uniform spacing constraint from the FFT-based PM solver is to employ an adaptive grid refinement strategy as denoted in  \cite{berger1989local}.}

\appendix
\input{appendix}

\section*{Acknowledgments}
This work was supported by computational time granted from the Greek Research \& Technology Network (GRNET) in the National HPC facility - ARIS - under project  "SHIPFLOW" with ID  pr010039.


\bibliography{main}

\end{document}

%% file: appendix.tex
\appendix
\section{Convergence of the hybrid solver in space and time \label{sec:appendix}}
In order to validate the hybrid method a convergence analysis is next carried out. The case considered, concerns the diffusion of a Lamb-Osseen vortex. The specific test-case was inspired by \cite{Gillis2019} from  which the analysis procedure is also followed herein.  


The vortex is centered at $\vec x_c$ having the following initial vorticity distribution:

\begin{equation}
     \omega = \frac{\Gamma}{\pi} \frac{1}{\sigma^2 + 4 \nu t} \exp{\frac{-r^2}{\sigma^2 + 4 \nu t}}
\end{equation}
where $r=\vec x - \vec x_c$. Assuming laminar flow conditions, circumferential velocity ($u_\theta$) and pressure ($p$) assume the following analytic expressions:

\begin{align}
    u_\theta &= \frac{\Gamma}{2\pi r} \left[ 1-\exp{\frac{-r^2}{\sigma^2 + 4 \nu t }}\right]  \label{eq:apanal}\\
    p &= \rho u_\theta^2 \log(r) + p_\infty
\end{align}

As outlined in \cite{Gillis2019} the core of the Gaussian vorticity field can be cropped and replaced by a rotating cylinder of radius $R$ ($\sigma=R$). Consequently, only the flow outside the cylinder is considered. The angular velocity ($\Omega$) of the cylinder is defined so that on the  the wall of the cylinder ($r=R$)  the circumferential velocity is the same as \autoref{eq:apanal} and so:

\begin{align}
    \Omega = \frac{\Gamma}{2 \pi R^2} \left[ 1 - \exp{\frac{-R^2}{R^2 + 4  \nu t}}\right]  
\end{align}

The results correspond to $R=0.5$, $\Gamma=\pi$ and $Re=\frac{U D}{\nu}=200$ while for the convergence analysis the $L_2$ and $L_\infty$ norms are used as in \cite{Gillis2019}:


\begin{equation}
    L_2 = \frac{R}{\Gamma}\sqrt{\sum_{cells} \left(w_{anal} - w_{comp}\right)^2h^2},\,
    L_\infty =  \frac{R^2}{\Gamma} \max_{cells} |w_{anal} - w_{comp}|
\end{equation}

Both the PM solver as well as the CFD one are initialized using the analytical expressions for $t=0$. 


In order to check convergence, space refinement is carried out with respect to both solvers while the levels of refinement are given in the following table.

\begin{table}[H]
    \centering
    \begin{tabular}{c|c|c|c}
    Level of Refinement  & CFD grid cells & PM spacing & $\Delta t$\\
    \hline
    0 & 6084   & 0.08 & 0.008\\
    1 & 18960  & 0.04 & 0.002\\
    2 & 75684  & 0.02 & 0.0005\\
    3 & 306240 & 0.01 & 0.000125\\
    \end{tabular}
    \caption{Caption}
    \label{tab:agrids}
\end{table}

The distance of the first cell from the cylinder wall ($S_B$) was kept the same for all grids as well as the location of the external boundary $S_E$. The time-step for each level of the spatial refinement was selected so that the PM Fourier number remained the same, $r=\frac{\nu \Delta t}{h}=0.00625$.

\begin{figure}[H]
	\centering
	\begin{minipage}{0.45\linewidth}
		\includegraphics[width=\textwidth]{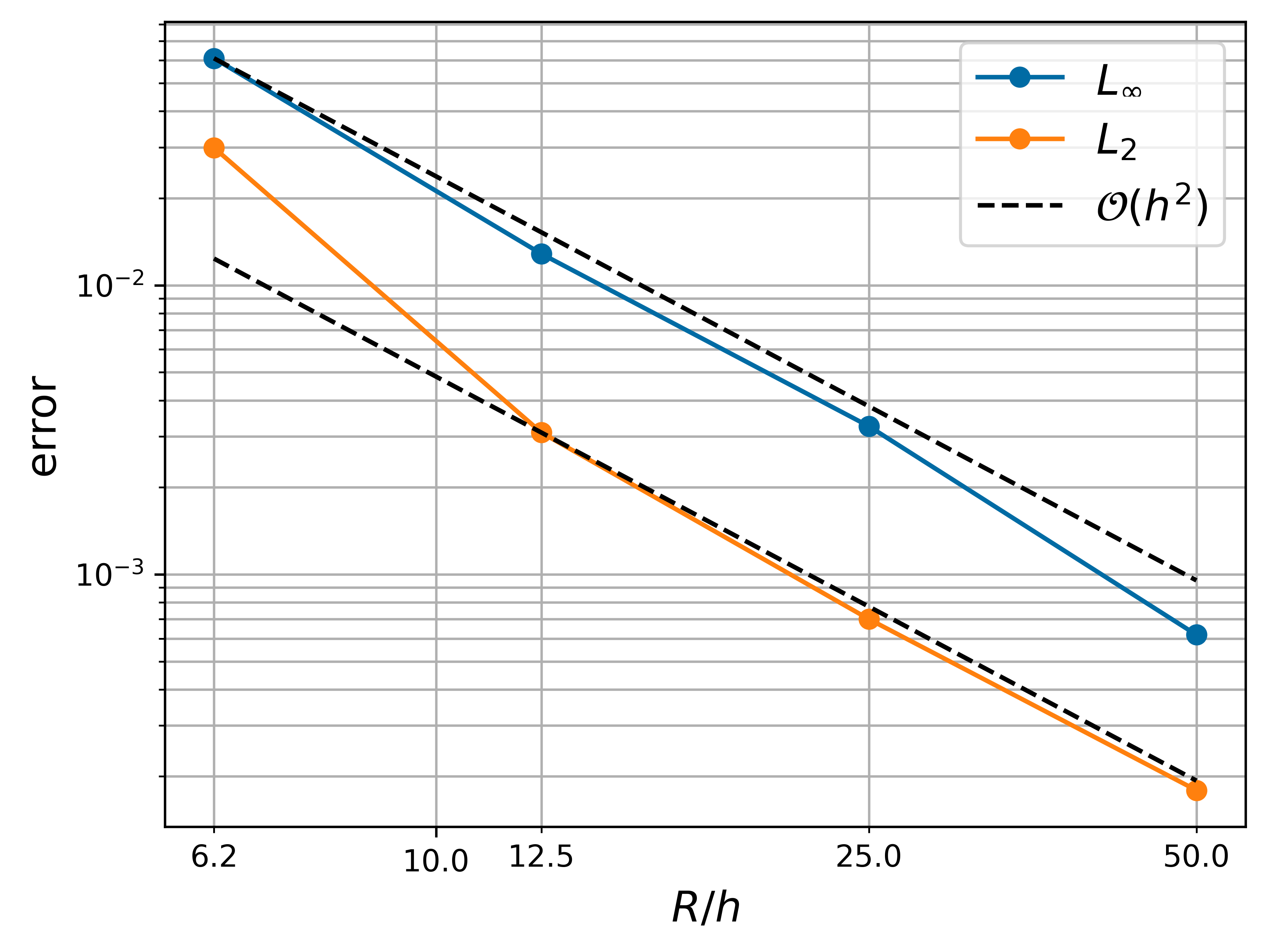}
	\end{minipage} \hfill
	\begin{minipage}{0.45\linewidth}
		\includegraphics[width=\textwidth]{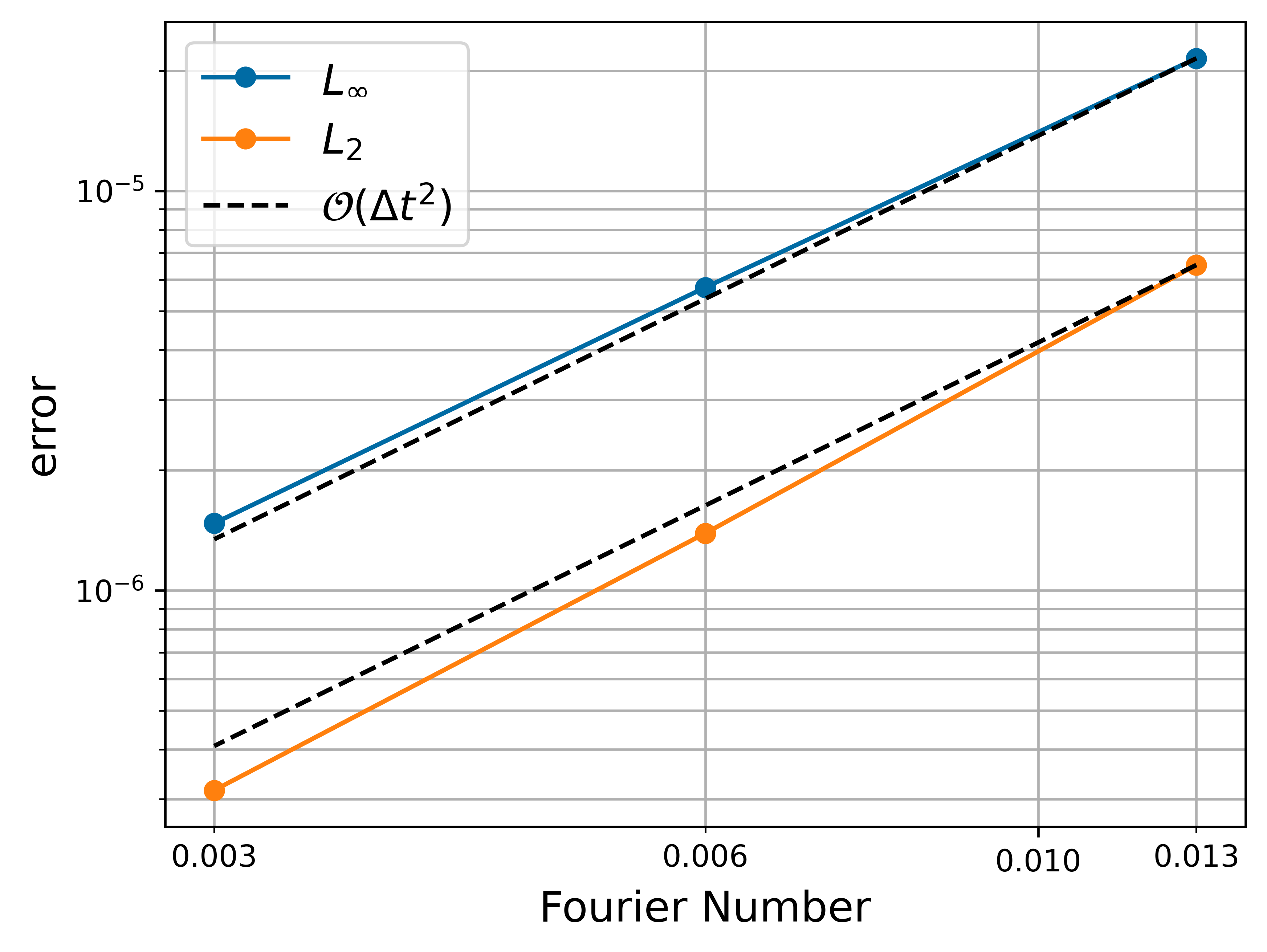}
	\end{minipage}
	\caption{Spatial (left) and temporal (right) convergence of the hybrid solver. The results indicate the t solver is $2nd$ order in time and space\label{fig:aerrors}}
\end{figure}

With respect to time convergence, a self-convergence strategy is employed, since the spatial error is much larger than the temporal one (see \cite{Gillis2019}). To this end,a simulation is conducted for a very small Fourier Number $r=0.0003$, which is used as the reference solution. It is stressed, here that since the particles carry mass, pressure and dilatation apart from vorticity, the time-step restrictions are stricter with respect to \cite{Gillis2019} and consequently  smaller time-steps are required. For the temporal study, the 1st level of spatial  refinement is used($h=0.04$). The temporal convergence can be seen in Figure \ref{fig:aerrors}-right).  

Results for the spatial and temporal convergence can be found in Figure \ref{fig:aerrors} and Table \ref{tab:aconv}.

Detailed results regarding the $L_\infty$ and $L_2$ error local convergence rates ($r$) (see \cite{Gillis2019}) are presented below.

\begin{table}[H]
    \centering
    \begin{tabular}{c|c|c|c|c|c|}
    \hline
     R/h & $h$ &$L_\infty$ & $r_\infty$ & $L_2$ & $r_2$\\
    \hline
     6.25 & 0.08 & 6.1004e-02  &   -     & 2.992e-02 &  -        \\
     12.5 & 0.04 & 1.2848e-02  & 2.24735 & 3.093e-03 & 3.27413   \\
     25   & 0.02 & 3.2533e-03  & 1.98156 & 7.000e-04 & 2.14373   \\
     50   & 0.01 & 6.1786e-04  & 2.39656 & 1.791e-04 & 1.96648   \\
    \hline
     Fourier N. ($r$) & $\Delta_t$ &$L_\infty$ & $r_\infty$ & $L_2$ & $r_2$\\
    \hline
    0.01250 & 0.01250 & 2.148e-05 & -    & 6.526e-06  & -  \\
    0.00625 & 0.00625 & 5.734e-06 & 1.90 & 1.390e-06  & 2.23 \\
    0.00313 & 0.00313 & 1.472e-06 & 1.96 & 3.160e-07  & 2.14
     \end{tabular}
    \caption{Caption}
    \label{tab:aconv}
\end{table}
In addition to the error indicators between the analytical and the numerical solution, it is important to compare the Particle Mesh and CFD solutions especially near $S_E$ where the E-solver boundary conditions are specified in terms of density, velocity and pressure. The two solutions (PM and CFD) have been extracted  at 3 radial positions ($S_B,R=0.65,S_E$) and compared to the analytical solution. The corresponding results are presented in Table \ref{tab:atwosol}.

\begin{table}[h]
    \centering
    \begin{tabular}{c|c|c|c|c}
          Radial Position & \multicolumn{2}{c|}{error $u_\theta$ \%} &\multicolumn{2}{c}{error $p$ \%} \\
         \hline
          & CFD & PM & CFD & PM \\
          \hline
          $R=0.5(S_B)$  & 7e-04    & 47.3  & 0.033    & 0.18     \\  
          $R=0.65$      & 1.6e-02  & 0.29  & 0.015    & 0.014    \\  
          $R=0.88(S_E)$ & 2.1e-02  & 0.12  & 0.012    & 0.012     
    \end{tabular}
    \caption{Caption}
    \label{tab:atwosol}
\end{table}

Over $S_B$ the CFD error is, as expected, very low while that of the PM solver is high especially with respect to the velocity which derives from the fact that the Lagrangian solver does not  accurately enforce the wall boundary condition. However, as the distance from the wall increases, the quality of the PM solution rapidly improves arriving at very low errors over $S_E$. It is worth noticing that even at the intermediate distance the  agreement is very good.

%% file: main.bbl
\begin{thebibliography}{10}
\expandafter\ifx\csname url\endcsname\relax
  \def\url#1{\texttt{#1}}\fi
\expandafter\ifx\csname urlprefix\endcsname\relax\def\urlprefix{URL }\fi
\expandafter\ifx\csname href\endcsname\relax
  \def\href#1#2{#2} \def\path#1{#1}\fi

\bibitem{Borazjani2008}
I.~Borazjani, L.~Ge, F.~Sotiropoulos, {Curvilinear immersed boundary method for
  simulating fluid structure interaction with complex 3D rigid bodies}, Journal
  of Computational Physics 227~(16) (2008) 7587--7620.
\newblock \href {https://doi.org/10.1016/j.jcp.2008.04.028}
  {\path{doi:10.1016/j.jcp.2008.04.028}}.

\bibitem{Papadakis2019}
G.~Papadakis, S.~G. Voutsinas, {A strongly coupled Eulerian Lagrangian method
  verified in 2D external compressible flows}, Computers and Fluids 195 (2019)
  104325.
\newblock \href {https://doi.org/10.1016/j.compfluid.2019.104325}
  {\path{doi:10.1016/j.compfluid.2019.104325}}.

\bibitem{Chorin1967}
A.~J. Chorin, {A Numerical Method for Solving Incompressible Visous Flow
  Problems}, in: Journal of Computational Physics, 1967, pp. 12--26.
\newblock \href {https://doi.org/10.1109/ET2ECN.2012.6470106}
  {\path{doi:10.1109/ET2ECN.2012.6470106}}.

\bibitem{govardhan2000modes}
R.~Govardhan, C.~Williamson, Modes of vortex formation and frequency response
  of a freely vibrating cylinder, Journal of Fluid Mechanics 420 (2000)
  85--130.

\bibitem{khalak1999motions}
A.~Khalak, C.~H. Williamson, Motions, forces and mode transitions in
  vortex-induced vibrations at low mass-damping, Journal of fluids and
  Structures 13~(7-8) (1999) 813--851.

\bibitem{Leontini2006}
J.~S. Leontini, M.~C. Thompson, K.~Hourigan, {The beginning of branching
  behaviour of vortex-induced vibration during two-dimensional flow}, Journal
  of Fluids and Structures 22~(6-7) (2006) 857--864.
\newblock \href {https://doi.org/10.1016/j.jfluidstructs.2006.04.003}
  {\path{doi:10.1016/j.jfluidstructs.2006.04.003}}.

\bibitem{Borazjani2009}
I.~Borazjani, F.~Sotiropoulos, {Vortex-induced vibrations of two cylinders in
  tandem arrangement in the proximity - Wake interference region}, Journal of
  Fluid Mechanics 621 (2009) 321--364.
\newblock \href {http://arxiv.org/abs/NIHMS150003} {\path{arXiv:NIHMS150003}},
  \href {https://doi.org/10.1017/S0022112008004850}
  {\path{doi:10.1017/S0022112008004850}}.

\bibitem{Griffith2017b}
M.~D. Griffith, J.~S. Leontini, {Sharp interface immersed boundary methods and
  their application to vortex-induced vibration of a cylinder}, Journal of
  Fluids and Structures 72 (2017) 38--58.
\newblock \href {https://doi.org/10.1016/j.jfluidstructs.2017.04.008}
  {\path{doi:10.1016/j.jfluidstructs.2017.04.008}}.

\bibitem{Griffith2017a}
M.~D. Griffith, D.~{Lo Jacono}, J.~Sheridan, J.~S. Leontini, {Flow-induced
  vibration of two cylinders in tandem and staggered arrangements}, Journal of
  Fluid Mechanics 833 (2017) 98--130.
\newblock \href {https://doi.org/10.1017/jfm.2017.673}
  {\path{doi:10.1017/jfm.2017.673}}.

\bibitem{chorin1973discretization}
A.~J. Chorin, P.~S. Bernard, Discretization of a vortex sheet, with an example
  of roll-up, Journal of Computational Physics 13~(3) (1973) 423--429.

\bibitem{koumoutsakos1995high}
P.~Koumoutsakos, A.~Leonard, High-resolution simulations of the flow around an
  impulsively started cylinder using vortex methods, Journal of Fluid Mechanics
  296 (1995) 1--38.

\bibitem{degond1989weighted}
P.~Degond, S.~Mas-Gallic, The weighted particle method for convection-diffusion
  equations. i. the case of an isotropic viscosity, Mathematics of computation
  53~(188) (1989) 485--507.

\bibitem{slaouti1992flow}
A.~Slaouti, P.~Stansby, Flow around two circular cylinders by the random-vortex
  method, Journal of Fluids and Structures 6~(6) (1992) 641--670.

\bibitem{Gillis2019}
T.~Gillis, Y.~Marichal, G.~Winckelmans, P.~Chatelain, {A 2D immersed interface
  vortex particle-mesh method}, Journal of Computational Physics~(May) (2019).
\newblock \href {https://doi.org/10.1016/j.jcp.2019.05.033}
  {\path{doi:10.1016/j.jcp.2019.05.033}}.

\bibitem{Eldredge2002b}
J.~D. Eldredge, T.~Colonius, A.~Leonard, {A Vortex Particle Method for
  Two-Dimensional Compressible Flow}, Journal of Computational Physics 179~(2)
  (2002) 371--399.
\newblock \href {https://doi.org/10.1006/jcph.2002.7060}
  {\path{doi:10.1006/jcph.2002.7060}}.

\bibitem{Eldredge2002}
J.~D. Eldredge, A.~Leonard, T.~Colonius, {A General Deterministic Treatment of
  Derivatives in Particle Methods }, Journal of Computational Physics 180~(2)
  (2002) 686--709.
\newblock \href {https://doi.org/10.1006/jcph.2002.7112}
  {\path{doi:10.1006/jcph.2002.7112}}.

\bibitem{Batchelor}
G.~Batchelor, {An Introduction to Fluid Mechanics}, Cambridge University Press,
  UK, 1967.

\bibitem{Chatelain2016a}
P.~Chatelain, M.~Duponcheel, D.-G. Caprace, Y.~Marichal, G.~Winckelmans,
  {Vortex Particle-Mesh simulations of Vertical Axis Wind Turbine flows: from
  the blade aerodynamics to the very far wake}, Journal of Physics: Conference
  Series 753 (2016) 032007.
\newblock \href {https://doi.org/10.1088/1742-6596/753/3/032007}
  {\path{doi:10.1088/1742-6596/753/3/032007}}.

\bibitem{Parmentier2018}
P.~Parmentier, G.~Winckelmans, P.~Chatelain, {A Vortex Particle-Mesh method for
  subsonic compressible flows}, Journal of Computational Physics 354 (2018)
  692--716.
\newblock \href {https://doi.org/10.1016/j.jcp.2017.10.040}
  {\path{doi:10.1016/j.jcp.2017.10.040}}.

\bibitem{ploumhans2002vortex}
P.~Ploumhans, G.~Winckelmans, J.~K. Salmon, A.~Leonard, M.~Warren, Vortex
  methods for direct numerical simulation of three-dimensional bluff body
  flows: application to the sphere at re= 300, 500, and 1000, Journal of
  Computational Physics 178~(2) (2002) 427--463.

\bibitem{caprace2021flups}
D.-G. Caprace, T.~Gillis, P.~Chatelain, Flups: A fourier-based library of
  unbounded poisson solvers, SIAM Journal on Scientific Computing 43~(1) (2021)
  C31--C60.

\bibitem{monaghan2005smoothed}
J.~J. Monaghan, Smoothed particle hydrodynamics, Reports on progress in physics
  68~(8) (2005) 1703.

\bibitem{CottetKoumou}
G.-H. Cottet, P.~Koumoutsakos, {Vortex methods: Theory and Practice}, Cambridge
  University Press, 2000.

\bibitem{Papadakis2014e}
G.~Papadakis, {Development of a hybrid compressible vortex particle method and
  application to external problems including helicopter flows}, Ph.D. thesis
  (2014).

\bibitem{Manolas2014}
D.~I. Manolas, V.~A. Riziotis, S.~G. Voutsinas, {Assessing the Importance of
  Geometric Nonlinear Effects in the Prediction of Wind Turbine Blade Loads},
  Journal of Computational and Nonlinear Dynamics 10~(4) (2014) 041008.
\newblock \href {https://doi.org/10.1115/1.4027684}
  {\path{doi:10.1115/1.4027684}}.

\bibitem{Sharman2005}
B.~Sharman, F.~S. Lien, L.~Davidson, C.~Norberg, {Numerical predictions of low
  Reynolds number flows over two tandem circular cylinders}, International
  Journal for Numerical Methods in Fluids 47~(5) (2005) 423--447.
\newblock \href {https://doi.org/10.1002/fld.812} {\path{doi:10.1002/fld.812}}.

\bibitem{Park1998}
J.~Park, K.~Kwon, H.~Choi, {Numerical solutions of flow past a circular
  cylinder at Reynolds numbers up to 160}, KSME International Journal 12~(6)
  (1998) 1200--1205.
\newblock \href {https://doi.org/10.1007/BF02942594}
  {\path{doi:10.1007/BF02942594}}.

\bibitem{Stalberg2006}
E.~St{\aa}lberg, A.~Br{\"{u}}ger, P.~L{\"{o}}tstedt, A.~V. Johansson, D.~S.
  Henningson, {High order accurate solution of flow past a circular cylinder},
  Journal of Scientific Computing 27~(1-3) (2006) 431--441.
\newblock \href {https://doi.org/10.1007/s10915-005-9043-y}
  {\path{doi:10.1007/s10915-005-9043-y}}.

\bibitem{Qu2013}
L.~Qu, C.~Norberg, L.~Davidson, S.-H. Peng, F.~Wang, {Quantitative numerical
  analysis of flow past a circular cylinder at Reynolds number between 50 and
  200}, Journal of Fluids and Structures 39 (2013) 347--370.
\newblock \href {https://doi.org/10.1016/j.jfluidstructs.2013.02.007}
  {\path{doi:10.1016/j.jfluidstructs.2013.02.007}}.

\bibitem{Posdziech2007}
O.~Posdziech, R.~Grundmann, {A systematic approach to the numerical calculation
  of fundamental quantities of the two-dimensional flow over a circular
  cylinder}, Journal of Fluids and Structures 23~(3) (2007) 479--499.
\newblock \href {https://doi.org/10.1016/j.jfluidstructs.2006.09.004}
  {\path{doi:10.1016/j.jfluidstructs.2006.09.004}}.

\bibitem{Williamson1996}
C.~H.~K. Williamson, {Vortex Dynamics in the Cylinder Wake}, Annu. Rev. Fluid.
  Mech (1996) 28--477\href
  {https://doi.org/10.1146/annurev.fluid.36.050802.122128}
  {\path{doi:10.1146/annurev.fluid.36.050802.122128}}.

\bibitem{McCorquodale2005}
P.~McCorquodale, P.~Colella, G.~T. Balls, S.~B. Baden, {A scalable parallel
  poisson solver in three dimensions with infinite-domain boundary conditions},
  Proceedings of the International Conference on Parallel Processing Workshops
  2005 (2005) 163--172.
\newblock \href {https://doi.org/10.1109/ICPPW.2005.17}
  {\path{doi:10.1109/ICPPW.2005.17}}.

\bibitem{berger1989local}
M.~J. Berger, P.~Colella, Local adaptive mesh refinement for shock
  hydrodynamics, Journal of computational Physics 82~(1) (1989) 64--84.

\end{thebibliography}
